\documentclass[useAMS,usenatbib]{mn2e}

\usepackage{graphicx}
\usepackage{threeparttable}
\usepackage{epsfig}
\usepackage{times}
\usepackage{natbib}
\usepackage{amsfonts}
\usepackage[usenames,dvipsnames]{color}
\usepackage{amsmath}
\usepackage{bbold}
\usepackage{lineno}

\onecolumn

\newcommand{\be}{\begin{equation}}
\newcommand{\ee}{\end{equation}}
\newcommand{\bea}{\begin{eqnarray}}
\newcommand{\eea}{\end{eqnarray}}
\def\beq{\begin{equation}}
\def\eeq{\end{equation}}

\newcommand{\ps}{{{\rm ps}}}

\newcommand{\sky}{{\rm{sky}}}

\newcommand{\fnl}{$f_{\rm nl~}$}
\newcommand{\sn}{{\rm sn}}
\newcommand{\gal}{{\rm gal}}
\newcommand{\Planck}{{\it Planck}~}

\newcommand{\fg}{{\mathtt{fg}}}
\newcommand{\CIB}{{\mathtt{CIB}}}
\newcommand{\CMB}{{\mathtt{CMB}}}
\newcommand{\Radio}{{\mathtt{Radio}}}
\newcommand{\Lens}{{\mathtt{Lens}}}
\newcommand{\LensISW}{{\mathtt{ISW-Lens}}}
\newcommand{\LensCIB}{{\mathtt{CIB-Lens}}}
\newcommand{\LensRadio}{{\mathtt{Radio-Lens}}}
\newcommand{\raw}{{\mathtt{raw}}}
\newcommand{\clean}{{\mathtt{clean}}}
\newcommand{\obsv}{{\mathtt{obs}}}
\newcommand{\prim}{{\mathtt{prim}}}
\newcommand{\comb}{{\mathtt{comb}}}
\newcommand{\noise}{{\mathtt{noise}}}
\newcommand{\targ}{{\mathtt{targ}}}

\newcommand{\hn}{\hat{\mathbf n}}
\newcommand{\hk}{\hat{\mathbf k}}
\newcommand{\vk}{{\mathbf k}}

\newcommand{\aj}{{AJ}}
\newcommand{\apj}{{ApJ}}
\newcommand{\apjs}{{ApJS}}

\newcommand{\mnras}{{MNRAS}}
\newcommand{\aap}{{A\&A}}
\newcommand{\physrep}{{Phys. Rep.}}
\newcommand{\jcap}{{J. Cosmol. Astropart. Phys.}}
\newcommand{\advas}{{Adv. Astron.}}
\newcommand{\nar}{{New Astron. Rev.}}

\begin{document}
\title[CIB-lensing bispectrum] 
{The CIB-lensing bispectrum: impact on primordial non-Gaussianity and detectability for the \Planck mission} 
\author[A. Curto et al.]{A. Curto$^{1,}$$^{2,}$$^{3}$
  \thanks{E-mail:curto@ifca.unican.es,acurto@mrao.cam.ac.uk} M. Tucci$^{4}$
 M. Kunz$^{4,}$$^{5}$ E. Mart\'{\i}nez-Gonz\'alez$^1$
\\
$^{1}$Instituto de F\'isica de Cantabria, CSIC-Universidad de Cantabria, Avda. de los Castros s/n, 39005 Santander, Spain\\
$^{2}$Astrophysics Group, Cavendish Laboratory, Madingley Road, Cambridge CB3 0H3, U.K.\\
$^{3}$Kavli Institute for Cosmology Cambridge, Madingley Road, Cambridge, CB3 0HA, U.K.\\
$^{4}$D\'epartement de Physique Th\'eorique and Center for Astroparticle Physics, Universit\'e de Gen\`eve,~~24 quai Ansermet, CH--1211 Gen\`eve 4, Switzerland\\
$^{5}$African Institute for Mathematical Sciences, 6 Melrose Road, Muizenberg, 7945, South Africa
}
%
\date{\today}
\pagerange{\pageref{firstpage}--\pageref{lastpage}} \pubyear{2014}
\label{firstpage}
\maketitle
\begin{abstract}
  We characterize the Cosmic Infrared Background (CIB)--lensing
  bispectrum which is one of the contributions to the three-point
  functions of Cosmic Microwave Background (CMB) maps in harmonic
  space. We show that the CIB--lensing bispectrum has a considerable
  strength and that it can be detected with high significance in the
  \Planck high--frequency maps. We also present forecasts of the
  contamination on different shapes of the primordial non-Gaussianity
  \fnl parameter produced by the CIB--lensing bispectrum and by the
  extragalactic point sources bispectrum in the \Planck
  high--resolution CMB anisotropy maps. The local, equilateral and
  orthogonal shapes are considered for 'raw' single--frequency (i.e.,
  without applying any component separation technique) and
  foreground--reduced \Planck temperature maps. The CIB--lensing
  correlation seems to mainly affect orthogonal shapes of the
  bispectrum -- with $ \Delta f_{\rm nl}^{\rm (ort)} =-21$ and $ -88$
  for the 143 and 217\,GHz bands respectively -- while point sources
  mostly impact equilateral shapes, with $\Delta f_{\rm nl}^{\rm (eq)}
  =160, 54$ and 60 at 100, 143 and 217\,GHz.  However, the results
  indicate that these contaminants do not induce any relevant bias on
  \Planck \fnl estimates when foreground--reduced maps are considered:
  using SEVEM for the component separation we obtain $ \Delta f_{\rm
    nl}^{\rm (ort)} =10.5$ due to the CIB--lensing and $ \Delta f_{\rm
    nl}^{\rm (eq)}=30.4$ due to point sources, corresponding to $
  0.3\sigma$ and $ 0.45\sigma$ in terms of the \Planck {2013} $f_{\rm
    nl}$ uncertainty.  The component separation technique is, in fact,
  able to partially clean the extragalactic source contamination and
  the bias is reduced for all the shapes. We have further developed
  single- and multiple-frequency estimators based on the Komatsu,
  Spergel \& Wandelt (2005) formalism that can be implemented to
  efficiently detect this signal.
\end{abstract}

\begin{keywords}
methods: data analysis -- cosmic microwave background -- extragalactic points sources -- radio and far--IR: galaxies
\end{keywords}

%
\section{Introduction}
\label{section1_introduction}
Primordial non--Gaussianity (NG) in the cosmic microwave background
(CMB) radiation has emerged as one of the key tests for the physics of
the early Universe, as different models of e.g.\ inflation predict
slightly different deviations from Gaussian primordial fluctuations
\citep[see
e.g.][]{bartolo2004,bartolo2010,yadav2010,liguori2010,martinez2012}. The
latest constraints by the {\it Planck\footnote{{\it Planck}
    (http://www.esa.int/Planck) is a project of the European Space
    Agency --ESA-- with instruments provided by two scientific
    consortia funded by ESA member states with contributions from
    NASA.}}  satellite put strong constraints on the amount of
primordial NG that is present in the data
\citep{planck_xxiv_2013}. But the precision of the {\it Planck} data
requires great care concerning the subtraction of astrophysical
contributions to the observed CMB anisotropies (so--called
foregrounds). It is important to check all possible contributions for
their expected level of contamination of the primordial NG estimate
both on sky maps and on foreground-cleaned maps.  At least the
non-negligible foreground contributions should then be estimated
jointly with the primordial ones, which requires the construction of
an estimator also for the foregrounds.

Conventionally the CMB anisotropies $\Delta T(\hn)$, being a real-valued random field
on the sky sphere, are expanded in spherical harmonics,
\be
\Delta T(\hn) = \sum_{\ell m} a_{\ell m} Y_{\ell m}(\hn) 
\ee
and schematically we can write the coefficients $a_{\ell m}$ as a superposition of different
contributions,
\be
a_{\ell m} = \tilde{a}^{(\CMB)}_{\ell m} + a^{(\fg)}_{\ell m} + n_{\ell m} \, . \label{eq:contributions}
\ee
Here the first term on the right hand side $\tilde{a}_{\ell m}$ is the
primordial contribution, lensed by the intervening large--scale
structure. The second term is the contribution due to foregrounds --
in general there are both Galactic and extragalactic contributions,
but in this paper we will neglect the former and use the term
`foreground' to denote the extragalactic contribution only. For the
extragalactic foreground radiation we expect that radio sources
dominate at low frequencies, and dusty star--forming galaxies creating
the cosmic infrared background (CIB) at high frequencies. The final
contribution is instrumental noise, obviously uncorrelated with the
CMB and the foreground contributions, that we assume to be Gaussian.

The main tool to study primordial NG is the angular bispectrum, the
three--point function of the $a_{\ell m}$,
\beq
\langle a_{\ell_1 m_1} a_{\ell_2 m_2} a_{\ell_3 m_3} \rangle = G_{\ell_1 \ell_2 \ell_3}^{m_1 m_2 m_3} b_{\ell_1 \ell_2 \ell_3} \, ,
\label{bispectrum_three_alm}
\eeq
where the Gaunt integral \citep{komatsu2001}
\be
G_{\ell_1 \ell_2 \ell_3}^{m_1 m_2 m_3} = \int d^2\hn Y_{\ell_1 m_1}(\hn) Y_{\ell_2 m_2}(\hn) Y_{\ell_3 m_3}(\hn)
\label{eq:gaunt_integral}
\ee
takes care of rotational symmetry, and where the non-trivial
contribution to the three-point function is encoded in the reduced
bispectrum $b_{\ell_1 \ell_2 \ell_3}$.

The product of three $a_{\ell m}$ as written in Eq.\
(\ref{eq:contributions}) will not only contain a primordial
contribution. In addition there are additional elements that involve
non-primordial terms, some of them already studied in previous works,
such as the `foreground' contribution given schematically by $\langle
a^{{(\fg)} 3}_{\ell m} \rangle$ \citep[see
e.g.][]{lacasa2013,penin2013}, a contribution from the correlation
between the lensing of the CMB and the integrated Sachs--Wolfe (ISW)
effect contained in the $\langle \tilde{a}^{(\CMB) 3}_{\ell m}
\rangle$ term \citep[see e.g.][]{Mangilli:2013sxa} and finally a
contribution from the correlation between the lensing of the CMB and
extragalactic foregrounds. This article is focusing on the last
correlation, already detected with a significance of 42$\sigma$ by
\cite{planck_xviii_2013} considering statistical errors only
(19$\sigma$ when systematics are included). The CIB--lensing
correlation arises as the large--scale structure (LSS) both lenses the
CMB and emits the `foreground' (radio or CIB) radiation. The main
contribution here is expected due to the CIB--lensing correlation, as
in radio galaxies the clustering signal is highly diluted by the
broadness of their luminosity function and of their redshift
distribution \citep[e.g.,][]{tof05}.

In this paper we focus on the CIB--lensing bispectrum, for two
reasons.  Firstly, in order to ensure that the CMB constraints on
primordial NG are accurate, we need to check that the additional
contributions are under control. In \cite{planck_xxiv_2013} the
ISW--lensing contribution was fit simultaneously with the primordial
contribution, and was shown to be small. \cite{lacasa2012a} and
\cite{curto2013} studied the bispectrum of unresolved point sources
and concluded that it is small enough to neglect. However, the
situation of the CIB--lensing contribution was so far not investigated
in detail, and this paper aims to close this gap. Secondly, probing
the non--Gaussianity due to the large--scale structure is not only
important for assessing the contamination of the primordial NG, but is
also interesting in its own right. The LSS contains important
information on the late--time evolution and content of the universe,
as well as on the formation and evolution of galaxies. Studying the
CIB--lensing correlation is thus not only important to assess the
reliability of the constraints on primordial $f_{\rm nl}$, but also
potentially useful for cosmology and astrophysics.

The outline of the paper is as follows: in Section \ref{section2} we
discuss the CIB--lensing contribution as well as the CIB model that we
will use. In Section \ref{section3_delta_fnl} we estimate the bias on
\fnl for local, equilateral and orthogonal configurations. We perform
the calculation both for raw frequency maps and for linear
combinations of maps that remove most of the astrophysical
foregrounds, following the SEVEM component separation method used by
the \Planck collaboration \citep[see, e.g.][]{planck_xii_2013}. We
then construct in Sections \ref{section:detectability} and
\ref{section5} an optimal estimator for measuring the CIB--lensing
bispectrum and assess the level at which we expect to be able to
detect it with {\it Planck} data before presenting our conclusions in
Section \ref{section6}. In the four attached appendices we provide
more details on our model of CIB anisotropies power spectra
(including a new parameter fit of the {\it Planck} measurement of
  CIB power spectra) and on the calculation of power spectra and
bispectra.

%
\section{Modelling the CIB and CIB--lensing power spectra}
\label{section2}
\begin{figure*}
\centering
\includegraphics[width=80mm]{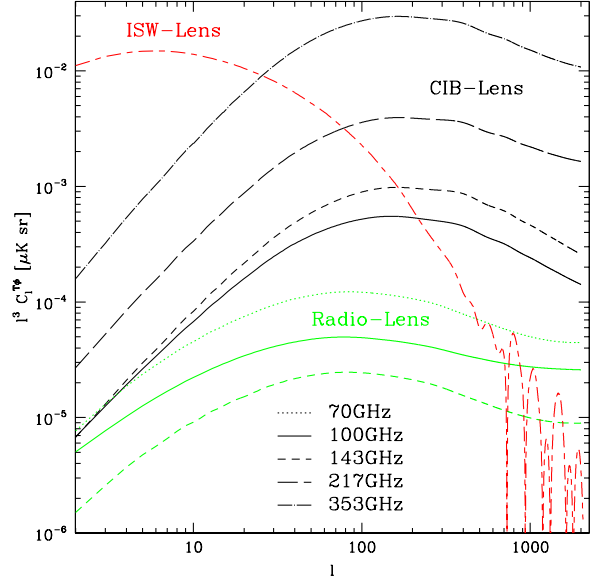}
\caption{Predicted power spectra of the lensing correlation with ISW (red
  curve), CIB (black curves) and radio sources (green curves) at the
  \Planck frequencies relevant for cosmological analysis.}
\label{f1}
\end{figure*}
\begin{figure}
\includegraphics[width=90mm]{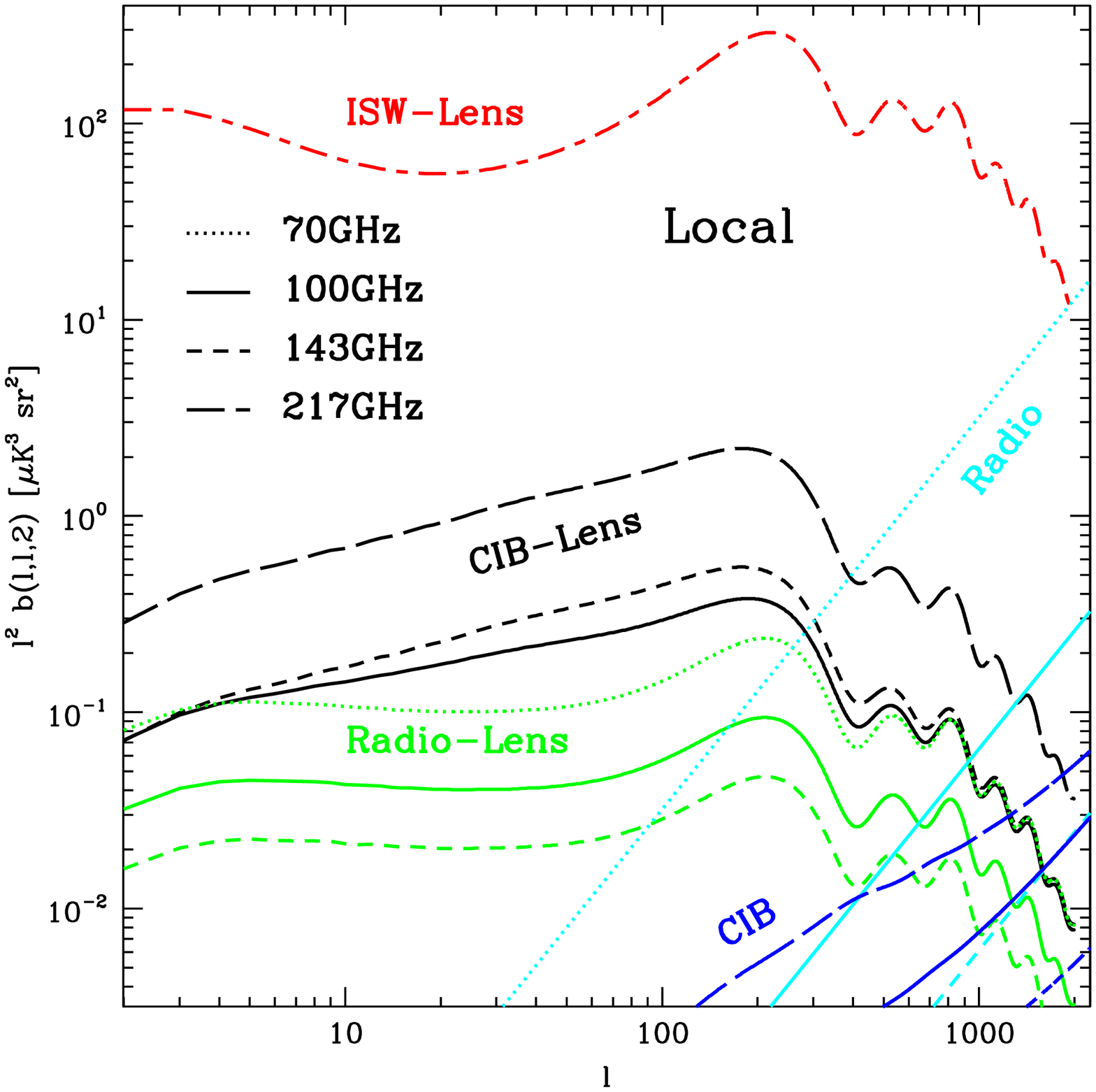}
\includegraphics[width=90mm]{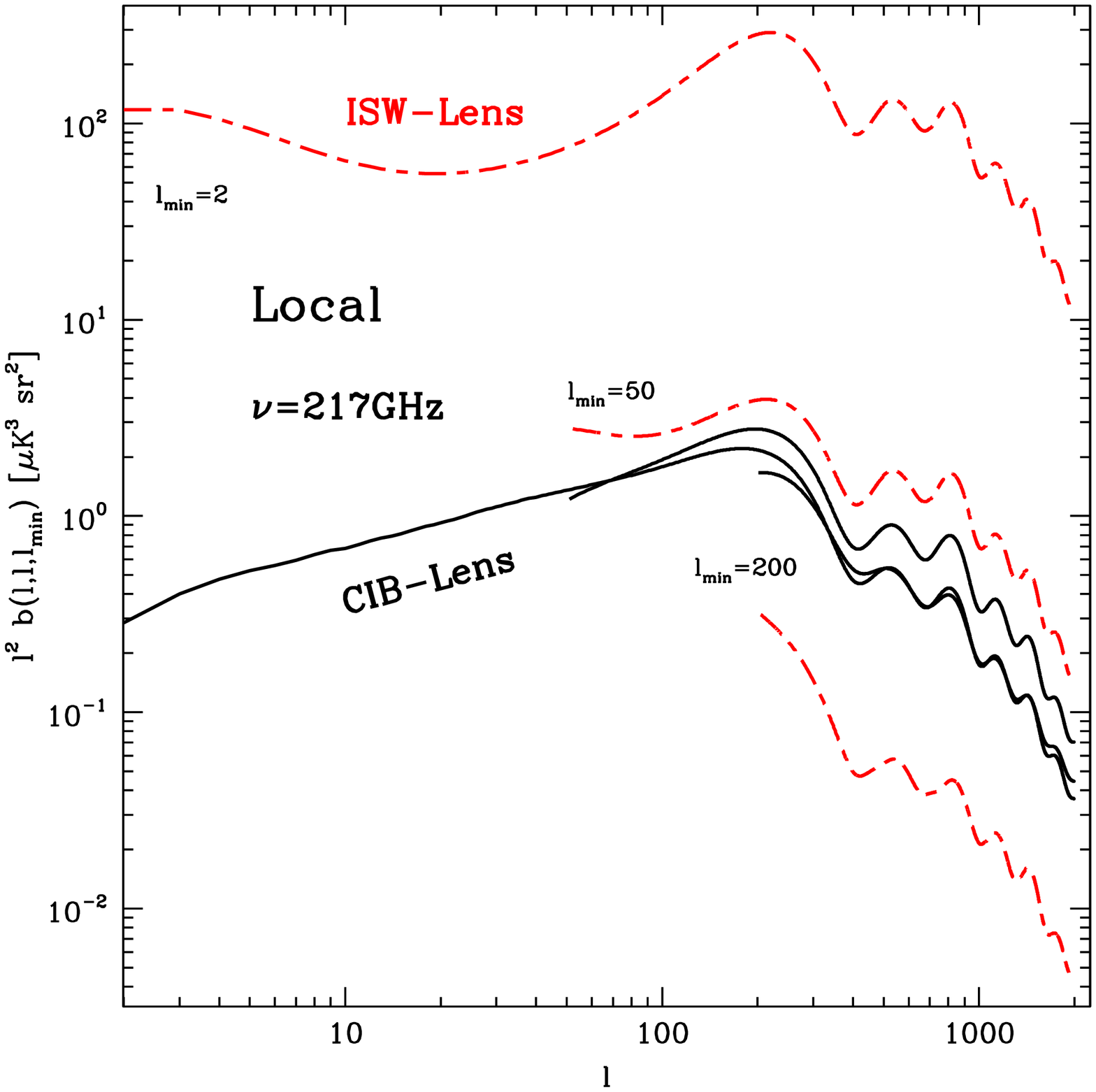}
\includegraphics[width=90mm]{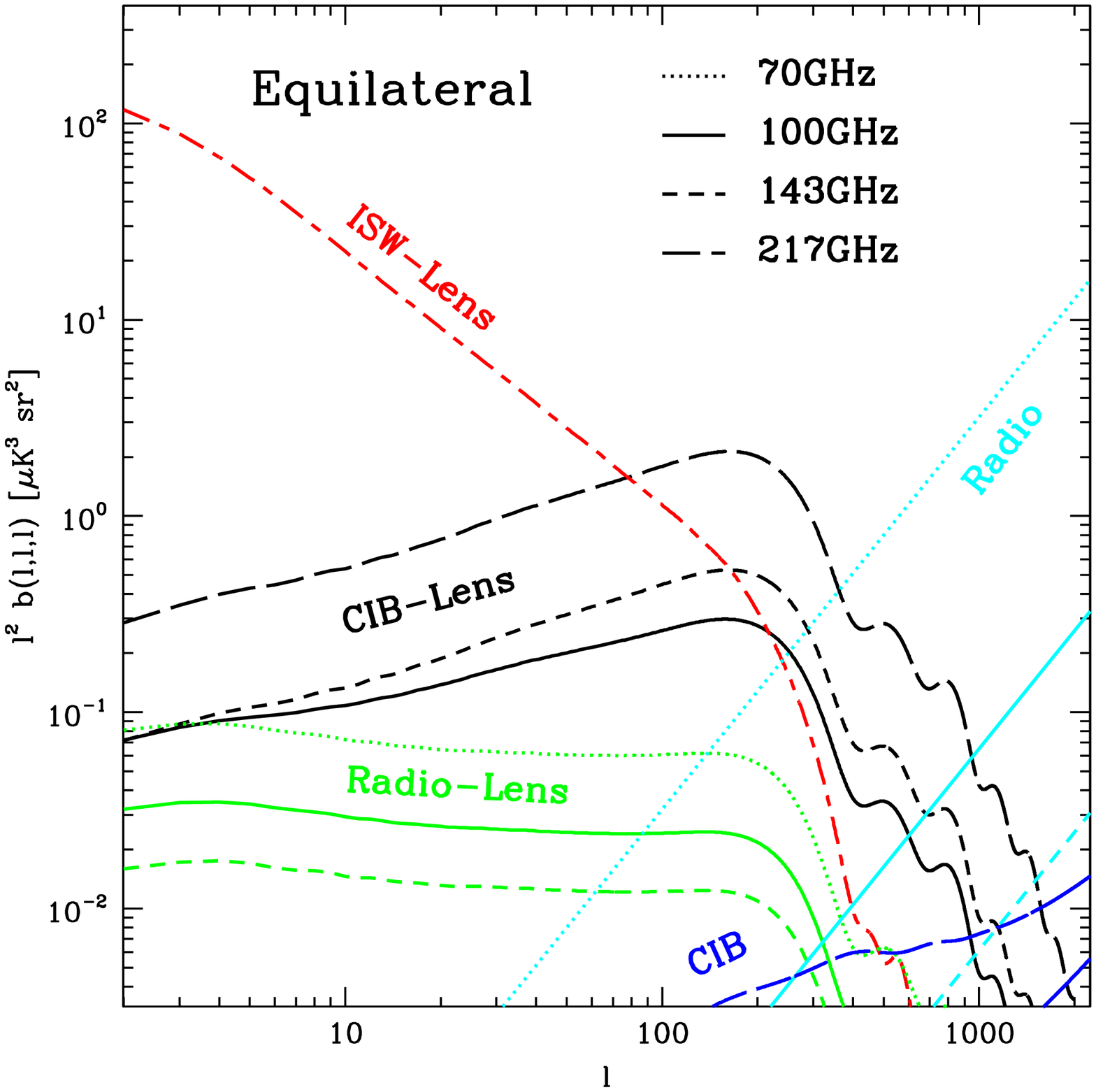}
\includegraphics[width=90mm]{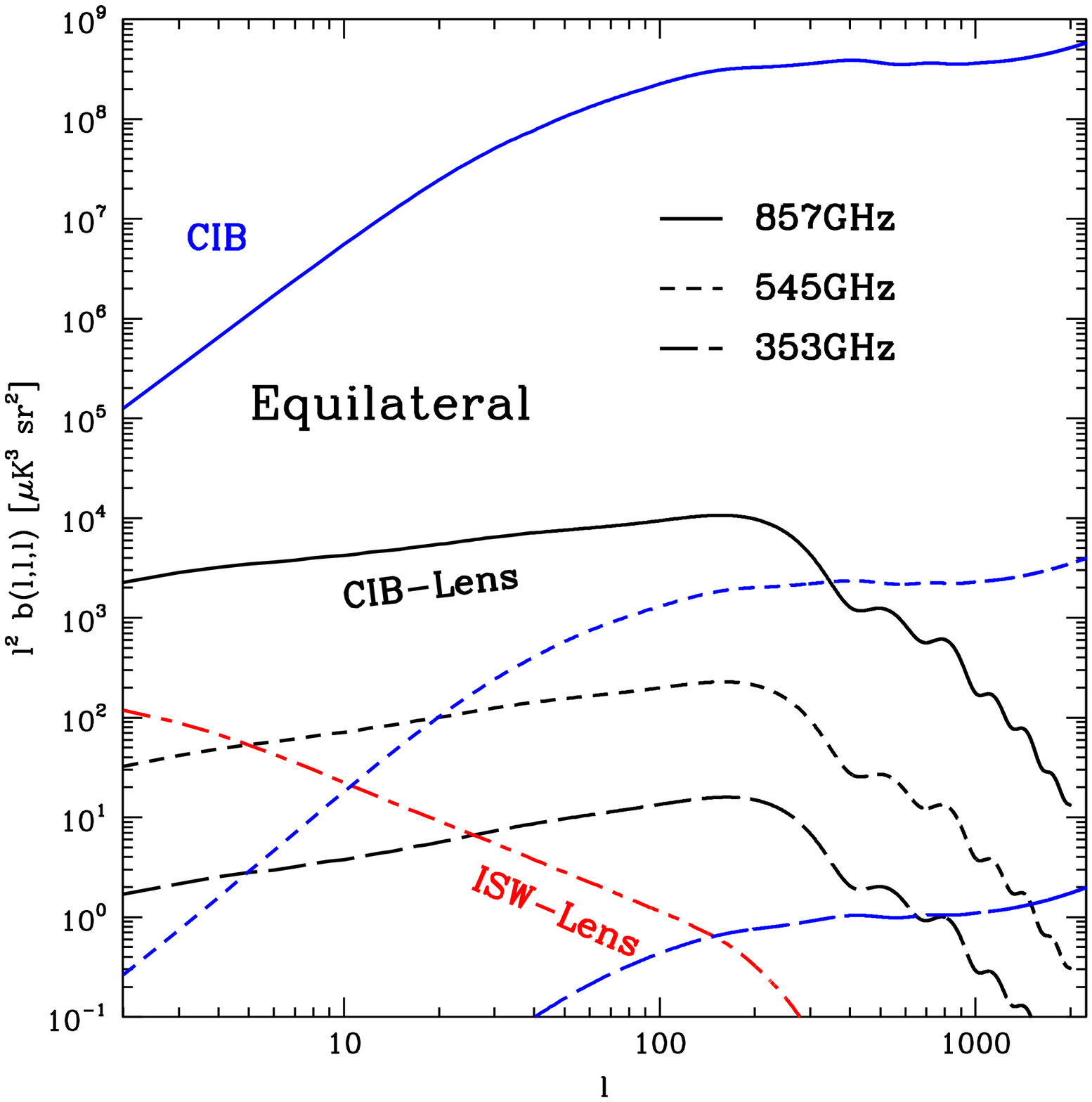}
\caption{Predicted bispectra for the local ({\it upper panels}) and
  equilateral ($\ell_1=\ell_2=\ell_3$; {\it lower panels})
  configurations. Contributions are as in Figure\,\ref{f1}, plus
  bispectra from radio sources (cyan curves) and IR galaxies (blue
  curves). Frequencies are indicated inside the plots. In the {\it
    upper left panel}, ``local'' bispectra are plotted at
  different frequencies (from 70 to 217\,GHz) 
  and for $\ell_1=\ell_2$ and $\ell_{3}=\ell_{\rm min}=2$; in
  the {\it upper right panel}, they are plotted at 217\,GHz
  for $\ell_3=\ell_{\rm min}=2,~50,~200$.}
\label{f2}
\end{figure}
The interplay between the CMB gravitational lensing and CIB intensity
fluctuations are studied in this work in terms of the cross-bispectrum
(the so called CIB--lensing bispectrum), its detectability levels and
the bias on the primordial non-Gaussianity through the \fnl
parameter. The observed temperature fluctuations, neglecting other
foreground sources, can be expanded at first order as \citep[see
e.g.][]{Goldberg:1999xm,hu2000}:
\beq
\Delta T(\hn) = \Delta T_{\CMB}(\hn + \nabla \phi (\hn)) + 
\Delta T_{{\CIB}} (\hn) \simeq \Delta T_{\CMB} (\hn) + \nabla 
\big(\Delta T_{\CMB} (\hn)\big) \nabla \phi (\hn)  + 
\Delta T_{\CIB} (\hn)\,,
\eeq
where $\phi (\hn)$ is the lensing potential, $\Delta T_{\CMB}(\hn)$
are the primordial CMB anisotropies and $\Delta T_{\CIB}(\hn)$ are the
anisotropies due to the CIB. Going into the spherical harmonic space,
the observed anisotropies are:
\beq
a_{\ell m} = \tilde{a}^{(\CMB)}_{\ell m} + a^{(\CIB)}_{\ell m} =
a^{(\CMB)}_{\ell m} +\sum_{\ell' m' \ell'' m''}(-1)^m 
G_{\ell \ell' \ell''}^{m m' m''}
\Big[\frac{\ell'(\ell'+1)-\ell(\ell+1)+\ell''(\ell''+1)}{2}
a^{(\CMB)}_{\ell' m'}\phi_{\ell'' m''} \Big] + a^{(\CIB)}_{\ell m},
\label{almcmb}
\eeq
where $\tilde{a}^{(\CMB)}_{\ell m}$, $a^{(\CMB)}_{\ell m}$,
$a^{(\CIB)}_{\ell m}$ and $\phi_{\ell m}$ are the spherical harmonic
coefficients of the observed/primordial CMB, CIB and gravitational
potential anisotropies, respectively, and $G_{\ell \ell' \ell''}^{m m' m''}$
is the Gaunt coefficient (see Eq.\ \ref{eq:gaunt_integral}).

The angular power spectra of CIB fluctuations and of their
cross--correlation with the CMB lensing are typically written in the
Limber approximation as \citep[e.g.,][see also Appendix
  \ref{appendix_cib_lensing} for a full derivation]{son03}:
\bea
C_{\ell}^{(\CIB)}(\nu,\nu') =\langle a^{(\CIB)*}_{\ell
  m}a^{(\CIB)}_{\ell m}\rangle & = & \int_0^{\chi_*}{d\chi \over
  \chi^2}\, W_{\nu}^{(\CIB)}(\chi)W_{\nu'}^{(\CIB)}(\chi)\,
P_{\rm gg}(k=\ell/\chi,\chi)\,; \nonumber \\
C_{\ell}^{(\LensCIB)}(\nu) =\langle\phi^*_{\ell
  m}a^{(\CIB)}_{\ell m}\rangle & = & \int_0^{\chi_*} {d\chi \over
  \chi^2}\,W_{\nu}^{(\CIB)}(\chi) W^{(\Lens)}(k,\chi)\,P_{\delta
  {\rm g}}(k=\ell/\chi,\chi)\,.
\label{s2e1}
\eea
The integral is over the comoving distance $\chi$ along the line of
sight, and extends up to the comoving distance of the last scattering
surface $\chi=\chi_*$ (in practice the integral is computed up to
redshift 7 because of the negligible contribution of CIB
fluctuations at higher redshifts). The $W^{(\CIB)}(\chi)$ and
$W^{(\Lens)}(k,\chi)$ functions are the redshift weights for CIB
fluctuations and for the lensing potential $\phi$, respectively,
\beq
W_{\nu}^{(\CIB)}(\chi)=a(\chi)\,\bar{j}_{\nu}(\chi)~~~~~~~~~~~~
W^{(\Lens)}(k,\chi)=3{\Omega_m \over a(\chi)}\bigg({H_0 \over ck}\bigg)^2
{\chi_*-\chi \over \chi_*\chi}\,, 
\label{s2e2}
\eeq
where $a(\chi)$ is the scale factor and $\bar{j}_{\nu}(\chi)$ is the
mean CIB emissivity at frequency $\nu$ as a function of $\chi$:
  \beq \bar{j}_{\nu}(\chi)=(1+z)\bigg({d\chi \over dz}\bigg)^{-1}
  \int_0^{S_c}\,S{d^2N \over dSdz}dS\,.
\label{ss2e3}
\eeq
Here $d^2N/dSdz$ denotes galaxies number counts per interval of
flux density and redshift, and $S_c$ is the flux limit above which
sources are subtracted or masked\footnote{Hereafter, we use as
  flux limit for the \Planck mission the 90\% completeness level of
  the \Planck Catalogue of Compact Sources, given in
  \citet{planck_xxviii_2013}.}. We compute the redshift evolution of
the CIB emissivity from the model of galaxy evolution of
\citet{bet11}\footnote{We use number counts of the so--called {\it
    mean model}, see the http://www.ias.u-psud.fr/irgalaxies/ web
  page.}. This is a backward evolution model based on parametric
luminosity functions for two populations of galaxies: normal and
starburst galaxies.
It uses spectral energy distribution templates for the two galaxy
populations taken from the \citet{lag04} library. The model is
described by 13 free parameters and the best--fit values are computed
using observational number counts and luminosity functions from
mid--infrared to millimetre wavelengths \citep{bet11}. This CIB model
was previously used by \citet{pen12} before and then applied to
\Planck results \citep{planck_xviii_2011,planck_xxx_2013} in order to
compute CIB and CIB--lensing power spectra.

In Eq.\ (\ref{s2e1}), $P_{\rm gg}(k,\chi)$ and $P_{\delta {\rm
    g}}(k,\chi)$ are respectively the 3D power spectrum of galaxies
and of the cross--correlation between galaxies and the dark matter
(DM) density field. In the context of the halo model
\citep{sch91,sel00,sco01,coo02}, the power spectra are the sum of the
contribution of the clustering in one single halo (1--halo term) and
in two different halos (2--halo term):
\bea
 P_{\rm gg}(k)=P_{\rm gg}^{1h}(k)+P_{\rm gg}^{2h}(k) & &
 P_{\delta {\rm g}} (k)=P_{\delta {\rm g}}^{1h}(k)+
 P_{\delta {\rm g}}^{2h}(k) \nonumber \\
 P_{\rm gg}^{1h}(k)=\int\,dM\,n(M){\langle N_{\rm gal}(N_{\rm gal}-1)\rangle
\over \bar{n}^2_{\rm gal}}u^2(k,M) & &
 P_{\delta {\rm g}}^{1h}(k)=\int\,dM\,n(M){M \over \bar{\rho}}
{\langle N_{\rm gal}\rangle \over \bar{n}_{\rm gal}}u^2(k,M) \nonumber \\
 P_{\rm gg}^{2h}(k)=P_{\rm lin} (k)\bigg[\int\,dM\,n(M)b(M){\langle
N_{\rm gal}\rangle \over \bar{n}_{\rm gal}}u(k,M)\bigg]^2 & &
 P_{\delta {\rm g}}^{2h}(k)=P_{\rm lin}(k)\,\int\,dM_1\,n(M_1)b(M_1)
{M_1 \over \bar{\rho}}u(k,M)\,\times \nonumber \\
& &  ~~~~~~~~~~~~~~~~~~ \times \int\,dM_2\,n(M_2)b(M_2){\langle
N_{\rm gal}\rangle \over \bar{n}_{\rm gal}}u(k,M_2)
\label{b0}
\eea
The main inputs required for the calculations of $ P_{\rm gg}(k)$ and
$ P_{\delta {\rm g}}(k)$ are: (i) the mass function $ n(M)$ of DM
halos --we use the mass function fit of \citet{tin08} with its
associated prescription for the halo bias, $ b(M)$
\citep[see][]{tin10}--; (ii) the distribution of DM within halos, $
u(k,M)$, --we use the NFW \citep{nav97} profile--; and (iii) the Halo
Occupation Distribution (HOD), that is a statistical description of
how DM halos are populated with galaxies. We model the HOD using a
central-satellite formalism \citep[e.g.,][]{kra04,zhe05}: it
introduces a distinction between central galaxies, which lies at the
centre of the halo, and satellite galaxies that populate the rest of
the halo and are distributed in proportion to the halo mass
profile. The mean number of galaxies in a halo of mass $ M$ is thus
written as $ \langle N_{\gal} \rangle=\langle N_{\rm cen}
\rangle+\langle N_{\rm sat} \rangle$. Following \citet{tin10b}, the
mean occupation functions of central and satellite galaxies are:
\beq
 \langle N_{\rm cen} \rangle={1 \over 2}\Bigg[1+{\rm erf}\Bigg({\log
  M-\log M_{\rm min}
\over \sigma_{\log M}}\Bigg)\Bigg]\,,
\label{b1}
\eeq
and
\beq
 \langle N_{\rm sat} \rangle={1 \over 2}\Bigg[1+{\rm erf}\Bigg({\log
  M-\log 2M_{\rm min}
\over \sigma_{\log M}}\Bigg)\Bigg]\Bigg({M \over M_{\rm sat}}
\Bigg)^{\alpha_{\rm sat}}\,,
\label{b2}
\eeq
where $ M_{\rm min}$, $ \alpha_{\rm sat}$, $ M_{\rm sat}$ and
$ \sigma_{\log M}$ are free parameters. Within this parametrisation,
most of the halos with $ M\ga M_{\rm min}$ contain a central
galaxy. For satellite galaxies the mass threshold is chosen to be
twice $ M_{\rm min}$, so that halos with a low probability of having a
central galaxy are unlikely to contain a satellite galaxy. The number
of satellite galaxies grows with a slope $ \alpha_{\rm sat}$ for
high--mass halos. Moreover, assuming a Poisson distribution for
$N_{\rm sat}$, we can write $ \langle N_{\rm gal}(N_{\rm gal}-1)\rangle=2\langle
N_{\rm sat}\rangle+\langle N_{\rm sat}\rangle^2$ \citep{zhe05}. Finally, in
Eq.\,(\ref{b0}), $ \bar{\rho}$ is the background density, $ \bar{n}_{\rm gal}$
is the mean number of galaxies given by $ \int dMn(M)\langle
N_{\rm gal}\rangle$, and $ P_{\rm lin}$ is the linear DM power spectrum.

Following the analysis in \citet{planck_xviii_2011}, we restrict the
free HOD parameters to only $ M_{\rm min}$ and $ \alpha_{\rm sat}$ by
imposing $ M_{\rm sat}=3.3M_{\rm min}$ and $ \sigma_{\log M}=0.65$.
Moreover, because of the uncertainty of the evolution model of
galaxies at high redshift, the effective mean emissivity $ j_{\rm eff}$
at $ z>3.5$ is also constrained from data as an extra free parameter
\citep[see Appendix \ref{appendix_hod_parameters},
and][]{planck_xviii_2011}.  We find the best--fit values of the model
parameters (i.e., $ M_{\rm min}$, $ \alpha_{\rm sat}$ and $ j_{\rm eff}$)
using the recent \Planck measurements of the CIB power spectra
\citep{planck_xxx_2013}: the results are in good agreement with values
of \citet{planck_xviii_2011}. Details and results of the analysis are
provided in Appendix \ref{appendix_hod_parameters}. As shown in
Figures\,\ref{fb1}--\ref{fb2} of Appendix
\ref{appendix_hod_parameters}, the model is able to reproduce in a
quite remarkable way \Planck measurements both for the auto-- and
cross-- CIB spectra and for the CIB--lensing spectra. 
\subsection{Computing CIB and CIB--lensing spectra at very large scales}
At the very large scales, i.e. at $\ell\la10$, the Limber
approximation used in Eq.\ (\ref{s2e1}) is not further valid.  Here we
provide the general expression for CIB and CIB--lensing power spectra,
that we use for the angular scales ranging from $\ell=2$ to 40:
\bea
C^{(\CIB)}_{\ell}(\nu,\nu') & = & {2 \over \pi}\int\,dk\,k^2\,
\int_0^{\chi_*}d\chi\,W_{\nu}^{(\CIB)}(\chi)j_{\ell}(k\chi)
P^{1/2}_{\rm gg}(k,\chi)
\int_0^{\chi_*}d\chi'\,W_{\nu'}^{(\CIB)}(\chi')j_{\ell}(k\chi')
P^{1/2}_{\rm gg}(k,\chi')
\nonumber \\
C^{(\LensCIB)}_{\ell}(\nu) & = & {2 \over \pi}\int\,dk\,k^2\,
\int_0^{\chi_*}d\chi\,W_{\nu}^{(\CIB)}(\chi)j_{\ell}(k\chi)
P^{1/2}_{\rm gg}(k,\chi)\,
\int_0^{\chi_*}d\chi'\,W^{(\Lens)}(k,\chi')j_{\ell}(k\chi')
P^{1/2}_{\delta\delta}(k,\chi')\,,
\label{ss2e1}
\eea 
where $j_{\ell}(x)$ are the spherical Bessel functions and
$P_{\delta\delta}(k,\chi)$ is the power spectrum of the DM density
field respectively. The expression for the CIB--lensing correlation
power spectrum has been derived in Appendix
\ref{appendix_cib_lensing}, following the procedure developed in
\citet{lewis2006} for the ISW--Lensing power spectrum (the CIB
spectrum can be derived in a similar way). We have verified that in
the Limber approximation power spectra are typically overestimated by
a factor $1.5$ at $\ell=2$.
\subsection{Correlation of radio sources with the CMB lensing}
Intensity fluctuations produced by extragalactic radio sources at
cm/mm wavelengths are dominated by the shot--noise term due to bright
objects. The contribution from the radio sources clustering is
expected to be significant for faint objects, i.e. for flux densities
$S\la10$\,mJy \citep{gon05,tof05}. This is actually observed in
low--frequency surveys like the NVSS survey \citep{con98}, which have
been found to be fair tracers of the underlying density field at
redshifts $z\la2$ \citep[see,
e.g.,][]{bou05,vie06,planck_xix_2013}. Therefore, although small, we
expect some level of correlation between the signal from radio
sources and the CMB lensing potential, which is primarily induced by
dark matter halos at $1\la z\la3$.

In order to estimate this contribution we use the same formalism
  as for the CIB. The radio--lensing power spectrum is given
therefore by
\beq
C_{\ell}^{(\LensRadio)}(\nu)=\int_0^{\chi_*} {d\chi \over
  \chi^2}\,W_{\nu}^{(\Radio)}(\chi) W^{(\Lens)}(k,\chi)\,P_{\delta
  {\rm g}}(k=\ell/\chi,\chi)~~~~~~~{\rm with}~~~
W_{\nu}^{(\Radio)}(\chi)=a(\chi)\bar{j}_{\nu}(\chi)\,.
\label{ss2e2}
\eeq 
The mean emissivity of radio sources is computed from
Eq. (\ref{ss2e3}) with number counts $d^2N/dSdz$ provided by the model
described in \citet{tuc11}. We estimate the integral starting
  from $S=10^{-5}\,$Jy, which nearly corresponds to the limit of
validity of the model.  At lower flux densities, we expect number
counts to have a break at $\sim \mu$Jy, and that the contribution from
fainter sources, although maybe not completely negligible, should not
affect the conclusions of our analysis.

The power spectrum $P_{\delta {\rm g}}(k,\chi)$ is computed as the sum
of the 1--halo and 2--halo terms, see Eq.\ (\ref{b0}). Unlike the CIB,
we assume that the mean number of galaxies per halo, $\langle N_{\rm
  gal} \rangle$, is equal to 1 if the halo mass is larger than some
threshold $M_{\rm min}$, and otherwise is zero. This choice is
motivated by the fact that we are taking into account only the most
powerful radio objects that are typically associated to the centre of
dark matter halos. This assumption also agrees with results from
\citet{mar13} for the NVSS survey: they found that the average number
of galaxies within a halo of mass $M$ can be described by a step
function with the mass threshold in the range $ 12.3\la\log(M_{\rm
  min}/M_{\odot})<12.4$. We take $ \log(M_{\rm
  min}/M_{\odot})=12.34$. We have also verified that our results are
only weakly dependent on the value of $M_{\rm min}$.  The
radio-lensing power spectra for this parametrization are shown in
Fig.\ \ref{f1}.  \citet{cha12} studied the HOD of AGNs using
cosmological hydrodynamic simulations: they found that the mean
occupation function can be modelled as a softened step function for
central AGNs (same as our Eq.\,\ref{b1}) and as a power law for
satellite AGNs. Using their occupation functions for redshift $z=1$
and for the brightest sources ($L_{\rm bol}\ge10^{42}$\,erg\,s$^{-1}$;
see their Table\,2), we find radio--lensing power spectra in very good
agreement with the ones shown in Fig.\,\ref{f1}. On the other hand,
their occupation functions for fainter AGNs give significantly lower
power spectra.

We want to stress however that our estimates for the radio--lensing
power spectra should be taken only as indicative of the level of the
signal, due to the large uncertainties in modelling radio sources. It
is outside of the aim of this work to provide more accurate predictions
for this component.
\subsection{Forecasts for non--primordial bispectra}
Due to the strong clustering of Infrared (IR) galaxies, CIB
  fluctuations produce a non--constant bispectrum that we compute with
  the following prescription \citep[see
  e.g.][]{arg03,lacasa2012a,curto2013}:
\beq
b^{(\CIB)}_{\ell_1\ell_2\ell_3}=b^{(\CIB)}_{\sn}\,
\sqrt{{C^{(\CIB)}_{\ell_1}C^{(\CIB)}_{\ell_2}C^{(\CIB)}_{\ell_3} \over
(C^{(\CIB)}_{\sn})^3}}\,,
\label{s2e4}
\eeq
where $C^{(\CIB)}_{\sn}$ and $b^{(\CIB)}_{\sn}$ are the shot--noise
contributions to CIB power spectra and bispectra (see
Appendix\,\ref{appendix_sn}).

Additionally the coupling of the weak lensing of the CMB with CIB
anisotropies leads also to a bispectrum that is, in the reduced form,
given by \citep{Goldberg:1999xm,coo00,lewis2011}
\beq
b^{(\LensCIB)}_{\ell_1\ell_2\ell_3}={\ell_1(\ell_1+1)-\ell_2(\ell_2+1)
  +\ell_3(\ell_3+1) \over
  2}\,\tilde{C}^{(\CMB)}_{\ell_1}\,C^{(\LensCIB)}_{\ell_3} +(5~\rm{perm})\,,
\label{s2e3}
\eeq 
where $\tilde{C}^{(\CMB)}_{\ell}$ is the lensed CMB power
spectrum\footnote{We have computed the lensed and unlensed power
spectra and the CMB-lensing cross-spectrum using the cosmological
parameters that best fit the combined WMAP and \Planck 2013 data
\citep[referred as 'Planck+WP+highL' in][]{planck_xvi_2013} using the
latest version of CAMB \citep{lewis2000}.}. Eq.\ (\ref{s2e3}) was
derived for the first time by \citet{Goldberg:1999xm} for a generic
tracer of the matter distribution expanding lensed CMB temperature
fluctuations $\Delta T(\hn) = \Delta T(\hn+\nabla\phi (\hn))$ to
the first order in $\phi$. In the original form they used the unlensed
$C^{(\CMB)}_{\ell}$ in the right-hand part of the
equation. \citet{lewis2011} showed instead that, when higher--order
terms are taken into account, the correct equation requires to use the
lensed CMB power spectrum.

Bispectra induced by the correlation of the CMB lensing potential with
a tracer of the matter distribution differ only for the shape of the
cross--power spectrum of the lensing and the matter tracer. In
  Fig.\ \ref{f1} we compare the cross--power spectra for the case of
  ISW and extragalactic sources. We see that the ISW--lensing
  power spectrum is the most relevant one on scales larger than few
degrees (it is 2--3 order of magnitude larger than other spectra at
$\ell=2$), but rapidly decreases at $\ell\ga100$. On these scales the
CIB--lensing power spectrum dominates even at frequencies as low as
100--143\,GHz. On the contrary, as expected, the radio--lensing
correlation produces just a sub-dominant contribution at all the
angular scales and therefore it will not be taken into account in the
following analysis.

The ISW--lensing correlation is found to be a significant contaminant
for \Planck mainly on local primordial NG: \citet{planck_xxiv_2013}
estimated a bias on \fnl of 7.1, 0.4 and $-$22 (1.22, 0.01, $-$0.56 in
$\sigma$ units) for the local, equilateral and orthogonal shape,
respectively. In Fig. \,\ref{f2} we compare the non--primordial
bispectra for the local and the equilateral configurations at
frequencies between 70 and 217\,GHz (the orthogonal and equilateral
configurations are very similar). We can see that for the local shape
the ISW--lensing bispectrum is about two orders of magnitude larger
than the CIB--lensing contribution for $ \ell_{\rm
  min}=2$. However, whereas the latter changes very moderately with
$ \ell_{\rm min}$, the ISW--lensing bispectrum is strongly reduced
increasing $ \ell_{\rm min}$, e.g., by a factor $ \sim10^2$ and
$ 10^3$ for $ \ell_{\rm min}=50$ and 200 respectively. For the
other shapes, the ISW--lensing bispectrum decreases rapidly with the
angular scale $ \ell$, and the dominant contributions come from the
CIB--lensing correlation at $ \ell\ga100$ and from extragalactic
sources at very small scales ($ \ell\ga1000$). Fig. \,\ref{f2}
shows that the CIB and its correlation with the CMB lensing could be a
non-negligible contaminant in NG studies, and motivates the following
deeper analysis.

Finally, in Fig.\ \ref{f2} we also consider the different
contributions for the equilateral shape at the highest \Planck
frequencies. This is interesting in terms of a possible detection of
the CIB-lensing bispectrum. Due to its strong signal at these
frequencies, it should be detectable with high significance for
\Planck at $\nu\ge217$\,GHz. However, IR galaxies give rise themselves
to a strong contribution at high frequencies and they can be therefore
a strong contaminant for the detection of the CIB--lensing
bispectrum. We discuss later how to tackle this problem.
\section{The CIB-lensing bispectrum bias on the primordial non-Gaussianity}
\label{section3_delta_fnl}
\begin{figure*}
\includegraphics[width=90mm]{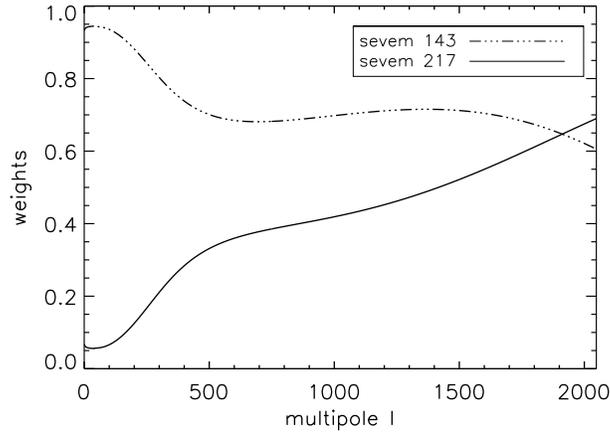}
\caption{The SEVEM component separation weights $w^{ (\nu_i)}_\ell$
  for the combination of 143 and 217 GHz maps. Please notice that the
  weights include the deconvolution/convolution process to reach the
  final 5 arcmin resolution.}
\label{f3}
\end{figure*}
To continue with our study of the bispectra presented above, we
consider three scenarios for the estimation of the \fnl bias due to
the CIB-lensing correlation: (i) raw per-frequency maps, which in
particular contain CMB lensed signal plus CIB and radio point source
contributions plus instrumental noise, (ii) foreground-reduced (clean)
maps per frequency and (iii) a combination of clean maps.  Galactic
foregrounds are not taken into account. We use \Planck ``ideal''
instrumental characteristics -- i.e. isotropic noise, spherically
symmetric beams, full sky coverage -- that are summarized in
Table\,\ref{t1}.  As a representative component separation technique
already used by the \Planck collaboration, and for reason of
simplicity, we select SEVEM
\citep{leach2008,fernandezcobos2011,planck_xii_2013}. This cleaning
technique is based on a template fitting approach. The templates used
by SEVEM are constructed using only \Planck data and there are no
assumptions on the foregrounds or noise levels. The templates are
constructed by taking the difference of two close \Planck frequency
maps previously smoothed to a common resolution\footnote{E.g. the
  44-70 template would be constructed by subtracting the 44 and 70 GHz
  maps previously smoothed to a common beam defined as the product of
  the beams of the two maps in spherical harmonics space,
  $b^{\nu=44}_{\ell}$ and $b^{\nu=70}_{\ell}$.}. This template is
therefore clean of CMB signal. The SEVEM foreground-reduced map at a
given frequency is computed by subtracting from the raw map at that
frequency a linear combination of selected templates. The linear
coefficients are computed by minimising the variance of the final
map. SEVEM is linear and therefore the $a_{\ell m}$ coefficients of a
cleaned map can be written as a linear combination of the coefficients
of the raw maps involved in the cleaning process. The SEVEM cleaned
maps used in this paper are the 143 and 217 GHz maps, also considered
in the non-Gaussianity analyses performed in
\citet{planck_xxiii_2013,planck_xxiv_2013}.  These two maps are
computed with 4 templates -- two corresponding to the LFI channels,
namely the 30-44 and 44-70 templates, and two corresponding to the HFI
channels, namely the 545-353 and the 857-545 templates. These
templates take into account different Galactic and extra-Galactic
foregrounds at low and high frequencies and their residual amplitude
present in the data is minimised.

The spherical harmonic coefficients of maps for the three
cases mentioned above are given by:
\begin{itemize}
\item (i) \Planck raw maps per frequency 
\beq
a^{(\raw,{\nu})}_{\ell m}=\big[\tilde{a}^{(\prim)}_{\ell m}+a^{(\CIB,\nu)}_{\ell m}+a^{(\Radio,\nu)}_{\ell m}\big]b^{(\nu)}_\ell + a^{(\noise)}_{\ell m},
\label{alm_raw_maps}
\eeq
\item (ii) \Planck clean maps per frequency
\beq
a^{(\clean,\nu)}_{\ell m}=\sum_{i=1}^{9}f^{(\nu)}_{\nu_i} a^{(\raw,\nu_i)}_{\ell m},
\label{sevem_alm_per_freq}
\eeq
\item (iii) \Planck combined clean map
\beq
a^{(\comb)}_{\ell m}=\sum_{i=5}^{6}{w^{(\nu_i)}_\ell a^{(\clean,\nu_i)}_{\ell m}}.
\label{sevem_alm_comb}
\eeq
\end{itemize}
where $b^{(\nu)}_\ell$ is the beam for each frequency channel $\nu$,
$f^{(\nu)}_{\nu_i}$ are the SEVEM component separation weights per
frequency $\nu$, and $w^{(\nu_i)}_\ell$ are the SEVEM component
separation weights for the combined map. We use the raw maps in the
frequency range between 100 and 353 GHz. Frequencies lower than 100
GHz have a negligible IR contribution. At frequencies higher than 353
GHz, the CIB signal is clearly dominant over the CMB and the
corresponding cosmic variance completely masks the CIB--lensing
signal. The low and high frequency maps are nonetheless useful as
templates to clean the central frequency maps in Eq.\
(\ref{sevem_alm_per_freq}) where the sum runs over all nine {\it
  Planck} frequencies from frequency 1 = 30 GHz to frequency 9 = 857
GHz.  These maps are only produced for 143 GHz (frequency 5) and 217
GHz (frequency 6).  The weights for the SEVEM templates needed to
construct the foreground-reduced maps are given in Table \ref{t1b} and
are based on \citet[][]{planck_xii_2013}.  Finally the SEVEM combined
map is computed using the weights given in Fig. \ref{f3} following
Eq. (\ref{sevem_alm_comb}), reaching a resolution of 5 arcmin.
\subsection{Power spectrum}
The power spectrum for the three considered cases is:
\begin{itemize}
\item (i) \Planck raw maps per frequency 
\beq
C^{(\nu)}_{\ell} = \big[b^{(\nu)}_\ell\big]^2\big[\tilde{C}^{(\CMB)}_{\ell}+C^{(\CIB,\nu)}_{\ell}+C^{(\Radio,\nu)}_{\ell}\big]+C^{(\noise,\nu)}_{\ell},
\label{power_spectrum_i}
\eeq
\item (ii) \Planck cleaned maps per frequency
\beq
C^{(\clean,\nu)}_{\ell} = \sum_{\{i,j\}=1}^{9}f^{(\nu)}_{\nu_i}f^{(\nu)}_{\nu_j}b^{(\nu_i)}_\ell b^{(\nu_j)}_\ell\big[\tilde{C}^{(\CMB)}_{\ell}+C^{(\CIB,\nu_i,\nu_j)}_{\ell}+C^{(\Radio,\nu_i,\nu_j)}_{\ell}\big]+\sum_{i=1}^{9}\big[f^{(\nu)}_{\nu_i}\big]^2C^{(\noise,\nu_i)}_{\ell},
\label{power_spectrum_ii}
\eeq
\item (iii) \Planck combined cleaned maps
\beq
C^{(\comb)}_{\ell} =\sum_{\{i,j\}=1}^{9}g^{(\nu_i)}_{\ell}g^{(\nu_j)}_{\ell}b^{(\nu_i)}_\ell b^{(\nu_j)}_\ell\big[\tilde{C}^{(\CMB)}_{\ell}+C^{(\CIB,\nu_i,\nu_j)}_{\ell}+C^{(\Radio,\nu_i,\nu_j)}_{\ell}\big]+\sum_{i=1}^{9}\big[g^{(\nu_i)}_{\ell}\big]^2C^{(\noise,\nu_i)}_{\ell},
\label{power_spectrum_iii}
\eeq
\end{itemize}
where
\beq
g^{(\nu)}_{\ell}\equiv\sum_{i=5}^{6}w_{\ell}^{(\nu_i)}f_{\nu}^{(\nu_i)},
\eeq
and $C^{(\noise,\nu)}_{\ell}$ is the \Planck instrumental noise power
spectrum at frequency $\nu$.
\begin{table*}
  \begin{center}
    \caption{\Planck instrumental characteristics based on published information \citep[see Tables 2 and 6 in][]{planck_i_2013}.\label{t1}}
    \begin{tabular}{c|ccccccccc}
      \hline
      \hline
      Channel index & 1 & 2 & 3 & 4 & 5 & 6 & 7 & 8 & 9 \\
      Frequency (GHz) & 30 & 44 & 70 & 100 & 143 & 217 & 353 & 545 & 857 \\
      Beam FWHM (arcmin) & 33 & 28 & 13 & 10 & 7 & 5 & 5 & 5 & 5 \\
      $\sigma_{noise}$ per pixel ($\mu K$) & 9.2 & 12.5 & 23.2 & 11.0 & 6.0 & 12.0 & 43.0 & 897.5 & 37178.6 \\
      $N_{side}$ & 1024 & 1024 & 1024 & 2048 & 2048 & 2048 & 2048 & 2048 & 2048 \\
      \hline
      \hline
    \end{tabular}
  \end{center}
\end{table*}
\begin{table}
  \begin{center}
    \caption{Linear coefficients and templates used to clean
      individual frequency maps with SEVEM \citep[see Table C1
      in][]{planck_xii_2013}.\label{t1b}}
    \begin{tabular}{c|cccc}
      \hline
      \hline
      Template & 30-44 & 44-70 &  545-353 &  857-545 \\
      $f^{(143)}_{\nu_i}$  & -2.14$\times 10^{-2}$ & -1.23$\times 10^{-1}$ &  -7.52$\times 10^{-3}$ & 6.67$\times 10^{-5}$ \\
      \hline
      Template & 30-44 & 44-70 &  545-353 & 857-545 \\
      $f^{(217)}_{\nu_i}$ & 1.03$\times 10^{-1}$ & -1.76$\times 10^{-1}$ & -1.84$\times 10^{-2}$ & 1.21$\times 10^{-4}$ \\
      \hline
      \hline
    \end{tabular}
  \end{center}
\end{table}

\begin{table*}
  \center
  \caption{The expected bias $\Delta f_{\rm nl}$ produced by the
    CIB-lensing bispectrum for $\ell_{\rm max}~=~$ 2000 for the raw \Planck 
    frequency maps between 100 and 353 GHz, the cleaned maps at 143 and 
    217 GHz and the combination of the two previous cleaned maps. The
      expected uncertainties on \fnl are also reported.
      \label{t2}}
  \begin{tabular}{c|ccccccc}
    \hline
    \hline
    Frequency (GHz) & 100 & 143 & 217 & 353 &  SEVEM 143 & SEVEM 217 & SEVEM combined\\
    \hline
    Local $\Delta f_{\rm nl}$ & 0.34 &       0.73 &       3.06 &      13.89 &      -0.48 &      -0.28 &      -0.39 \\
    Local $\sigma(f_{\rm nl})$  & 6.65 &       5.25 &       5.55 &      14.06 &       5.73 &       5.87 &       5.70 \\
    Local $\Delta f_{\rm nl}/\sigma(f_{\rm nl})$  & 0.05 &       0.14 &       0.55 &       0.99 &      -0.08 &      -0.05 &      -0.07 \\
    \hline
    Equilateral $\Delta f_{\rm nl}$ &  0.37 &      -4.56 &     -16.47 &     150.23 &       1.17 &       0.09 &       1.99 \\
    Equilateral $\sigma(f_{\rm nl})$  & 76.39 &      68.28 &      71.00 &     134.60 &      71.08 &      72.22 &      70.97 \\
    Equilateral $\Delta f_{\rm nl}/\sigma(f_{\rm nl})$  &  0.00 &      -0.07 &      -0.23 &       1.12 &       0.02 &       0.00 &       0.03 \\ 
    \hline
    Orthogonal $\Delta (f_{\rm nl})$ & -8.45 &     -21.31 &     -87.92 &    -233.57 &      12.55 &       5.78 &      10.54 \\
    Orthogonal $\sigma(f_{\rm nl})$  & 39.29 &      33.19 &      34.73 &      76.33 &      35.17 &      35.82 &      35.04 \\ 
    Orthogonal $\Delta f_{\rm nl}/\sigma(f_{\rm nl})$  & -0.22 &      -0.64 &      -2.53 &      -3.06 &       0.36 &       0.16 &       0.30 \\ 
    \hline
    \hline
  \end{tabular}
\end{table*}
\begin{figure*}
\includegraphics[width=58mm,height=58mm]{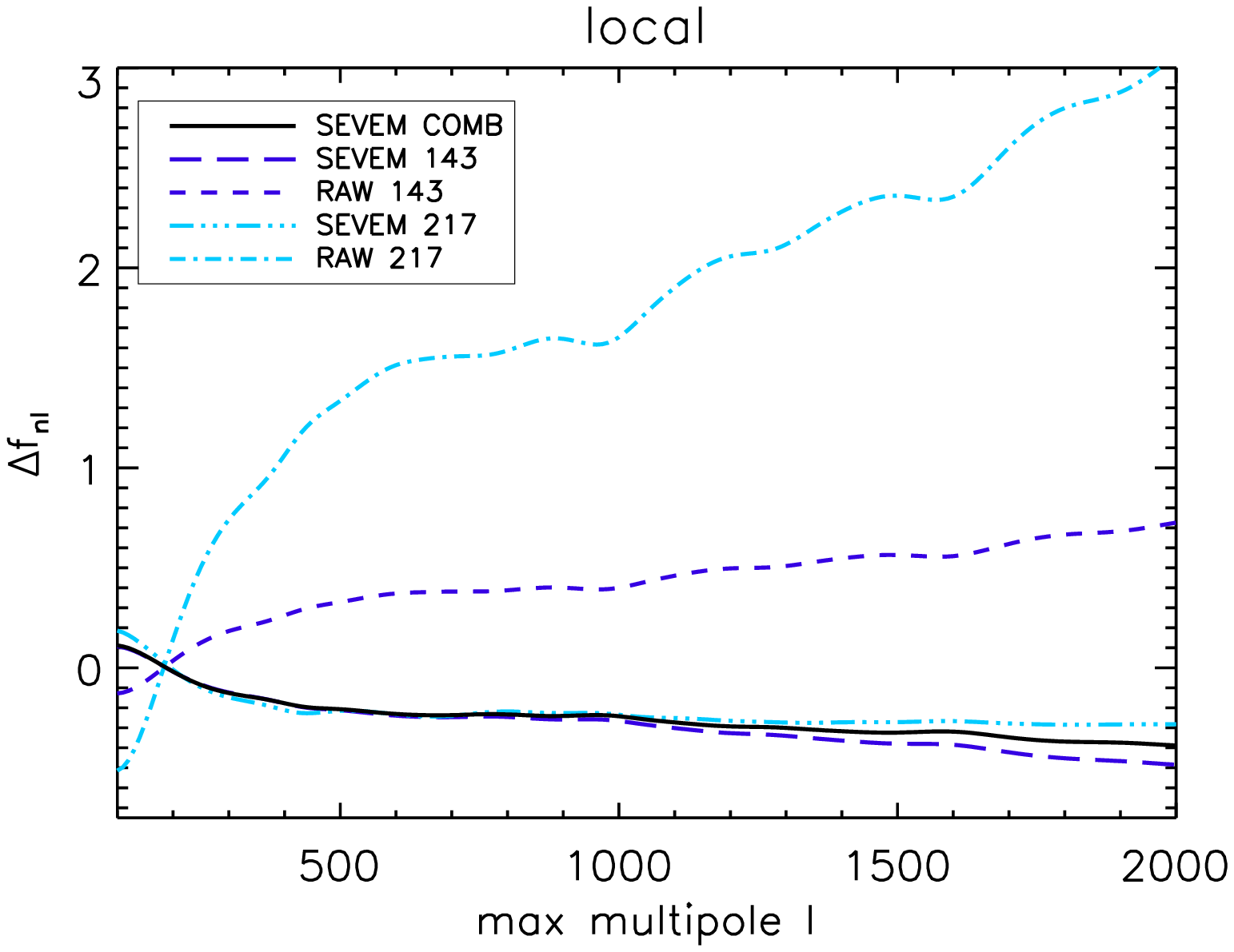}
\includegraphics[width=58mm,height=58mm]{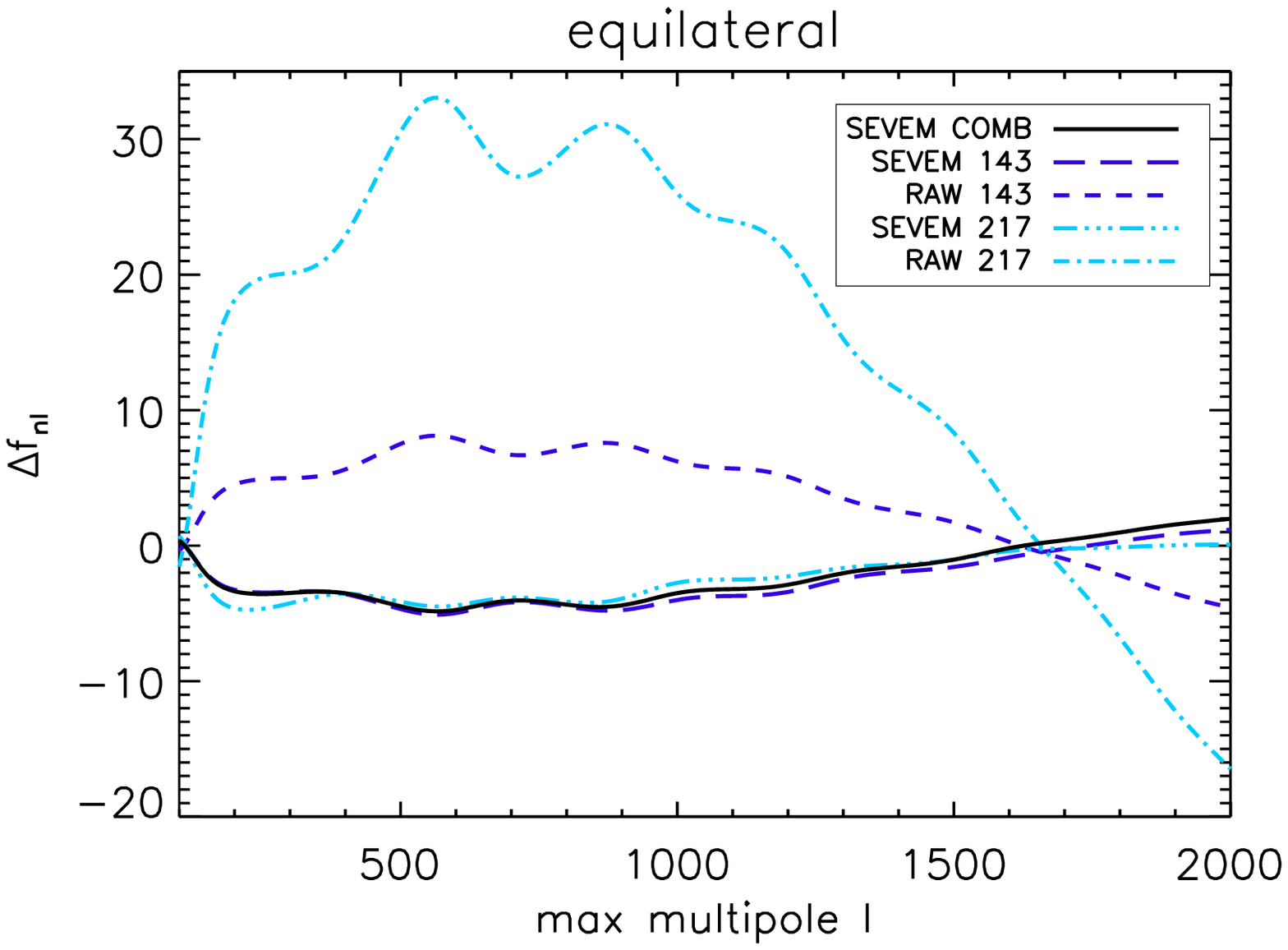}
\includegraphics[width=58mm,height=58mm]{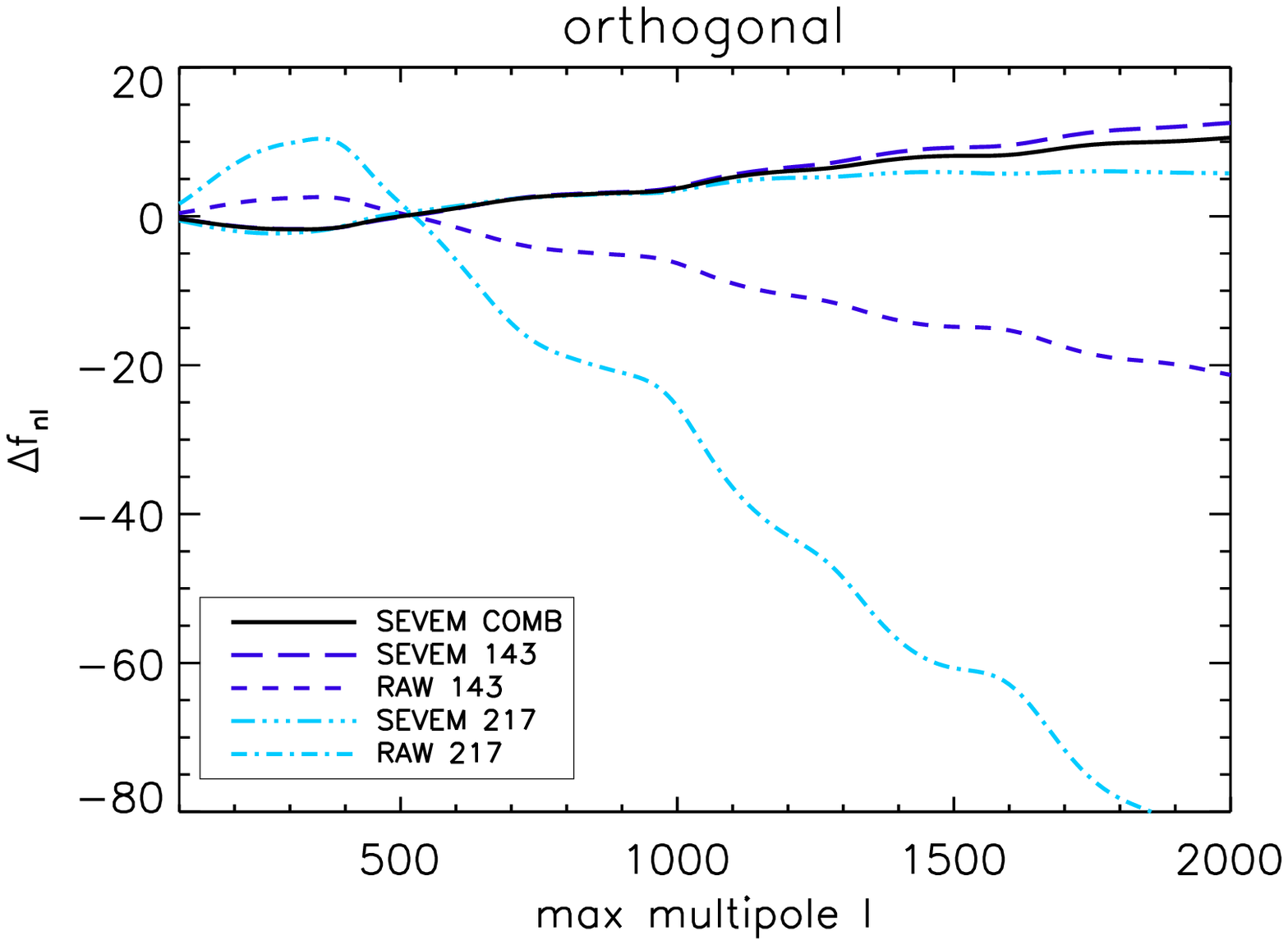}
\caption{The bias $\Delta f_{\rm nl}$ produced by the CIB-lensing
  bispectrum as a function of $\ell_{\rm max}$ for \Planck
  cosmological frequencies (143 and 217 GHz). Solid lines correspond
  to the cleaned combined maps. Dash lines correspond to the raw map
  at 143 GHz and long dash lines correspond to the cleaned map at 143
  GHz. Dash-and-dot lines correspond to the raw map at 217 GHz and
  dash-and-3-dots lines correspond to the cleaned map at 217 GHz. From
  left to right, we plot results for the local, equilateral and
  orthogonal \fnl shapes. }
\label{f4}
\end{figure*}
\subsection{Bispectrum}
We derive the CIB--lensing, CMB and point sources bispectra for
the three considered types of maps.
\begin{itemize}
\item (i) \Planck raw maps per frequency
\beq
b^{(\LensCIB, \nu)}_{\ell_1 \ell_2 \ell_3}=\frac{\ell_1(\ell_1+1)-\ell_2(\ell_2+1)+\ell_3(\ell_3+1)}{2}C^{(\LensCIB,\nu)}_{\ell_3}\tilde{C}^{(\CMB)}_{\ell_1}b^{(\nu)}_{\ell_1} b_{\ell_2}^{(\nu)}b_{\ell_3}^{(\nu)}+(5~perm),
\label{bispectrum_i}
\eeq
\beq
b^{(\CMB,\nu)}_{\ell_1 \ell_2 \ell_3}=b_{\ell_1}^{(\nu)} b_{\ell_2}^{(\nu)}b_{\ell_3}^{(\nu)}b^{(\CMB)}_{\ell_1 \ell_2 \ell_3},~~~~~ b^{(\ps,\nu)}_{\ell_1 \ell_2 \ell_3}= b_{\ell_1}^{(\nu)}b_{\ell_2}^{(\nu)}b_{\ell_3}^{(\nu)}b^{(\ps,\nu)}. \label{prim_radio_bispectrum_i}
\eeq
\item (ii) \Planck cleaned maps per frequency 
\beq
b^{(\LensCIB,\clean,\nu)}_{\ell_1 \ell_2 \ell_3}=\sum_{ijk=1}^{9}f^{(\nu)}_{\nu_i}f^{(\nu)}_{\nu_j}f^{(\nu)}_{\nu_k}b^{(\nu_i)}_{\ell_1} b_{\ell_2}^{(\nu_j)}b_{\ell_3}^{(\nu_k)}b^{(\LensCIB,\nu_i\nu_j\nu_k)}_{\ell_1 \ell_2 \ell_3},
\label{bispectrum_ii}
\eeq
\beq
b^{(\CMB,\clean,\nu)}_{\ell_1 \ell_2 \ell_3}=\sum_{ijk=1}^{9}f^{(\nu)}_{\nu_i}f^{(\nu)}_{\nu_j}f^{(\nu)}_{\nu_k}b_{\ell_1}^{(\nu_i)} b_{\ell_2}^{(\nu_j)}b_{\ell_3}^{(\nu_k)}b^{(\CMB)}_{\ell_1 \ell_2 \ell_3},~~~~~b^{(\ps,\clean,\nu)}_{\ell_1 \ell_2 \ell_3}=\sum_{ijk=1}^{9}f^{(\nu)}_{\nu_i}f^{(\nu)}_{\nu_j}f^{(\nu)}_{\nu_k}b_{\ell_1}^{(\nu_i)} b_{\ell_2}^{(\nu_j)}b_{\ell_3}^{(\nu_k)}b^{(\ps,\nu_i,\nu_j,\nu_k)}.
\label{prim_radio_bispectrum_ii}
\eeq
\item (iii) \Planck combined cleaned maps
\beq
b^{(\LensCIB,\comb)}_{\ell_1 \ell_2 \ell_3}=\sum_{ijk=1}^{9}g^{(\nu_i)}_{\ell_1}g^{(\nu_j)}_{\ell_2}g^{(\nu_k)}_{\ell_3}b^{(\nu_i)}_{\ell_1} b_{\ell_2}^{(\nu_j)}b_{\ell_3}^{(\nu_k)}b^{(\LensCIB,\nu_i\nu_j\nu_k)}_{\ell_1 \ell_2 \ell_3},
\label{bispectrum_iii}
\eeq
\beq
b^{(\CMB,\comb)}_{\ell_1 \ell_2 \ell_3}=\sum_{ijk=1}^{9}g^{(\nu_i)}_{\ell_1}g^{(\nu_j)}_{\ell_2}g^{(\nu_k)}_{\ell_3}b_{\ell_1}^{(\nu_i)} b_{\ell_2}^{(\nu_j)}b_{\ell_3}^{(\nu_k)}b^{(\CMB)}_{\ell_1 \ell_2 \ell_3},~~~~~ b^{(\ps,\comb)}_{\ell_1 \ell_2 \ell_3}=\sum_{ijk=1}^{9}g^{(\nu_i)}_{\ell_1}g^{(\nu_j)}_{\ell_2}g^{(\nu_k)}_{\ell_3}b_{\ell_1}^{(\nu_i)} b_{\ell_2}^{(\nu_j)}b_{\ell_3}^{(\nu_k)}b^{(\ps,\nu_i,\nu_j,\nu_k)}.
\label{prim_radio_bispectrum_iii}
\eeq
\end{itemize}
$b^{(\CMB)}_{\ell_1 \ell_2 \ell_3}$ is the primordial bispectrum
\citep[see e.g.][for the equations for the local, equilateral and
  orthogonal shapes]{curto2013}. The term
$b^{(\LensCIB,\nu_i\nu_j\nu_k)}_{\ell_1 \ell_2 \ell_3}$ can be
straightforwardly computed as a generalisation of the CIB-lensing
bispectrum in Eq. (\ref{s2e3}) for three different frequencies
\begin{eqnarray}
\nonumber
& b^{(\LensCIB,\nu_i\nu_j\nu_k)}_{\ell_1 \ell_2 \ell_3} \equiv  \frac{\ell_1(\ell_1+1)-\ell_2(\ell_2+1)+\ell_3(\ell_3+1)}{2}C^{(\LensCIB,\nu_k)}_{\ell_3}\tilde{C}^{(\CMB)}_{\ell_1} + \frac{\ell_1(\ell_1+1)-\ell_3(\ell_3+1)+\ell_2(\ell_2+1)}{2}C^{(\LensCIB,\nu_j)}_{\ell_2}\tilde{C}^{(\CMB)}_{\ell_1}&\\
\nonumber
& +\frac{\ell_2(\ell_2+1)-\ell_1(\ell_1+1)+\ell_3(\ell_3+1)}{2}C^{(\LensCIB,\nu_k)}_{\ell_3}\tilde{C}^{(\CMB)}_{\ell_2}+\frac{\ell_3(\ell_3+1)-\ell_1(\ell_1+1)+\ell_2(\ell_2+1)}{2}C^{(\LensCIB,\nu_j)}_{\ell_2}\tilde{C}^{(\CMB)}_{\ell_3}&\\
& +\frac{\ell_2(\ell_2+1)-\ell_3(\ell_3+1)+\ell_1(\ell_1+1)}{2}C^{(\LensCIB,\nu_i)}_{\ell_1}\tilde{C}^{(\CMB)}_{\ell_2}+\frac{\ell_3(\ell_3+1)-\ell_2(\ell_2+1)+\ell_1(\ell_1+1)}{2}C^{(\LensCIB,\nu_i)}_{\ell_1}\tilde{C}^{(\CMB)}_{\ell_3}.&
\end{eqnarray}
Finally, the total point sources bispectrum
$b^{(\ps,\nu_i,\nu_j,\nu_k)}_{\ell_1 \ell_2 \ell_3}$ is computed from
the standard prescription in terms of radio and CIB shot noise
bispectra (see Appendix\,\ref{appendix_sn} for details on how we
estimate shot noise bispectra of point sources when multiple
frequencies are considered)
\begin{align}
b^{(\ps,\nu_i,\nu_j,\nu_k)}_{\ell_1 \ell_2 \ell_3} =
b^{(\Radio,\nu_i,\nu_j,\nu_k)}_{\sn} 
+ b^{(\CIB,\nu_i,\nu_j,\nu_k)}_{\sn} \sqrt{\frac{C^{(\CIB,\nu_i)}_{\ell_1}C^{(\CIB,\nu_j)}_{\ell_2}C^{(\CIB,\nu_k)}_{\ell_3}}{C^{(\CIB,\nu_i)}_{\sn}C^{(\CIB,\nu_j)}_{\sn}C^{(\CIB,\nu_k)}_{\sn}}}.
\label{radio_bispectrum_raw}
\end{align}
Considering weak levels of non-Gaussianity, the bias induced in the
primordial non-Gaussianity \fnl due to a given target bispectrum
$B^{(\targ)}_{\ell_1 \ell_2 \ell_3}$ is given by \citep[see
    e.g.][]{lewis2011,lacasa2012a}:
\beq
\Delta f_{\rm nl}  = \sigma^2\big(f_{\rm nl}\big) \times \sum_{\ell_1 \le \ell_2 \le \ell_3}^{\ell_{\rm max}}
\frac{b^{(\targ)}_{\ell_1 \ell_2 \ell_3} b^{(\prim)}_{\ell_1 \ell_2 \ell_3}  }{\sigma^2\big(b^{(\obsv)}_{\ell_1 \ell_2 \ell_3}\big)},
\label{bias_fnl_lens_cib}
\eeq
where $\sigma^2\big(b^{(\obsv)}_{\ell_1 \ell_2 \ell_3}\big)$ is the
variance of the total observed bispectrum \citep{komatsu2001}
\begin{align}
\sigma^2\big(b^{(\obsv)}_{\ell_1 \ell_2 \ell_3}\big) & \equiv \langle b^{(\obsv)}_{\ell_1 \ell_2 \ell_3}b^{(\obsv)}_{\ell_1 \ell_2 \ell_3} \rangle - \langle b^{(\obsv)}_{\ell_1 \ell_2 \ell_3}\rangle \langle b^{(\obsv)}_{\ell_1 \ell_2 \ell_3} \rangle \simeq \frac{1}{I_{\ell_{1}  \ell_{2}  \ell_{3}}^2}\Delta_{\ell_1 \ell_2 \ell_3}C_{\ell_1}C_{\ell_2}C_{\ell_3},\\
& \Delta_{\ell_1 \ell_2 \ell_3} = 1 + 2\delta_{\ell_1
  \ell_2}\delta_{\ell_2 \ell_3} + \delta_{\ell_1 \ell_2} +
\delta_{\ell_2 \ell_3} + \delta_{\ell_1 \ell_3},
\label{variance_bispectrum_i}
\end{align}
$C_{\ell}$ is the total power spectrum of the map including CMB,
CIB, radio sources and instrumental noise spectra and 
\begin{eqnarray}
I_{\ell_{1}  \ell_{2}  \ell_{3}}\equiv \sqrt{\frac{\left(2\ell_1+1\right)\left(2\ell_2+1\right)\left(2\ell_3+1\right)}{4 \pi}}\left(\begin{array}{ccc}
\ell_1 & \ell_2 & \ell_3 \\ 0 & 0 & 0
\end{array}\right).
\label{angle_averaged_bisp2}
\end{eqnarray}
Finally $\sigma^{2}\big(f_{\rm nl}\big)$ is the expected variance of
the \fnl parameter, given in terms of its Fisher matrix
\beq 
\sigma^{-2}\big(f_{\rm nl}\big) = \sum_{\ell_1 \le
  \ell_2 \le \ell_3}^{\ell_{\rm max}} \frac{\big(b^{(\prim)}_{\ell_1 \ell_2
    \ell_3}I_{\ell_{1}  \ell_{2}  \ell_{3}}\big)^2} {\sigma^2\big(b^{(\obsv)}_{\ell_1 \ell_2 \ell_3}\big)}.
\label{eq:fnl_error_point_sources}
\eeq
The values of $\sigma^{2}\big(f_{\rm nl}\big)$ for the different maps
are given in Table\,\ref{t2}. For the SEVEM combined map they agree
well with the values published by
\citet{planck_xxiv_2013}\footnote{During the process of
  publication of this article the Planck Collaboration published new
  scientific results, in particular the first results on $f_{nl}$ with
  temperature and polarisation maps \citep{planck_xix_2015}. No
  significant changes have been reported in the new Planck article
  regarding the results with temperature only.}. Note that at 143 and
217 GHz, the component separation technique slightly increases the
error on $f_{\rm nl}$. This is due to the extra noise added by the
template subtraction and it might be seen as the price to pay to have
CMB cleaned maps.

\subsection{\fnl bias due to CIB--lensing and extragalactic
  sources bispectra}
\begin{figure*}
\includegraphics[width=58mm,height=58mm]{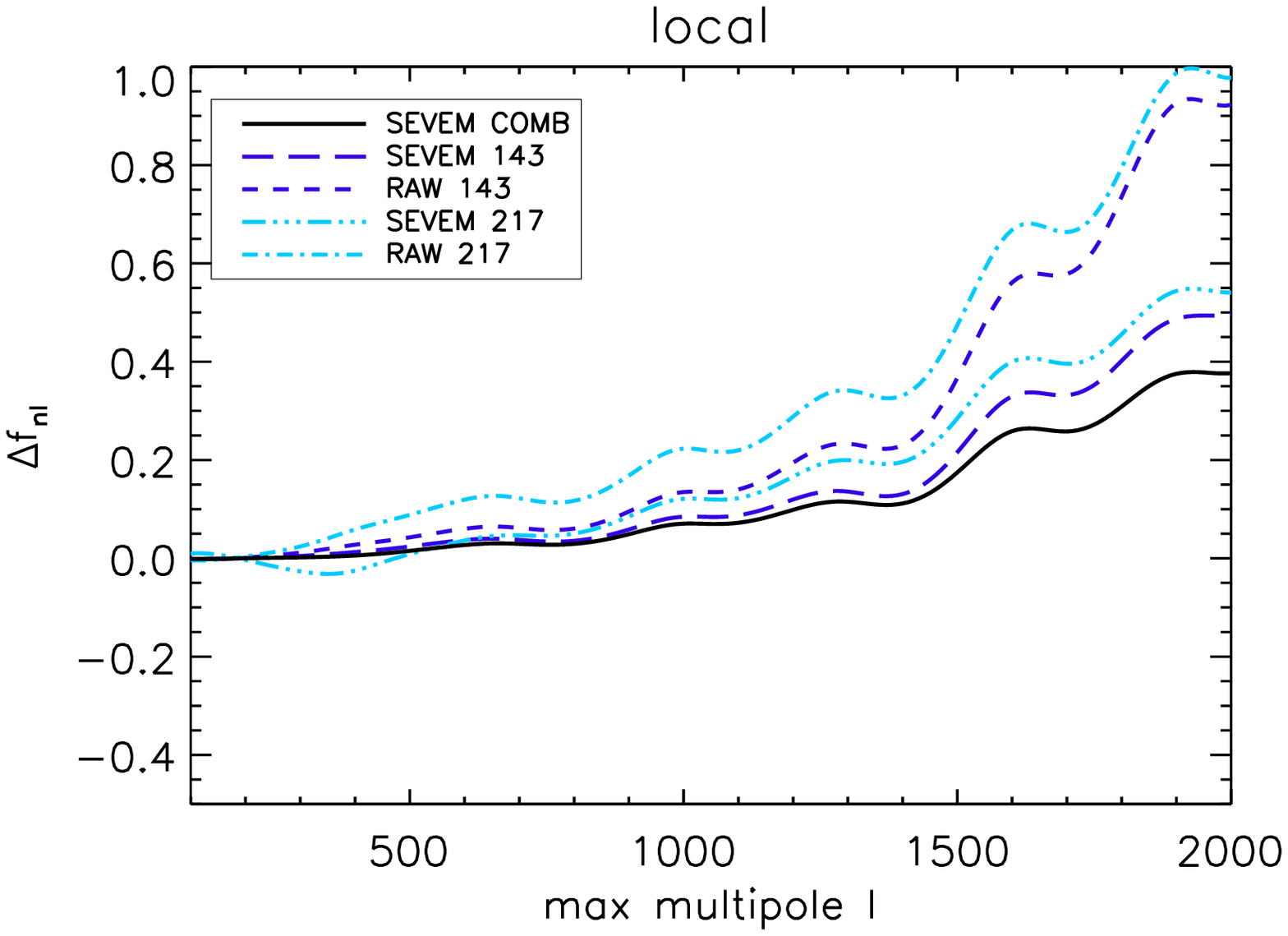}
\includegraphics[width=58mm,height=58mm]{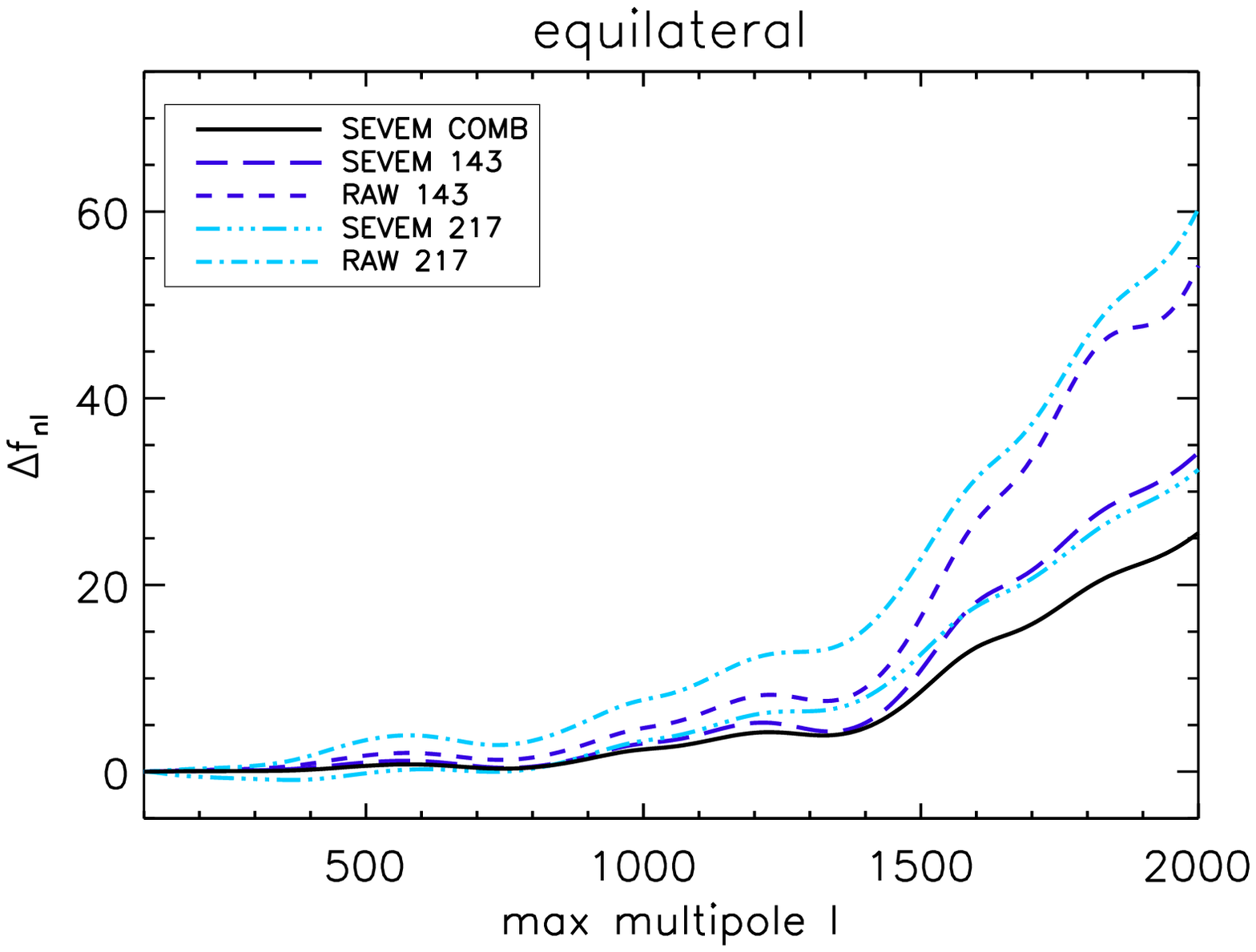}
\includegraphics[width=58mm,height=58mm]{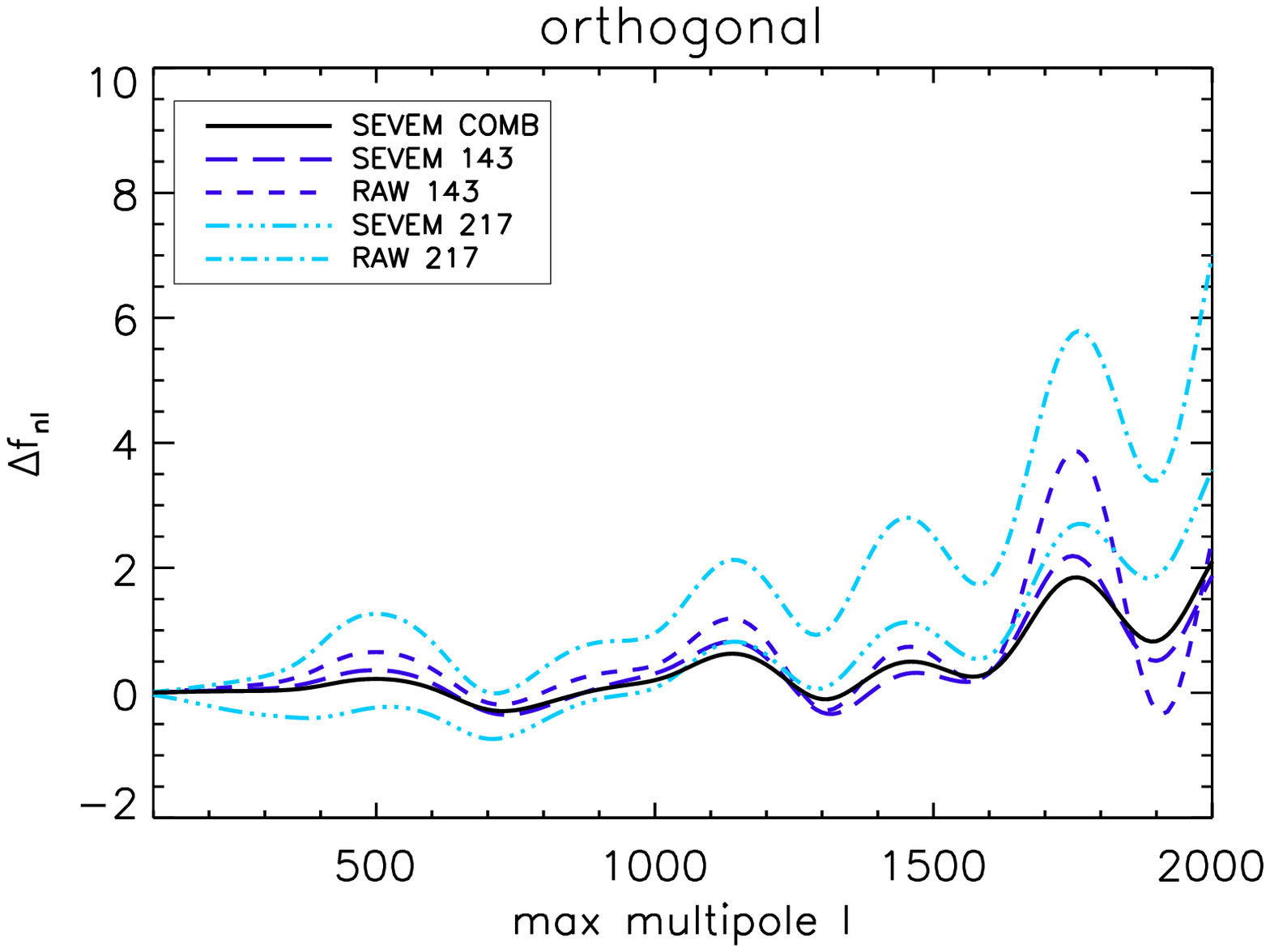}
\caption{
  The bias $\Delta f_{\rm nl}$ produced by the point sources
  bispectrum as a function of $\ell_{\rm max}$ for \Planck
  cosmological frequencies (143 and 217 GHz). Solid lines correspond
  to the cleaned combined maps. Dash lines correspond to the raw map
  at 143 GHz and long dash lines correspond to the cleaned map at 143
  GHz. Dash-and-dot lines correspond to the raw map at 217 GHz and
  dash-and-3-dots lines correspond to the cleaned map at 217 GHz.
  From left to right, we plot results for the local, equilateral and
  orthogonal \fnl shapes. }
\label{f4b}
\end{figure*}

The bias $\Delta f_{\rm nl}$ induced by the CIB-lensing correlation
and unresolved extragalactic sources is estimated for the three types
of maps previously described and for the local, equilateral and
orthogonal \fnl shapes in the frequency range between 100 and
353\,GHz. Results are presented in Table\,\ref{t2} and Figure \ref{f4}
for the CIB-lensing, and Table\,\ref{t3} and Figure \ref{f4b} for the
unresolved extragalactic sources.

For the frequency range considered here, the CIB-lensing bispectrum
causes negligible bias in the primordial local and equilateral shapes
with respect to the uncertainty on $f_{\rm nl}$ estimated by Eq.\
(\ref{eq:fnl_error_point_sources}). The bias is also negligible for
the orthogonal shape in the 100 and 143 GHz raw maps but reaches 2 and
3 $\sigma$ detection levels for the 217 and 353 GHz raw maps
respectively. The orthogonal bias is again negligible for the
foreground-reduced maps at 143 and 217 GHz and the combined map (see
Table \ref{t2} and Fig. \ref{f4}). These results are explained as
follows. Regarding the local shape, the CMB primordial signal peaks in
squeezed configurations, such as $ (\ell_1, \ell_2, \ell_3)=(1000,
1000, 2)$. However in this regime, the CIB-lensing bispectrum loses
most of its amplitude (see top-left panel of Fig. \ref{f2}). Regarding
the equilateral shape, the CMB primordial signal is spread in
configurations such that $ \ell_1 = \ell_2 = \ell_3=\ell$ and becomes
strongest at high resolution. The CIB-lensing bispectrum has some
peaks in equilateral configurations (see bottom-left panel of
Fig. \ref{f2}) but they are located at low $ \ell$ and therefore they
do not significantly couple with the CMB primordial equilateral
signal. Finally regarding the orthogonal shape, the CMB primordial
signal is peaked in configurations such as $ \ell_2 = \ell_3 = 2
\ell_1 $ and $ \ell_2 = \ell_3 = \ell_1$ \citep{martinez2012}. They
couple with the CIB-lensing signal producing an increasing bias for
multipoles $\ell > 500$ (see Fig. \ref{f4}). This explains the bias
predicted for the raw channels at high frequency. The process of
cleaning through the component separation subtracts part of the CIB
signal and the bias for this shape is reduced in cleaned maps to about
30 per cent of the {\it Planck} uncertainty.

The unresolved extragalactic sources present levels of detection
greater than $2\sigma$ at 353 GHz for the local, equilateral and
orthogonal shapes due to the higher amplitude of IR sources at this
frequency. There is a $2\sigma$ detection at 100 GHz for the
equilateral shape which can be explained as a trace of the radio
sources. The bias is again negligible for the foreground-reduced maps
as the component separation technique is able to significantly reduce
their contamination (see Table \ref{t3} and Fig. \ref{f4b}).  These
results are well in agreement with previous analyses from
\citet{lacasa2012a} and \citet{curto2013}.  These results are
explained as follows. Regarding the local shape, the bispectrum of
extragalactic sources does not have a significant signal in squeezed
configurations at low frequencies (see top-left panel of
Fig. \ref{f2}) and therefore there we do not expected significant
correlations with the CMB primordial local bispectrum. However the
high frequency channels contain a significant contribution from IR
sources in squeezed configurations that couple with the local CMB
bispectrum. Regarding the equilateral shape, the extragalactic radio
sources have significant signal in equilateral configurations ($
\ell_1 = \ell_2 = \ell_3 $) at high multipoles (see bottom-left panel
of Fig. \ref{f2}) explaining the deviation seen in Table
{\ref{t3}}. The bispectrum of IR sources is dominant in equilateral
configurations and at high multipoles explaining the large bias
predicted at 353 GHz. Finally regarding the orthogonal shape, only the
high frequency IR source bispectrum has a strong signal in
configurations that couple with the CMB primordial orthogonal
bispectrum, especially for high multipoles, explaining the large bias
predicted at 353 GHz.
\begin{table*}
  \center
  \caption{\Planck expected $\Delta f_{\rm nl}$ bias due to unresolved
    point sources for the local, equilateral and orthogonal \fnl
    shapes for 
    $\ell_{\rm max} =$2000.}
    \label{t3}
  \begin{tabular}{c|ccccccc}
    \hline 
    \hline
    Frequency (GHz) & 100 & 143 & 217 & 353 &  SEVEM 143 & SEVEM 217 & SEVEM combined\\
    \hline
    %
    Local $\Delta f_{\rm nl}$ & 2.95 &       0.92 &       0.98 &      29.30 &       0.50 &       0.54 &       0.35 \\
    Local $\Delta f_{\rm nl}/\sigma(f_{\rm nl})$  & 0.44 &       0.18 &       0.18 &       2.08 &       0.09 &       0.09 &       0.07 \\
    \hline 
    Equilateral $\Delta f_{\rm nl}$ & 160.07 &      54.20 &      60.32 &    1648.70 &      34.19 &      32.36 &      30.38 \\
    Equilateral $\Delta f_{\rm nl}/\sigma(f_{\rm nl})$  & 2.10 &       0.79 &       0.85 &      12.25 &       0.48 &       0.45 &       0.45 \\
    \hline 
    Orthogonal $\Delta f_{\rm nl}$ & 10.53 &       2.45 &       7.00 &     553.01 &       1.87 &       3.56 &       2.94 \\
    Orthogonal $\Delta f_{\rm nl}/\sigma(f_{\rm nl})$  & 0.27 &       0.07 &       0.20 &       7.24 &       0.05 &       0.10 &       0.09 \\
    %
    \hline
    \hline
  \end{tabular}
\end{table*}
\section{Detectability of the CIB--Lensing bispectrum}
\label{section:detectability}
In this section we develop statistical tools to detect the
CIB--lensing bispectrum using an alternative approach to the widely
known technique based on the cross-correlation of lensing potential
reconstruction and temperature maps used for example in
\citet{planck_xviii_2013} and \citet{planck_xix_2013}. The
cross-correlation approach followed in these publications used a
lensing reconstruction based on quadratic combination of \Planck CMB
maps \citep{planck_xvii_2013}. The estimators that we propose in this
article are directly defined in terms of cubic combinations of \Planck
maps where the CMB signal is dominant (100 to 217 GHz) and \Planck
maps where the CIB signal is dominant (217 to 857 GHz). Both
approaches are linearly dependent and should result in similar levels
of efficiency to detect the targeted CIB--lensing signal. The
advantage of the new approach defined here is the application for the
first time of a battery of well-known, efficient and optimal
estimators widely used in the primordial bispectrum estimation to
detect the CIB--lensing bispectrum. This approach has already been
applied to the ISW--lensing estimation by \citet{Mangilli:2013sxa}.
\subsection{Single frequency bispectrum estimator}
The optimal estimator for the amplitude of the CIB--lensing
bispectrum, assuming small departures of non-Gaussianity, for the
ideal, full--sky and isotropic instrumental noise is
\beq 
\hat{A}^{(\LensCIB)}=\big(F^{-1}\big)\hat{S}^{(\LensCIB)} 
\label{cib_lens_estimator}
\eeq
where 
\beq
\hat{S}^{(\LensCIB)}=\sum_{2\le \ell_{1} \le \ell_{2} \le \ell_{3}}^{\ell_{\rm max}}\frac{b^{(\obsv)}_{\ell_{1}  \ell_{2}  \ell_{3}}b^{(\LensCIB)}_{\ell_{1}  \ell_{2}  \ell_{3}}}{\sigma^2\big(b^{(\obsv)}_{\ell_{1}  \ell_{2}  \ell_{3}}\big)},
\label{s_estimator_cib_lensing}
\eeq
\beq
F=\sum_{2\le \ell_{1} \le \ell_{2} \le \ell_{3}}^{\ell_{\rm max}}\frac{b^{(\LensCIB)}_{\ell_{1}  \ell_{2}  \ell_{3}}b^{(\LensCIB)}_{\ell_{1}  \ell_{2}  \ell_{3}}}{\sigma^2\big(b^{(\obsv)}_{\ell_{1}  \ell_{2}  \ell_{3}}\big)},
\label{fisher_cib_l}
\eeq
and the observed bispectrum $b^{(\obsv)}_{\ell_{1} \ell_{2} \ell_{3}}$
is based on cubic combinations of \Planck data in the $a_{\ell m}$
decomposition on the sphere. Another interesting quantity is the
expected bispectrum signal-to-noise ratio as a function of the largest
scale mode $\ell_{\rm min}$ \citep{lewis2011}:
\beq
F_{\ell_{\rm min}}=\sum_{\ell_{\rm min}\le \ell_{1} \le \ell_{2} \le \ell_{3}}^{\ell_{\rm max}}\frac{b^{(\LensCIB)}_{\ell_{1}  \ell_{2}  \ell_{3}}b^{(\LensCIB)}_{\ell_{1}  \ell_{2}  \ell_{3}}}{\sigma^2\big(b^{(\obsv)}_{\ell_{1}  \ell_{2}  \ell_{3}}\big)},
\label{fisher_cib_lmin}
\eeq
This quantity helps to find the multipole configurations where the
bispectrum signal peaks. 

The signal-to-noise ratio of the CIB--lensing signal of this estimator
$\sqrt{F}$ is summarised in Table \ref{t4} for the frequency range
between 100 and 857 GHz and $\ell_{\rm max}=$2000 and different sky
fractions available: 100\% (full sky), 30.4\%\footnote{This is the
  percentage of available sky used in the main results of
  \citet{planck_xviii_2013}.} and 10\%. We compute this ratio
considering the CIB--lensing signal alone (Eq. \ref{fisher_cib_l}) and
the joint Fisher matrix for the four non-primordial bispectra used in
this article, i.e, CIB--lensing, ISW--lensing, CIB, and extragalactic
point sources, by using the generalised Fisher matrix between the
bispectra $i$ and $j$:
\beq
F_{ij}=\sum_{2\le \ell_{1} \le \ell_{2} \le \ell_{3}}^{\ell_{\rm max}}\frac{b^{(i)}_{\ell_{1}  \ell_{2}  \ell_{3}}b^{(j)}_{\ell_{1}  \ell_{2}  \ell_{3}}}{\sigma^2\big(b^{(\obsv)}_{\ell_{1}  \ell_{2}  \ell_{3}}\big)}.
\label{fisher_cib_l_ij}
\eeq
The first case is simply $F_{indep}=F_{ii}$ whereas the second is
$F_{joint}=1/\left(F^{-1}\right)_{ii}$ with $i$ being the
CIB--lensing case.  The significance level for the detection of the
CIB--lensing signal with this estimator increases from
approx. 0.5$\sigma$ at 100 GHz to 9.8$\sigma$ at 353 GHz considering
the full sky available. In a more realistic scenario with a 10\% sky
available for the CIB maps, the CIB--lensing signal would be detected
with a maximum precision of 1$\sigma$ at 353 GHz. At higher
frequencies, the CIB spectrum dominates in the denominator in
Eq. (\ref{fisher_cib_l}) leading to low significance levels of
detection for the CIB--lensing bispectrum.
\begin{table*}
  \begin{center}
    \caption{
      Signal-to-noise ratio $\sqrt{F}$ of the amplitude of the
      CIB--lensing bispectrum for the \Planck raw maps at frequencies
      between 100 and 857 GHz using ideal conditions (isotropic
      instrumental noise) from the Fisher matrix in
      Eq. (\ref{fisher_cib_l}) and several sky fractions available.}
    \label{t4}
    \begin{tabular}{c|cccccc}
      \hline
      \hline
      Frequency (GHz) & 100 & 143 & 217 & 353 & 545 & 857 \\
      \hline
      $F_{indep}^{1/2} \left(f_{sky}=1\right)$ &  0.55 &         1.80 &         7.57 &        10.19 &         0.25 &         0.00 \\ 
      $F_{indep}^{1/2} \left(f_{sky}=0.304\right)$ & 0.17 &         0.55 &         2.30 &         3.10 &         0.08 &         0.00 \\ 
      $F_{indep}^{1/2} \left(f_{sky}=0.100\right)$ & 0.06 &         0.18 &         0.76 &         1.02 &         0.03 &         0.00 \\ 
\hline
      $F_{joint}^{1/2} \left(f_{sky}=1\right)$ &  0.52 &         1.66 &         6.87 &         9.80 &         0.25 &         0.00 \\ 
      $F_{joint}^{1/2} \left(f_{sky}=0.304\right)$ & 0.16 &         0.51 &         2.09 &         2.98 &         0.07 &         0.00 \\ 
      $F_{joint}^{1/2} \left(f_{sky}=0.100\right)$ & 0.05 &         0.17 &         0.69 &         0.98 &         0.02 &         0.00 \\
      \hline
      \hline
    \end{tabular}
  \end{center}
\end{table*}
\subsection{Asymmetric estimator for CMB--CIB correlated maps}
\label{s4ss3}
\begin{figure*}
\includegraphics[width=58mm,height=58mm]{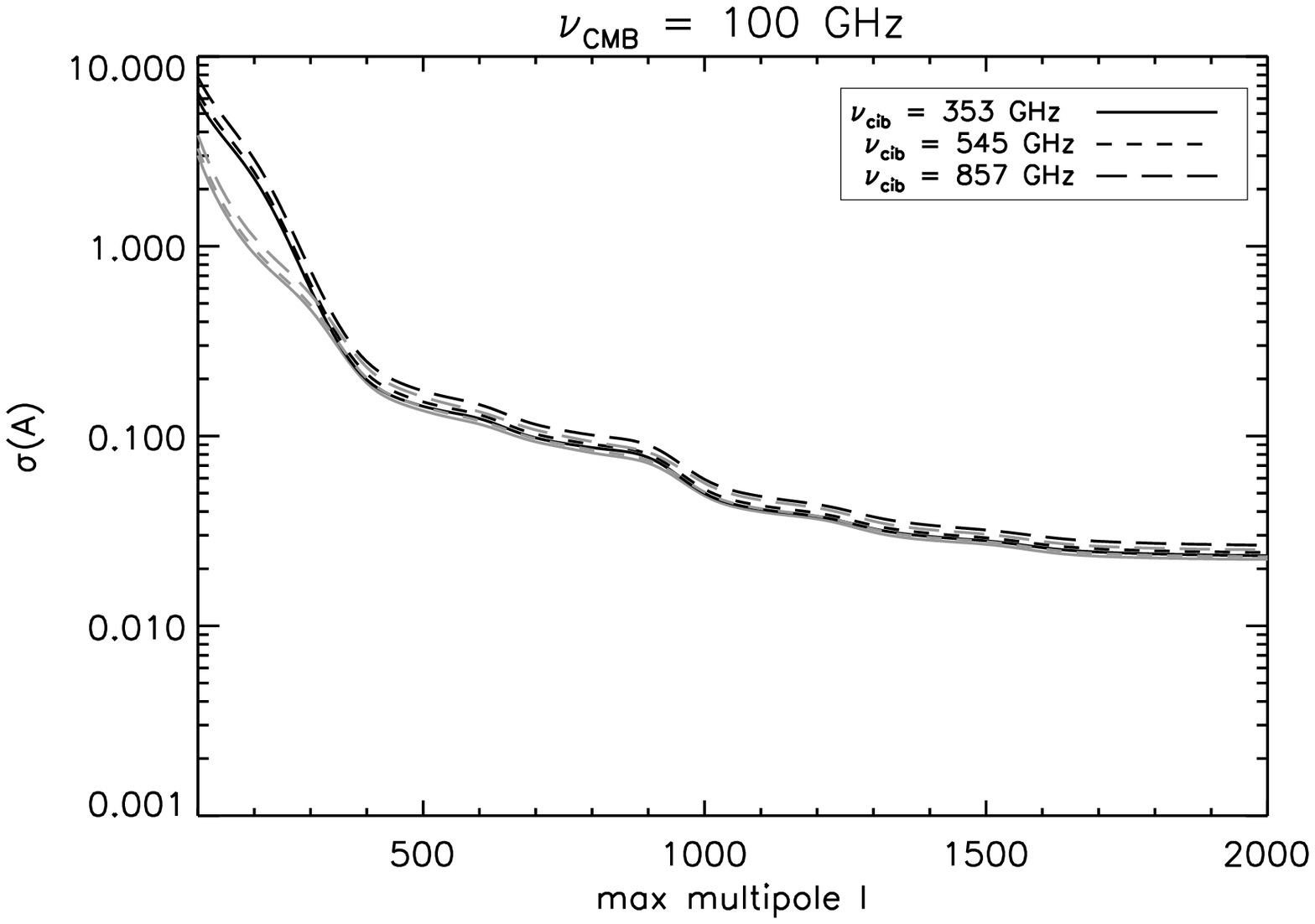}
\includegraphics[width=58mm,height=58mm]{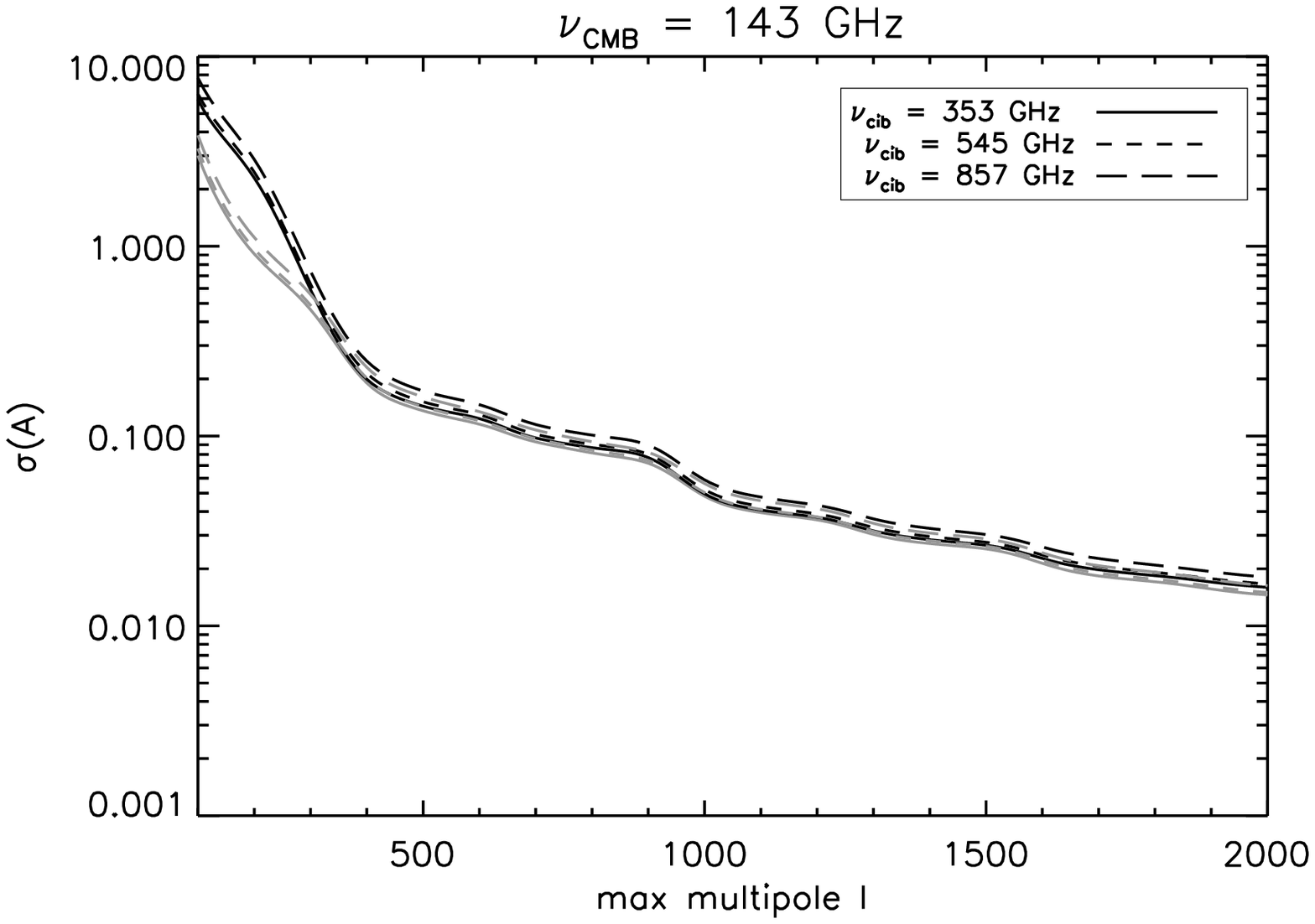}
\includegraphics[width=58mm,height=58mm]{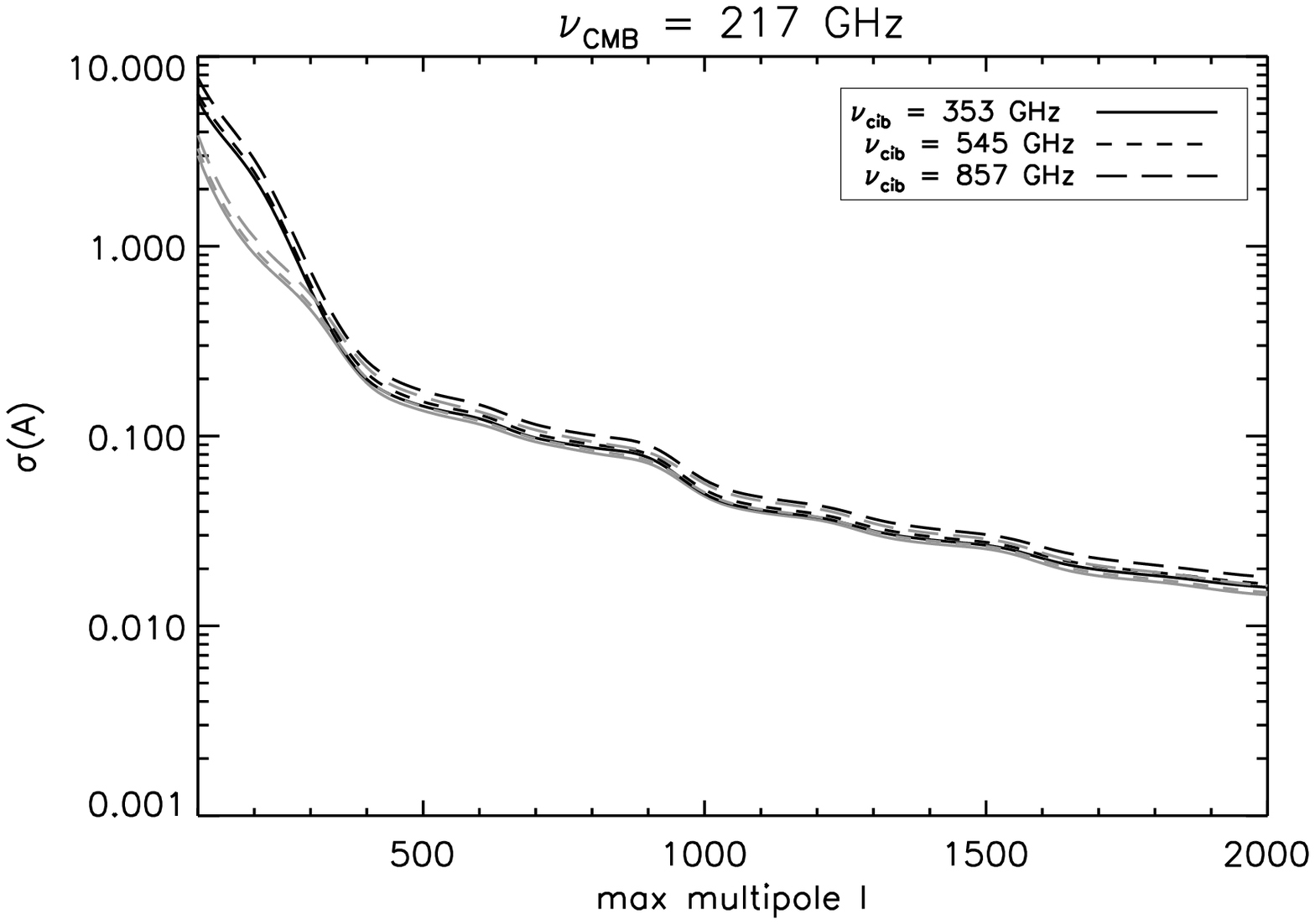}
\includegraphics[width=58mm,height=58mm]{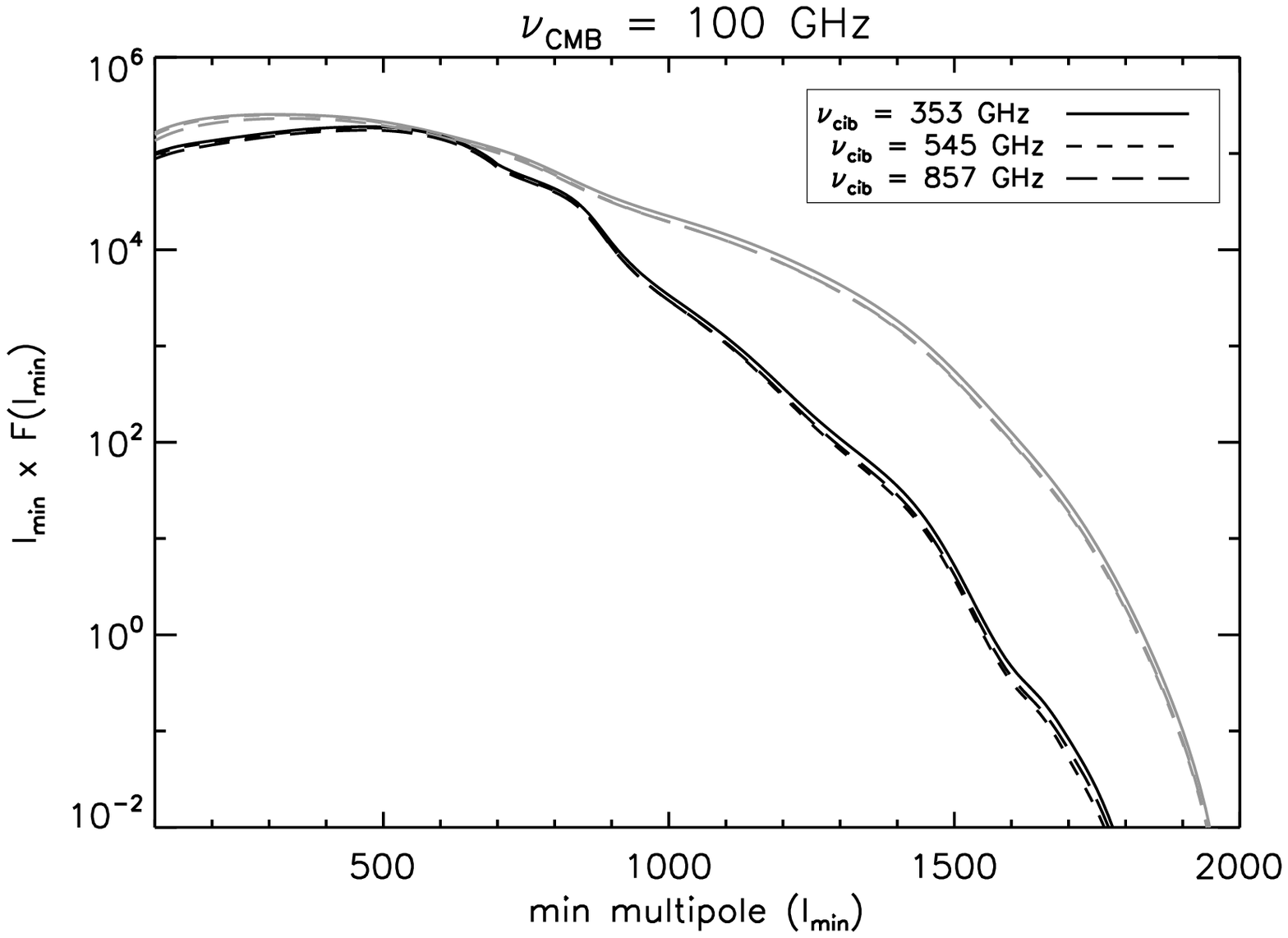}
\includegraphics[width=58mm,height=58mm]{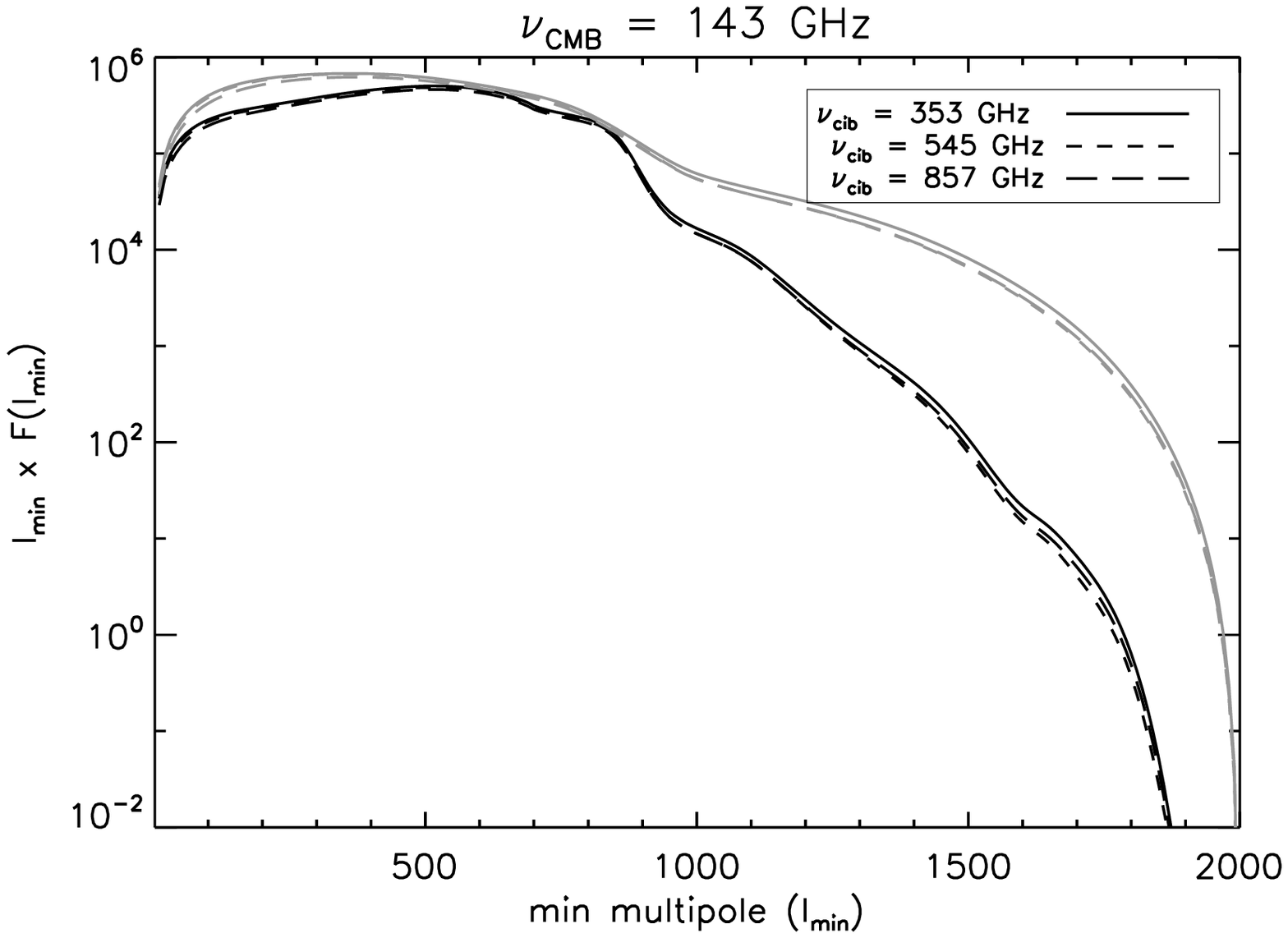}
\includegraphics[width=58mm,height=58mm]{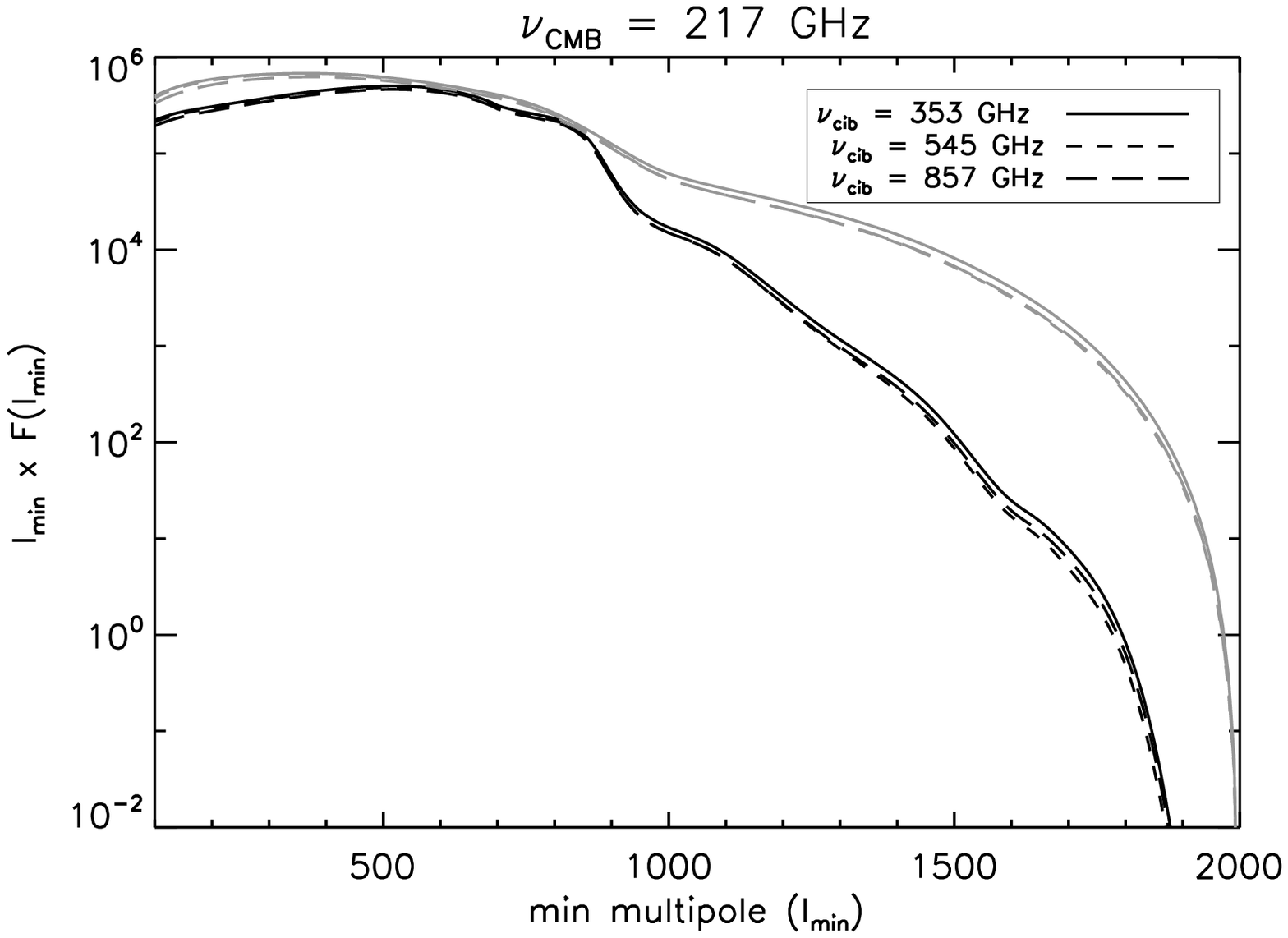}
\caption{{\it Top:} the uncertainty for the CIB--lensing bispectrum
  amplitude $\sigma(A)$ as a function of $\ell_{\rm max}$ for cleaned
  CMB maps (at 100, 143 and 217\,GHz) combined with CIB dominated maps
  at 353, 545 and 857\,GHz using the independent estimates (grey
  lines) and using the joint estimates (black lines). {\it Bottom: }
  the Fischer matrix $F_{\ell_{\rm min}}$ (multiplied by $\ell_{\rm
    min}$) as a function of $\ell_{\rm min}$ for the same cases as
  above. }
\label{f5}
\end{figure*}

At high \Planck frequencies the CIB bispectrum could strongly limit
our capability to detect the CIB--lensing signal, as well as the
ISW--lensing contribution could be a relevant ``noise'' at
100--217\,GHz. Therefore, a more feasible procedure to detect the
CIB--lensing bispectrum signal should correlate CMB
  signal-dominated maps and CIB signal-dominated maps. We define an
  estimator for the CIB-lensing signal by considering the asymmetric
configuration $\tilde{a}^{(\raw, \nu_\CMB)}_{\ell_1
  m_1}\tilde{a}^{(\raw, \nu_\CMB)}_{\ell_2 m_2}a^{(\raw,
  \nu_\CIB)}_{\ell_3 m_3}$
where $\nu_\CMB$ and $\nu_\CIB$ are frequency channels where the CMB
and CIB signal are significant respectively.  We consider $\nu_\CMB =
$ 100, 143 and 217 GHz and $\nu_\CIB = $ 217, 353, 545 and 857
GHz. The four non-primordial averaged bispectra considered in this
paper, namely the radio, CIB, CIB-lensing and ISW-lensing bispectra,
are written in this asymmetric configuration by:
\beq 
b^{(\Radio)}_{\ell_{1}  \ell_{2}  \ell_{3}}= b^{(\Radio,\nu_\CMB,\nu_\CMB,\nu_\CIB)}_{\sn}
\label{asym_radio}
\eeq
\beq 
 b^{(\CIB)}_{\ell_{1}  \ell_{2}  \ell_{3}}= b^{(\CIB,\nu_\CMB,\nu_\CMB,\nu_\CIB)}_{\sn} \sqrt{\frac{C^{(\CIB,\nu_\CMB)}_{\ell_1}C^{(\CIB,\nu_\CMB)}_{\ell_2}C^{(\CIB,\nu_\CIB)}_{\ell_3}}{C^{(\CIB,\nu_\CMB)}_{\sn}C^{(\CIB,\nu_\CMB)}_{\sn}C^{(\CIB,\nu_\CIB)}_{\sn}}} 
\label{asym_cib}
\eeq
\begin{align}
\nonumber
 &   b^{(\LensCIB)}_{\ell_{1}  \ell_{2}  \ell_{3}} = \\
&   \bigg[{\ell_1(\ell_1+1)-\ell_2(\ell_2+1) 
+\ell_3(\ell_3+1) \over 2}\,\tilde{C}^{(\CMB)}_{\ell_1}\,C^{(\LensCIB,\nu_\CIB)}_{\ell_3}
+{\ell_2(\ell_2+1)-\ell_1(\ell_1+1) 
+\ell_3(\ell_3+1) \over 2}\,\tilde{C}^{(\CMB)}_{\ell_2}\,C^{(\LensCIB,\nu_\CIB)}_{\ell_3}\bigg]\,,
\label{asym_cib_lens}
\end{align}
and 
\begin{align}
 b^{(\LensISW)}_{\ell_{1}  \ell_{2}  \ell_{3}} = 
\bigg[{\ell_1(\ell_1+1)-\ell_2(\ell_2+1) 
+\ell_3(\ell_3+1) \over 2}\,\tilde{C}^{(\CMB)}_{\ell_1}\,C^{(\LensISW)}_{\ell_3}
+(5~perm)\bigg] .
\label{asym_isw_lens}
\end{align}
The covariance matrix is nearly diagonal and can be approximated by
the following expression (see Appendix\,\ref{appendix_cov_bisp}) for
the CMB x CMB x CIB configurations:
\bea 
\sigma^2\big(b^{(\obsv)}_{\ell_{1} \ell_{2} \ell_{3}}\big)= \frac{1}{I_{\ell_{1} \ell_{2} \ell_{3}}^2}\tilde{C}^{(\CMB)}_{\ell_1}\tilde{C}^{(\CMB)}_{\ell_2}C^{(\CIB)}_{\ell_3}\left(1+\delta_{\ell_1 \ell_2}\right).
\label{cmbcmbcib_variance_asymm}
\eea
The CIB-lensing estimator for this configuration is now:
\beq
\hat{S}^{(\LensCIB)}=\sum_{2\le \ell_{1} \le \ell_{2}, \ell_{3}}^{\ell_{\rm max}}\frac{b^{(\obsv)}_{\ell_{1}  \ell_{2}  \ell_{3}}b^{(\LensCIB)}_{\ell_{1}  \ell_{2}  \ell_{3}}}{\sigma^2\big(b^{(\obsv)}_{\ell_{1} \ell_{2} \ell_{3}}\big)}.
\label{asymmetric_cmbcmbcib_s}
\eeq
The estimator is different with respect to the single map estimators
since it admits more configurations because we just have symmetry
  under permutations of $\ell_1$ and $\ell_2$ whereas $\ell_3$ is
  free.  The Fisher matrix for the four bispectra considered in this
  work is defined as \citep{komatsu2001}:
\beq
 F_{ij}=\sum_{2\le \ell_{1} \le \ell_{2}, \ell_{3}}^{\ell_{\rm max}}\frac{b^{(i)}_{\ell_{1}  \ell_{2}  \ell_{3}}b^{(j)}_{\ell_{1}  \ell_{2}  \ell_{3}}}{\sigma^2\big(b^{(\obsv)}_{\ell_{1} \ell_{2} \ell_{3}}\big)}
\label{fisher_cib_l_assym}
\eeq
where the indices $i$ and $j$ cover the following bispectra: (1) radio, (2)
CIB, (3) CIB-lensing and (4) ISW-lensing. We also define the Fisher
matrix in terms of the minimum multipole $\ell_{\rm min}$:
\beq
 \left(F_{ij}\right)_{\ell_{\rm min}}=\sum_{\ell_{\rm min}\le \ell_{1} \le \ell_{2}, \ell_{3}}^{\ell_{\rm max}}\frac{b^{(i)}_{\ell_{1}  \ell_{2}  \ell_{3}}b^{(j)}_{\ell_{1}  \ell_{2}  \ell_{3}}}{\sigma^2\big(b^{(\obsv)}_{\ell_{1} \ell_{2} \ell_{3}}\big)} \, .
\label{fisher_cib_l_assym2}
\eeq
The Cram\'er-Rao inequality states that the inverse of the Fisher
information matrix is a lower bound on the variance of any unbiased
estimator in the best optimal conditions. Therefore the variance of
the amplitude of each bispectra can be obtained by inverting the
Fisher matrix
\beq
 \sigma_{\rm joint}^2(A_i) = \left( F^{-1} \right)_{ii}.
\label{sigma2_assym}
\eeq
This approach performs a joint analysis including the correlations
among the four types of bispectra, in comparison to the independent
constraint that would have a lower variance:
\beq
 \sigma_{\rm indep}^2(A_i) = \left( F_{ii} \right)^{-1}.
\label{sigma2_assym_indep}
\eeq
{The signal-to-noise ratio is given by:}
\beq
 F_{\rm indep}^{1/2}(A_i) = F^{1/2}_{ii} ~~~~~~~~~~~~~~~~~~  F_{\rm joint}^{1/2}(A_i) = 1/\sqrt{\left(F^{-1}\right)_{ii}}.
\label{sigma2_assym_indep}
\eeq
The expected uncertainties for the CIB-lensing estimator, computed
both using the independent and the joint approach, are plotted in Fig.
\ref{f5} and is summarised in Table \ref{t5b} for $\ell_{\rm
  max}=$2000. In an ideal scenario where the CMB and the CIB maps are
completely separated in two full sky maps, we would have a
detectability level of approximately $63\sigma$ in the best
configurations given in Table \ref{t5b} for $\ell_{\rm
  max}=$2000. However available CIB maps cover only about 10\% of the
sky \citep{planck_xxx_2013} in the best case scenarios. For the
incomplete sky case, the detectability level of the bispectrum is
rescaled by the fraction of the available sky $f_{\sky}$ such that
$\sigma(A) \longrightarrow \sigma(A)/ \sqrt{f_{\sky}}$. As the
CIB--lensing bispectrum is not squeezed (see Fig.  \ref{f5}), this
case is not affected by the loss of low multipoles, which are
unobservable for small sky fractions, so this approximation is safe up
to the mentioned 10\% of the sky.  This would provide detectability
levels between 12$\sigma$ to 20$\sigma$ respectively for a mask with
10\% of the sky available using the joint estimator (see Table
\ref{t5b}).
\begin{table*}
  \begin{center}
    \caption{The signal-to-noise ratio $\sqrt{F}$ of the amplitude of
      the CIB--lensing bispectrum for clean full--sky maps using
        the joint (Eq. \ref{sigma2_assym}) and independent
        (Eq. \ref{sigma2_assym_indep}) Fisher matrices for the
      combinations the clean CMB at frequencies between 100 and 217
      GHz and the clean CIB at frequencies between 353 and 857 GHz
      using 100\%, (30.4\%), [10\%] of the sky..
      \label{t5b}}
    \begin{tabular}{c|c|cccc}
      \hline
      \hline
      Case & Frequency & $\nu_{\CIB}$ = 353 GHz &  $\nu_{\CIB}$ = 545 GHz &  $\nu_{\CIB}$ = 857 GHz \\
      \hline
      Joint & $\nu_{\CMB}$ = 100 GHz &    42.70 (     23.54)  [     13.50]  &      41.03 (     22.62)  [     12.98]  &      37.60 (     20.73)  [     11.89]  \\ 
      Joint & $\nu_{\CMB}$ = 143 GHz &  62.66 (     34.55)  [     19.82]  &      60.15 (     33.17)  [     19.02]  &      55.40 (     30.54)  [     17.52]  \\
      Joint & $\nu_{\CMB}$ = 217 GHz &  62.37 (     34.39)  [     19.72]  &      59.90 (     33.03)  [     18.94]  &      55.19 (     30.43)  [     17.45]  \\ 
      \hline
      Independent & $\nu_{\CMB}$ = 100 GHz & 44.65 (     24.62)  [     14.12]  &      43.21 (     23.83)  [     13.66]  &      39.71 (     21.90)  [     12.56]  \\ 
      Independent & $\nu_{\CMB}$ = 143 GHz &68.70 (     37.88)  [     21.73]  &      66.61 (     36.73)  [     21.06]  &      61.42 (     33.86)  [     19.42]  \\ 
      Independent & $\nu_{\CMB}$ = 217 GHz &68.51 (     37.77)  [     21.66]  &      66.42 (     36.62)  [     21.00]  &      61.26 (     33.78)  [     19.37]  \\ 
      \hline
      \hline
    \end{tabular}
  \end{center}
\end{table*}

We have compared our results with the estimates of the CIB--lensing
correlation published in \citet{planck_xviii_2013}. The statistical
estimator used in that work is based on a cross-correlation between
the lensing potential in harmonic space, $\phi_{\ell m}$, and a
temperature map. The amplitude of the detection is obtained with the
quadrature sum of the significance of the different multipole
bins. This amplitude takes into account effects such as correlations
among different bins but no systematic errors or point source
corrections. The final estimates presented in that paper are computed
using the lensing reconstruction at 143 GHz and the \Planck HFI
foreground reduced maps with a mask of 30.4\% of available sky. The
results obtained in this way and our predictions for the same
configuration are given in Table \ref{t7}. \citet{planck_xviii_2013}
provides two types of detection significances: one that only includes
the statistical errors only and another one that includes statistical
and systematic errors. Compared to the first one -- as we do not
consider systematic errors here -- the detection significance by
\citet{planck_xviii_2013} and our model are nearly equivalent for the
353 to 857 GHz range. Note that the values measured by
\citet{planck_xviii_2013} are quite sensitive to the systematic
effects (see Table \ref{t7}).  We have additionally repeated our
analysis without adding instrumental noise, i.e.\ for ideal
conditions, and have found $\sqrt{F}=$ 39, 38, 35 for the 353, 545 and
857 GHz bands. Both cases show a decreasing trend in the
signal-to-noise level of the CIB-lensing correlation as we increase
the frequency, due to the higher contamination of the CIB. We do not
see a peak at 545 GHz, and we think that the peak observed in
\citet{planck_xviii_2013} at 545 GHz might be explained by an unknown
systematic artefact present in the data.
\begin{table*}
  \begin{center}
    \caption{{ Significance of the amplitude of the CIB--lensing
        correlation using the lensing reconstruction at 143 GHz
        and the \Planck HFI foreground reduced maps with a mask of
        30.4\% of available sky. {\it Top line:} measurements by
        \citet{planck_xviii_2013}. {\it Bottom line:} our predictions
        for an optimal bispectrum estimator for the same
        configuration.
      \label{t7}}}
    \begin{tabular}{c|ccc}
      \hline
      \hline
       Case &     $\nu_{\CIB}$ = 353 GHz &    $\nu_{\CIB}$ = 545 GHz &    $\nu_{\CIB}$ = 857 GHz \\
      \hline
       Number of standard deviations$^a$ \citep{planck_xviii_2013} &    31 (24) &   42 (19) &   32 (16) \\
      \hline
       $\sqrt{F}$   &         35 &          33 &          31 \\ 
      \hline
      \hline
    \end{tabular}
\begin{tablenotes}
\small
\item  $^a$ Statistical error and statistical plus systematic errors in parenthesis.
\end{tablenotes}
  \end{center}
\end{table*}

The Wick expansions, used to compute the variance of the observed
bispectra (see Appendix\,\ref{appendix_cov_bisp}), are good
approximations when the departures from non-Gaussianity are
limited. Therefore, we could expect some contributions to the
covariance matrix of the bispectrum in
Eq.\,(\ref{cmbcmbcib_variance_asymm}) due to higher order moments
\citep[see Section 4.3 in][]{planck_xxx_2013}. We have computed higher
order contributions to the variance in Appendix
\ref{appendix_cov_bisp}. The result is that higher order moments do
not add a significant contribution to the covariance and that the
Wick expansions used in this paper hold. 
\section{KSW-based estimators for the CIB-lensing bispectrum}
\label{section5}
We present the formalism of an optimal estimator for the amplitude of
the bispectrum induced by the CIB--lensing correlation, based on the
estimator developed by \citet[][KSW]{Komatsu:2003iq} for the
primordial non-Gaussianity and extended to the ISW--lensing bispectrum
by \citet{Mangilli:2013sxa}. Here we consider the ideal case without
noise and beam function. The case of a realistic experiment can be
straightforward extended \citep[see,
e.g.,][]{Lacasa:2012zx,Mangilli:2013sxa}.
\subsection{Single frequency bispectrum estimator}
The optimal estimator $\hat{S}$ of the CIB--lensing bispectrum for a
single frequency map
is given by
Eq. (\ref{s_estimator_cib_lensing}).  Using the identity
\beq
\sum_{\ell_1 \le \ell_2 \le \ell_3}^{\ell_{\rm max}}F_{\ell_1  \ell_2  \ell_3} = \frac{1}{6}\sum_{\ell_1  \ell_2  \ell_3}^{\ell_{\rm max}}F_{\ell_1  \ell_2  \ell_3}\Delta_{\ell_1  \ell_2  \ell_3}
\eeq
for any given $F_{\ell_1\ell_2\ell_3}$ symmetric in $\ell_1$, $\ell_2$,
$\ell_3$, we can write the estimator $\hat{S}$ as:
\bea
\hat{S}^{\LensCIB} & = & {1 \over 6}\,\sum_{\ell_1\ell_2\ell_3}\,\sum_{m_1m_2m_3}
\left(\begin{array}{ccc}
  \ell_1 & \ell_2 & \ell_3\\ m_1 & m_2 & m_3
\end{array}\right)
{a_{\ell_1m_1}a_{\ell_2m_2}a_{\ell_3m_3} \over 
C_{\ell_1}C_{\ell_2}C_{\ell_3}}\,
\sqrt{(2\ell_1+1) (2\ell_2+1) (2\ell_3+1) \over 4\pi}
\left(\begin{array}{ccc}
\ell_1 & \ell_2 & \ell_3 \\ 0 & 0 & 0
\end{array}\right)
b^{(\LensCIB)}_{\ell_1\ell_2\ell_3}\nonumber \\
& = & {1 \over 6}\,\int\,d^2{\hn}\,\sum_{\ell_1\ell_2\ell_3}
\,\sum_{m_1m_2m_3}
{a_{\ell_1m_1}a_{\ell_2m_2}a_{\ell_3m_3} \over 
C_{\ell_1}C_{\ell_2}C_{\ell_3}}
Y_{\ell_1 m_1}({\hn}) Y_{\ell_2 m_2}({\hn}) 
Y_{\ell_3 m_3}({\hn})\,b^{(\LensCIB)}_{\ell_1\ell_2\ell_3}\,
\label{s5e4}
\eea
where $C_{\ell}$ is the total power spectrum in the single frequency
map. By including the bispectrum formula (Eq.\,\ref{s2e3}) into
Eq.\,(\ref{s5e4}) and factorizing the $\ell$ dependence, the integral
becomes
\bea
\hat{S}^{\LensCIB} & = & {1 \over 12}\int\,d^2{\hn}\,\sum_{\ell_1\ell_2\ell_3}
\sum_{m_1m_2m_3}\Bigg\{
\Bigg[\ell_1(\ell_1+1)a_{\ell_1m_1}Y_{\ell_1 m_1}({\hn})
{\tilde{C}^{\CMB}_{\ell_1} \over C_{\ell_1}}\Bigg]
\Bigg[{a_{\ell_2m_2}Y_{\ell_2 m_2}({\hn})
\over C_{\ell_2}}\Bigg]
\Bigg[a_{\ell_3m_3}Y_{\ell_3 m_3}({\hn})
{C^{(\LensCIB)}_{\ell_3} \over C_{\ell_3}}\Bigg] \nonumber \\
 & & -\Bigg[a_{\ell_1m_1}Y_{\ell_1 m_1}({\hn})
{\tilde{C}^{\CMB}_{\ell_1} \over C_{\ell_1}}\Bigg]
\Bigg[\ell_2(\ell_2+1){a_{\ell_2m_2}Y_{\ell_2 m_2}({\hn})
\over C_{\ell_2}}\Bigg]
\Bigg[a_{\ell_3m_3}Y_{\ell_3 m_3}({\hn})
{C^{(\LensCIB)}_{\ell_3} \over C_{\ell_3}}\Bigg] \nonumber \\
 & & +\Bigg[a_{\ell_1m_1}Y_{\ell_1 m_1}({\hn})
{\tilde{C}^{\CMB}_{\ell_1} \over C_{\ell_1}}\Bigg]
\Bigg[{a_{\ell_2m_2}Y_{\ell_2 m_2}({\hn})
\over C_{\ell_2}}\Bigg]
\Bigg[\ell_3(\ell_3+1)a_{\ell_3m_3}Y_{\ell_3 m_3}({\hn})
{C^{(\LensCIB)}_{\ell_3} \over C_{\ell_3}}\Bigg]+5perm.
\Bigg\}.
\label{s5e5}
\eea
Now, if we define the following filtered maps
\beq
P({\hn}) \equiv \sum_{\ell m}\,a_{\ell m}Y_{\ell m}({\hn})
{\tilde{C}^{(\CMB)}_{\ell} \over C_{\ell}} ~~~~~~~~~~
Q({\hn}) \equiv \sum_{\ell m}\,a_{\ell m}Y_{\ell m}({\hn})
{C^{(\LensCIB)}_{\ell} \over C_{\ell}} ~~~~~~~~~~
E({\hn}) \equiv \sum_{\ell m}\,{a_{\ell m}Y_{\ell m}({\hn})
\over C_{\ell}} \,,
\eeq
with the corresponding $\delta^2X$ maps (with $X=P,Q,E$) obtained by
substituting $a_{\ell m}$ with $\ell(\ell+1) a_{\ell m}$,
Eq.\,(\ref{s5e5}) becomes 
\beq
\hat{S}^{\LensCIB}={1 \over 2}\int\,d^2{\hn}\,\bigg[
\delta^2P({\hn})E({\hn})Q({\hn})-P({\hn})\delta^2
E({\hn})Q({\hn})+P({\hn})E({\hn})\delta^2Q({\hn})
\bigg]\,.
\label{s5e6}
\eeq
The expressions in this section assume that the full inverse
covariance matrix can be replaced by a diagonal covariance term,
$(C^{-1}a)_{\ell m}\rightarrow a_{\ell m}/C_{\ell}$. In a real
experiment, this approximation might not be valid. In fact, we have to
take into account that clean CIB maps can be obtained only in small
areas of the sky due to the Galactic dust contamination. Here we are
using this approximation just for simplicity. Moreover, when
rotational invariance is broken by, e.g., a Galactic mask or an
anisotropic noise, a linear term should be subtracted from the
estimator in Eq.\,(\ref{s5e4}) \citep[see, e.g.][]{Mangilli:2013sxa}.
This linear term correction for a generic bispectrum shape
$b_{\ell_1 \ell_2 \ell_3}$ is given by:
\beq
 \hat{S}_{lin}= -\frac{1}{2} \int d^2 \hat{n}   \sum_{\ell m} b_{\ell_1 \ell_2 \ell_3} 
\frac{a_{\ell_3m_3}}{C_{\ell_1}C_{\ell_2}C_{\ell_3}} Y_{\ell_1 m_1}({{\hat{n}}}) Y_{\ell_2 m_2}({{\hat{n}}})  Y_{\ell_3 m_3}({{\hat{n}}}).
\label{generic_linear_term}
\eeq
Using the explicit form of the CIB--lensing bispectrum, the linear
term correction for the single-frequency CIB--lensing estimator defined
in Eq. (\ref{s5e6}) is given by:
\begin{eqnarray}
   S_{lin}^{\LensCIB} &=&  \nonumber \\
&-&  \frac{1}{2} \int d^2 \hat{n}  \Big\{ Q( \hat{n}) \Big [ \langle P( \hat{n}) \delta^2E( \hat{n}) \rangle  -  \langle E( \hat{n}) \delta^2P( \hat{n}) \rangle  \Big ] - \delta^2Q( \hat{n}) \langle P( \hat{n}) E( \hat{n}) \rangle - E( \hat{n})  \Big [ \langle Q( \hat{n}) \delta^2P( \hat{n}) \rangle  -  \langle P( \hat{n}) \delta^2Q( \hat{n}) \rangle  \Big ] \nonumber \\ 
&+&    \delta^2E( \hat{n}) \langle P( \hat{n}) Q( \hat{n}) \rangle  
  -  \delta^2P( \hat{n}) \langle E( \hat{n}) Q( \hat{n}) \rangle +  P( \hat{n})  \Big [ \langle Q( \hat{n}) \delta^2E( \hat{n}) \rangle  -  \langle E( \hat{n}) \delta^2Q( \hat{n}) \rangle \Big ]  \Big\}.
  \label{linear_term_single_freq}
  \end{eqnarray}                                                         
  The averages in Eq. (\ref{linear_term_single_freq}) correspond to
  realistic Monte Carlo simulations which contain the instrumental
  properties (beams, noise, masks) and also the type of
  non-Gaussianity we are testing (in this case the CIB--lensing).

  In the above treatment we have also supposed that the CIB--lensing
  term is the only relevant non--Gaussian contribution. This is of
  course not the case when a single frequency map is used. In fact,
  the bispectrum from CIB sources and from ISW--lensing correlation
  can give stronger contributions as shown in Fig.\,\ref{f2}. A joint
  estimation of their amplitude can be applied
  \citep[see][]{Lacasa:2012zx} using estimators developed considering
  only one source of non--Gaussianity.
\subsection{Asymmetric estimator for CMB--CIB correlated maps}
The estimator for CMB--CIB correlated maps is defined in Eq.
(\ref{asymmetric_cmbcmbcib_s}). If we indicate the lensed CMB map and
the CIB map at frequency $\nu$ as
\beq
\bigg({\Delta T \over T}\bigg)^{(\CMB)}(\hn)=
\sum_{\ell m}\,\tilde{a}^{(\CMB)}_{\ell m}Y_{\ell m}({\hn})~~~~~~~~
T_{\nu}^{(\CIB)}({\hn})=\sum_{\ell m}\,a^{(\CIB)}_{\ell m}Y_{\ell m}({\hn})\,,
\label{s5e7}
\eeq
the estimator in Eq. (\ref{asymmetric_cmbcmbcib_s}) becomes
now:
\beq
\hat{S}^{\LensCIB *}={1 \over 2}\,\sum_{\ell_1\ell_2\ell_3}\,\sum_{m_1m_2m_3}
G^{\ell_1\ell_2\ell_3}_{m_1m_2m_3}
{\tilde{a}^{(\CMB)}_{\ell_1m_1}\tilde{a}^{(\CMB)}_{\ell_2m_2}
a^{(\CIB)}_{\ell_3m_3} \over
\tilde{C}^{(\CMB)}_{\ell_1}\tilde{C}^{(\CMB)}_{\ell_2}C^{(\CIB)}_{\ell_3}}\,
b^{(\LensCIB)*}_{\ell_1\ell_2\ell_3}\,,
\label{s5e8}
\eeq
where here the CIB--lensing bispectrum includes only two permutations
\beq
b^{(\LensCIB)*}_{\ell_1\ell_2\ell_3}={\ell_1(\ell_1+1)-\ell_2(\ell_2+1) 
+\ell_3(\ell_3+1) \over 2}\,\tilde{C}^{(\CMB)}_{\ell_1}\,C^{(\LensCIB)}_{\ell_3}
+{\ell_2(\ell_2+1)-\ell_1(\ell_1+1) 
+\ell_3(\ell_3+1) \over 2}\,\tilde{C}^{(\CMB)}_{\ell_2}\,C^{(\LensCIB)}_{\ell_3}\,.
\label{s5e8a}
\eeq
The factor $1/2$ in Eq.\,(\ref{s5e8}) is due to the fact the bispectrum
is now symmetric only in $\ell_1$ and $\ell_2$ and
$\sum_{\ell_1 \le \ell_2,\ell_3} \longrightarrow 
\frac{1}{2}\sum_{\ell_1  \ell_2  \ell_3}(1+\delta_{\ell_1  \ell_2})\,$.
By defining new filtered maps as: 
\beq
 P_{\CMB}({\hn})\equiv \bigg({\Delta T \over T}\bigg)^{(\CMB)}(\hn)=
\sum_{\ell m}\,\tilde{a}^{(\CMB)}_{\ell m} Y_{\ell m}({\hn}) ~~~~~
Q_{\CIB}({\hn})\equiv \sum_{\ell m}\,a^{(\CIB)}_{\ell m}Y_{\ell m}({\hn})
{C^{(\LensCIB)}_{\ell} \over C^{(\CIB)}_{\ell}}~~~~~
E_{\CMB}({\hn})\equiv\sum_{\ell m}\,{\tilde{a}^{(\CMB)}_{\ell m}Y_{\ell
 m}({\hn}) \over \tilde{C}^{(\CMB)}_{\ell}}\,.
\label{s5e9}
\eeq
the estimator can be written in terms of filtered maps as:
\beq
\hat{S}^{\LensCIB *}=-{1 \over 2}\int\,d^2{\hn}\,\bigg[
\delta^2P_{\CMB}({\hn})E_{\CMB}({\hn})Q_{\CIB}({\hn})
-P_{\CMB}({\hn})\delta^2E_{\CMB}({\hn})Q_{\CIB}({\hn})+
P_{\CMB}({\hn})E_{\CMB}({\hn})\delta^2Q_{\CIB}({\hn})
\bigg]\,.
\label{s5e10}
\eeq
The linear term correction of this estimator can be straightforwardly
computed following the same steps already developed for the
single-frequency estimator:
\begin{eqnarray}
   S_{lin}^{\LensCIB *} &=&  \nonumber \\
&-&  \frac{1}{2} \int d^2 \hat{n}  \Big\{ Q_{\CIB}( \hat{n}) \Big [ \langle P_{\CMB}( \hat{n}) \delta^2E_{\CMB}( \hat{n}) \rangle  -  \langle E_{\CMB}( \hat{n}) \delta^2P_{\CMB}( \hat{n}) \rangle  \Big ] - \delta^2Q_{\CIB}( \hat{n}) \langle P_{\CMB}( \hat{n}) E_{\CMB}( \hat{n}) \rangle \nonumber \\
 &-&  E_{\CMB}( \hat{n})  \Big [ \langle Q_{\CIB}( \hat{n}) \delta^2P_{\CMB}( \hat{n}) \rangle  -  \langle P_{\CMB}( \hat{n}) \delta^2Q_{\CIB}( \hat{n}) \rangle  \Big ] \nonumber \\ 
&+&    \delta^2E_{\CMB}( \hat{n}) \langle P_{\CMB}( \hat{n}) Q_{\CIB}( \hat{n}) \rangle  
  -  \delta^2P_{\CMB}( \hat{n}) \langle E_{\CMB}( \hat{n}) Q_{\CIB}( \hat{n}) \rangle \nonumber \\
&+&   P_{\CMB}( \hat{n})  \Big [ \langle Q_{\CIB}( \hat{n}) \delta^2E_{\CMB}( \hat{n}) \rangle  -  \langle E_{\CMB}( \hat{n}) \delta^2Q_{\CIB}( \hat{n}) \rangle \Big ]  \Big\}.
  \label{linear_term_double_freq}
  \end{eqnarray}                                                         
\section{Summary and Conclusions}
\label{section6}
In this paper we have investigated the NG signal arising from the CIB
and its correlation with the lensing signal imprinted in the CMB
temperature anisotropies, and we have estimated the bias they can
induce on the local, equilateral and orthogonal \fnl parameter using
\Planck data. The bias is computed for `raw' single--frequency
temperature maps, i.e. maps on which no component separation is
applied, and for maps cleaned by the SEVEM component separation
technique. For these maps, we have used the \Planck instrumental
characteristics and we have assumed they are free from Galactic
foregrounds. We have then studied the possibility to detect the
CIB--lensing bispectrum in the \Planck data.

CIB intensity fluctuations have been modelled following
\citet{planck_xviii_2011}. The parameters of the model have been
updated in order to have a better agreement with the recent \Planck
measurements of the CIB power spectra \citep{planck_xxx_2013}. We have
also considered the contribution from extragalactic radio sources and
from their correlation with CMB lensing. As expected,
radio--lensing power spectra and bispectra are found to be small and
negligible at the frequencies used for the cosmological analysis in
{\it Planck}.

Below we summarise and discuss our results.
\begin{itemize}
\item The bias $\Delta$\fnl induced by the CIB--lensing correlation is
  small but not negligible for orthogonal shapes in the ``raw'' 143
  and 217\,GHz {\it Planck} maps, approximately $ -21$ and $ -88$
  respectively. However, when maps are cleaned with a component
  separation technique, the bias is strongly reduced and becomes
  almost negligible for {\it Planck} results (the largest bias appears
  for the orthogonal shape and amounts to $0.3\, \sigma$).
\item We have estimated the bias produced by the intrinsic bispectrum
  of extragalactic sources. In agreement with the discussion in
  \citet{planck_xxiv_2013}, point sources turn out to be not a severe
  contaminant for \fnl studies with {\it Planck} foreground--reduced
  data, even though not completely negligible (the largest bias is for
  the equilateral shape amounting to $0.45\, \sigma$).  In ``raw''
  maps they produce a significant bias only for equilateral shapes,
  that is 160, 54 and 60 at 100, 143 and 217\,GHz respectively.
\item Our results confirm the capability and stress the importance of
  component separation techniques in removing extragalactic
  foregrounds as well. On the other hand, our results also
    predict that future experiments, with better sensitivity to the
  \fnl parameter, might have to consider extragalactic sources and the
  CIB--lensing correlation as further serious contaminants in some
  particular shapes.
\item The detection of the CIB--lensing bispectrum signal directly
  from {\it Planck} maps is not straightforward and it is feasible
  only for the 217\,GHz channel (with a significance of $
  \sim7.5\,\sigma$, assuming full sky coverage). Nevertheless, we have
  shown in this paper that a more efficient way to detect the
  bispectrum cross--correlates CMB maps with CIB maps at different
  frequencies. In this case the CIB--lensing bispectrum signal could
  be detected with very high significance ($ \la63\,\sigma$) if
  accurate CIB maps can be extracted at 353, 545 and 857\,GHz over a
  large area of the sky. \citet{planck_xxx_2013} were able to produce
  clean CIB maps just over $ \approx$ 10\% of the sky; in this case we
  still expect a high level of significance, of approximately {
    20}\,$\sigma$ or more. We have to note however that the cross
  correlation between CMB and CIB maps is not a simple procedure as
  possible residuals of the CIB (CMB) in ``clean'' CMB (CIB) maps
  could produce spurious signals that can be easily misinterpreted.
\item We have compared our predictions for the detectability levels of
  the CIB--lensing bispectrum on \Planck data with the results already
  obtained by the \Planck Collaboration using the cross-spectrum
  estimator \citep{planck_xviii_2013}. Our results are nearly
  equivalent to the ones presented in that work when we compare the
  case with $f_{sky}=30.4\%$, the lensing reconstruction at 143
  GHz and the CIB dominated bands of 353, 545 and 857 GHz (see Table
  \ref{t7}). This shows that both approaches are equivalent in terms
  of efficiency and both of them can be used to provide more robust
  results as different estimators might have different sensitivity to
  systematics.

\item Finally, we have developed an optimal estimator for the
  CIB--lensing bispectrum, based on the KSW formalism. Two different
  estimators have been constructed, one for use with single--frequency
  maps and a second one for separate CMB and CIB maps.
\end{itemize}
%


\section*{acknowledgements}
The authors acknowledge financial support from the Spanish Ministerio
de Econom\'ia y Competitividad project AYA-2012-39475-C02-01 and the
Consolider Ingenio-2010 Programme project CSD2010-00064, as well as
from the Swiss National Science Foundation. AC acknowledges the
Spanish Consejo Superior de Investigaciones Cient\'ificas (CSIC) and
the Spanish Ministerio de Educaci\'on, Cultura y Deporte for a
postdoctoral fellowship at the Cavendish Laboratory of the University
of Cambridge (UK). AC is thankful to Airam Marcos Caballero and Marina
Migliaccio for their useful comments that have helped in the
production of this paper. MK thanks Peter Wittwer for pointing out
helpful mathematical identities. The authors acknowledge useful
discussions and the feedback provided regarding the lensing to Anthony
Challinor and Antony Lewis. The authors acknowledge useful discussions
regarding the SEVEM component separation technique to Bel\'en Barreiro
and Patricio Vielva. The authors are grateful to the anonymous
  referee whose revision helped improving this article.  The authors
acknowledge the computer resources, technical expertise and assistance
provided by the Spanish Supercomputing Network nodes at Universidad de
Cantabria and Universidad Polit\'ecnica de Madrid. Some of the
calculations used the Andromeda cluster of the University of
Geneva. We have also used the software packages HEALPix
\citep{healpix} and CAMB \citep{lewis2000}.



\begin{appendix}
\section{CIB--lensing Power Spectrum at large scales}
\label{appendix_cib_lensing}
In this Appendix we derive the angular power spectrum for the
CIB--lensing correlation given in Eqs.\,(\ref{s2e1}) and
(\ref{ss2e1}). Starting from the angular correlation function of CIB
fluctuations and lensing potential, the procedure is equivalent to the
one presented in \citet{lewis2006} for the lensing power spectrum
\citep[see also, e.g.,][]{coo00}.

We write fluctuations of the CIB temperature in a direction $\hn$
at frequency $\nu$ as
\beq
T^{(\CIB)}_{\nu}(\hn)=\int d\chi\,a(\chi)\bar{j}_{\nu}(\chi)
{\delta j_{\nu}(\chi\hn,\chi) \over \bar{j}_{\nu}(\chi)}=
\int d\chi\,W_{\nu}^{(\CIB)}(\chi)\delta_{\gal}(\chi\hn,\chi,\nu)\,,
\label{a1}
\eeq
where $\chi$ is the comoving distance, $j_{\nu}(\chi)$ is the CIB
emissivity and $a(\chi)$is the scale factor. Here we have assumed that
fluctuations in CIB emissivity, $\delta j_{\nu}(z)/\bar{j}_{\nu}(z)$,
trace fluctuations in the number density of galaxies,
$\delta_{\gal}(z,\nu)$, with
$n_{\gal}=\bar{n}_{\gal}(1+\delta_{\gal})$. On the other hand, the
lensing potential along the line of sight is usually defined in terms
of the gravitational potential $\Phi$ by
\beq
\phi(\hn)=-2\,\int_0^{\chi_*} d\chi\,{\chi_*-\chi \over \chi\chi_*}
\Phi(\chi\hn,\chi)\,.
\label{a2}
\eeq
The gravitational potential field is related to the matter density
field $\delta_{\rm m}$\footnote{Here we assume that the anisotropic
  stress in the Universe is negligible so that the two Bardeen
  potentials $\Phi$ and $\Psi$ coincide. In general one needs to
  consider the Weyl potential $(\Phi+\Psi)/2$ which governs the
  lensing of light. However, the approximations made here are very
  good in a $\Lambda$CDM universe and for the redshift range
  considered.} by the Poisson equation and the lensing potential can
be written as
\beq
\phi(\hn)=\int_0^{\chi_*}d\chi\,\int\,{d^3\vk \over (2\pi)^{3/2}}
W^{(\Lens)}(k,\chi)\delta_{\rm m}(k,\chi) e^{i\chi\vk\cdot\hn}\,,
\label{a2b}
\eeq
where $\delta_{\rm m}(k,\chi)$ is the Fourier transform of the matter
density field. The functions $W^{(\CIB)}$ and $W^{(\Lens)}$ are the
same as in Eq. (\ref{s2e2}).

After introducing the Fourier transforms of CIB and matter
fluctuations, the cross--correlation between the lensing potential and
the CIB fluctuations is
\beq \langle T_{\nu}^{(\CIB)}(\hn)\phi(\hn') \rangle =\int_0^{\chi_*}
d\chi\,W_{\nu}^{(\CIB)}(\chi)\, \int_0^{\chi_*} d\chi'\,\int\, {d^3\vk
  d^3\vk' \over (2\pi)^3}W^{(\Lens)}(k',\chi') \langle
\delta_{\gal}(k,\chi,\nu)\delta^*_{\rm m}(k',\chi') \rangle
e^{i\chi\vk\cdot\hn}e^{-i\chi\vk'\cdot\hn'}\,,
\label{a4}
\eeq
where
\beq
\langle \delta_{\gal}(k,\chi,\nu)\delta^*_{m}(k',\chi') \rangle =
P_{\delta g}(k,\chi,\chi',\nu)\delta(\vk-\vk')
\label{a5}
\eeq
and $P_{\delta g}(k)$ is the power spectrum of the cross--correlation
between galaxy number density and matter fluctuations.

Using the relation
\beq
e^{i\chi\vk\cdot\hn}=4\pi\sum_{\ell m}\,i^{\ell}j_{\ell}(k\chi)
Y^*_{\ell m}(\hn)Y_{\ell m}(\hk)\,,
\label{a7}
\eeq
and the orthogonality of the spherical harmonics we have
\beq
\begin{split}
\langle T_{\nu}^{(\CIB)}(\hn)\phi(\hn') \rangle ={2 \over \pi}\sum_{\ell\ell' mm'}
\int_0^{\chi_*} d\chi\,W_{\nu}^{(\CIB)}\,(\chi)
\int_0^{\chi_*} d\chi'\,\times \\
\times\,\int\,dk\,k^2 W^{(\Lens)}(k,\chi') P_{\delta g}(k,\chi,\chi',\nu)
j_{\ell}(k\chi)j_{\ell'}(k\chi')Y_{\ell m}(\hn)Y^*_{\ell' m'}(\hn')
\delta_{\ell\ell'}\delta_{mm'}\,.
\end{split}
\label{a8}
\eeq
From the last equation it is straightforward to get the power
spectrum of the CIB--lensing correlation 
\beq
C^{(\LensCIB)}_{\ell}(\nu)={2 \over \pi}\int\,dk\,k^2\,
\int_0^{\chi_*}d\chi\,W_{\nu}^{(\CIB)}(\chi)j_{\ell}(k\chi)
\int_0^{\chi_*}d\chi'\,W^{(\Lens)}(k,\chi')j_{\ell}(k\chi')
P_{\delta g}(k,\chi,\chi',\nu).
\label{a9}
\eeq
In the HOD approach, at very large scales $P_{\delta {\rm g}}(k)$ is
dominated by the 2--halo term (at $\ell=40$ the 1--halo term in fact
contributes just for $\sim1\%$). Under this condition, both matter and
CIB fluctuations can be related to linear density perturbations
$\delta_{\rm lin}(\vk)$ through a transfer function
$\mathcal{T}_X(k,\chi)$ so that
$\delta_X(\vk,\chi)=\mathcal{T}_X(k,\chi) \delta_{\rm lin}(\vk)$, with
$X=\gal,{\rm m}$. Therefore $C^{(\LensCIB)}_{\ell}$ becomes
\beq
C^{(\LensCIB)}_{\ell}(\nu) = {2 \over \pi}\int\,dk\,k^2\,P_{\rm lin}(k)
\int_0^{\chi_*}d\chi\,W_{\nu}^{(\CIB)}(\chi)\mathcal{T}_{\gal}(k,\chi,\nu)
j_{\ell}(k\chi)\int_0^{\chi_*}d\chi'\,W^{(\Lens)}(k,\chi')
\mathcal{T}_{\rm m}(k,\chi')j_{\ell}(k\chi')
\label{a9a}
\eeq
Because the power spectrum of galaxies, $P_{\rm gg}(k)$, and of dark
matter perturbations, $P_{\delta\delta}(k)$, are
\beq
P_{\rm gg}(k,\chi,\nu)=\langle \delta_{\gal}(k,\chi,\nu)
\delta^*_{\gal}(k,\chi,\nu) \rangle =P_{\rm lin}(k)\mathcal{T}^2_{\gal}(k,\chi,\nu);
~~~~~
P_{\delta\delta}(k,\chi)=\langle \delta_{\rm m}(k,\chi)\delta^*_{\rm m}(k,\chi) \rangle 
=P_{\rm lin}(k)\mathcal{T}^2_{\rm m}(k,\chi)\,,
\label{a6}
\eeq
we find
\beq
C^{(\LensCIB)}_{\ell}(\nu) = {2 \over \pi}\int\,dk\,k^2\,
\int_0^{\chi_*}d\chi\,W_{\nu}^{(\CIB)}(\chi)j_{\ell}(k\chi)
\sqrt{P_{gg}(k,\chi,\nu)}\,
\int_0^{\chi_*}d\chi'\,W^{(\Lens)}(k,\chi')j_{\ell}(k\chi')
\sqrt{P_{\delta\delta}(k,\chi')}\,,
\label{a9b}
\eeq
as in Eq.\ (\ref{ss2e1}).

At high $\ell$, the cross--correlation power spectrum $P_{\delta {\rm
    g}}(k,\chi,\chi',)$ in Eq.\ (\ref{a9}) is expected to vary slowly
compared to spherical Bessel functions, and we can perform the
$k$-integration taking the power spectrum constant. The spherical
Bessel functions then give a delta-function,
\beq
\int\,dk\,k^2\,j_{\ell}(k\chi) j_{\ell}(k\chi')=
{2 \over \pi\chi^2}\delta(\chi-\chi')\, ,
\label{a10}
\eeq
which allows us to perform one of the $\chi$ integrations as well and
which fixes the scale $k\sim\ell/\chi$. We then obtain $C^{\LensCIB}_{\ell}$ in the Limber approximation
\beq
C^{(\LensCIB)}_{\ell}(\nu)=\int_0^{\chi_*}{d\chi \over \chi^2}
\,W_{\nu}^{(\CIB)}(\chi)W^{(\Lens)}(k,\chi)P_{\delta g}(k=\ell/\chi,\chi,\nu)\,.
\label{a11}
\eeq
\section{HOD model constraints from \Planck data}
\label{appendix_hod_parameters}
In this appendix we find the best--fit values for the HOD parameters
used in Section\,\ref{section2} to model CIB and CIB--lensing spectra.
The free parameters are simply $M_{\rm min}$ and $\alpha_{\rm sat}$,
after imposing $M_{\rm sat}=3.3M_{\rm min}$ and $\sigma_{\log
  M}=0.65$. Another free parameter is the effective mean emissivity
$j_{\rm eff}$. In order to isolate and constrain the high--z
contribution to the CIB that is poorly known from observations, we
make in fact the extra assumption that the mean emissivity of
galaxies, $j_{\rm eff}$, is constant at $z>3.5$
\citep{planck_xviii_2011}. More precisely, Eq.\ (\ref{s2e1}) is
rewritten as
\beq
C_{\ell}^{(\CIB)}(\nu,\nu')=\int_0^{3.5}{d\chi \over \chi^2}\,a^2(\chi)
\bar{j}_{\nu}(\chi) \bar{j}_{\nu'}(\chi) P_{\rm gg}(k=\ell/\chi,\chi)+
j_{\rm eff}(\nu) j_{\rm eff}(\nu') \int_{3.5}^7{d\chi \over
  \chi^2}\,a^2(\chi)P_{\rm gg}(k=\ell/\chi,\chi)\,.
\label{bb2}
\eeq
The three parameters of the model are fitted to the CIB spectra
measured in \citet{planck_xxx_2013}, using a Markov Chain Monte Carlo
(MCMC) method. For each \Planck frequency $\ge217\,$GHz we find the
values of $M_{\rm min}$, $\alpha_{\rm sat}$ and $j_{\rm eff}$ that
best fit the CIB power spectrum $C_{\ell}^{(\CIB)}(\nu)$ at the
corresponding frequency, and their associated uncertainties. For the
143\,GHz channel we prefer to use the 143$\times$857 and
143$\times$545 cross--power spectra, due to the large uncertainty in
the 143$\times$143 spectrum. Finally, at 100\,GHz we take the same HOD
values as at 143\,GHz and $j_{\rm eff}=4.15$\,Jy/Mpc/sr, that corresponds
to the average emissivity between $z=3.5$ and 7 according to the
\citet{bet11} model.

Because it is not possible to disentangle the shot--noise
contributions of radio and IR galaxies from observations, we fix them
on the basis of the \citet{tuc11} and \citet{bet11} models.

In Table\,B1 we report our results for the free parameters with the
corresponding reduced chi--squared $\chi_{\rm red}^2$ of the fits (for
8 degrees of freedom except at 143\,GHz where we have 16). Comparing
to the parameter values found in \citet{planck_xviii_2011}, we obtain
very similar results for the 217 and 353\,GHz channels. At 545 and
857\,GHz our best--fit values for $M_{\rm min}$ and $\alpha_{\rm sat}$
are slightly different from \Planck results, but still compatible at
1--$\sigma$. The only (small) discrepancy is on $j_{\rm eff}$: the
\Planck best--fits are around 300\,Jy/Mpc/sr at the two frequencies
and compatible with zero. \citet[][see their Table
3]{planck_xviii_2013} also estimated the mean CIB emissivity in the
redshift bin $3<z\le7$ and found $417\pm251$\,Jy/Mpc/sr at 545\,GHz
and $609\pm359$\,Jy/Mpc/sr at 857\,GHz. According to our results,
high--redshift galaxies seem to give a bigger contribution at these
frequencies, especially at 857\,GHz.

\begin{figure}
\includegraphics[width=90mm]{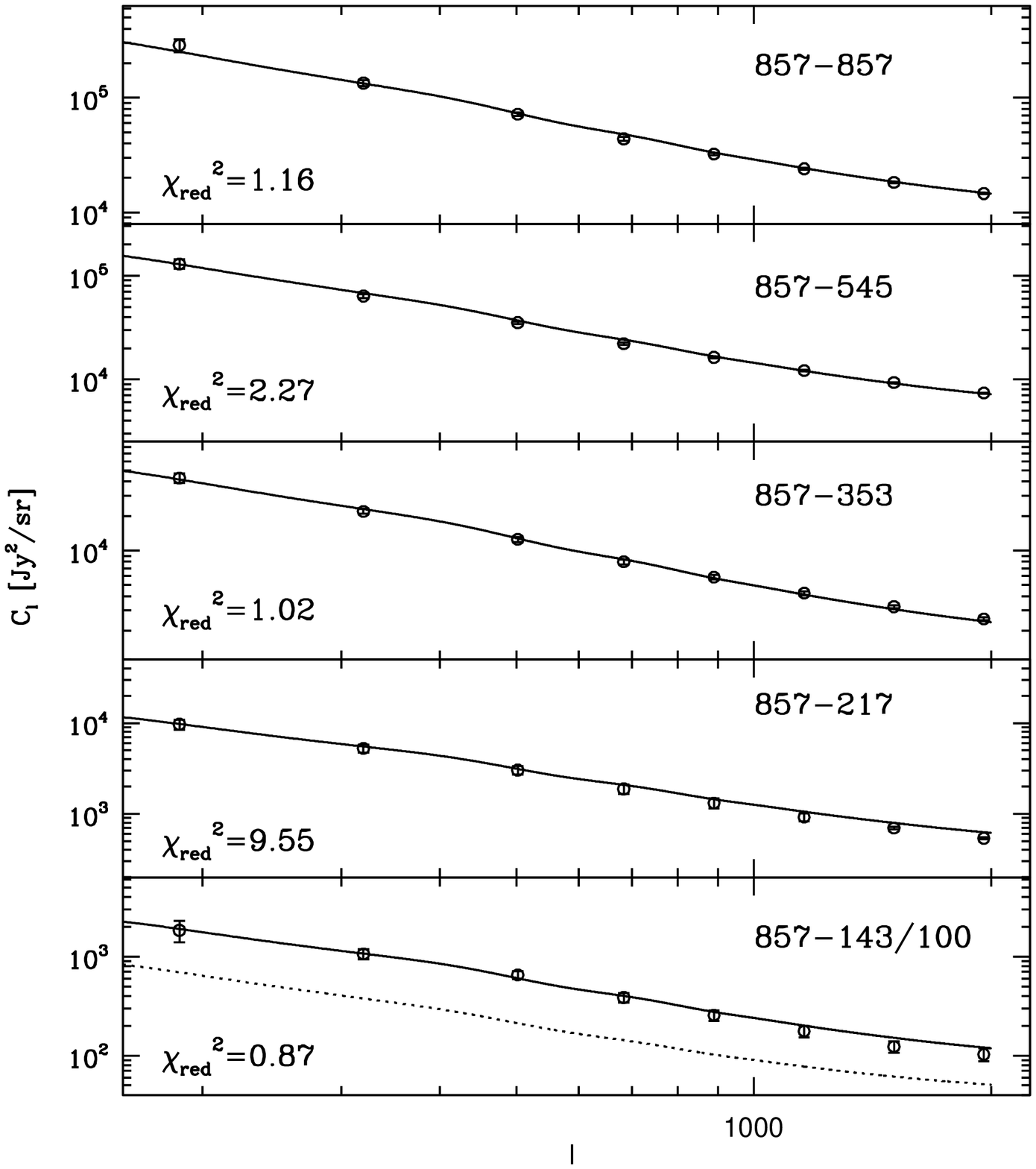}
\includegraphics[width=90mm]{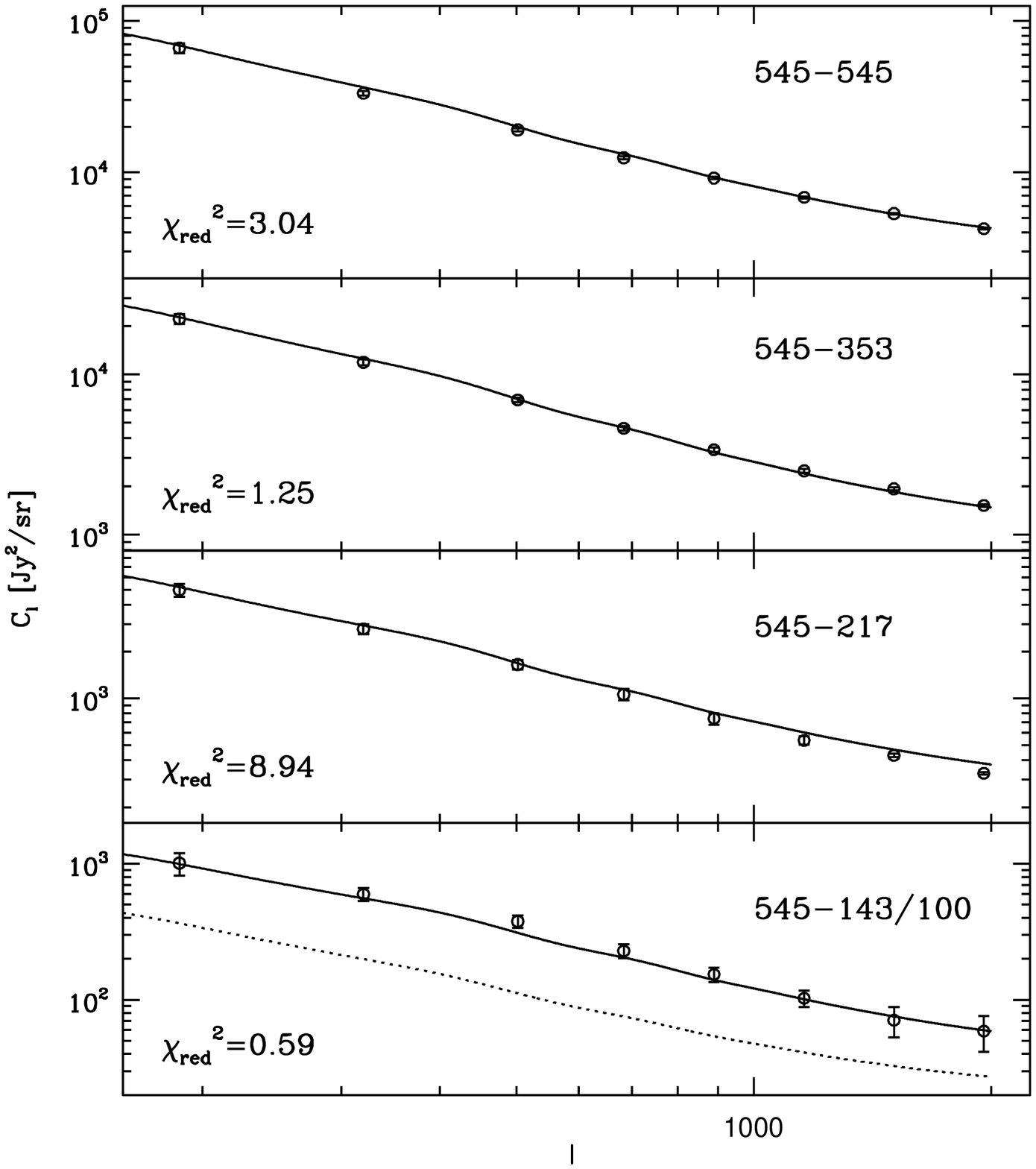}
\includegraphics[width=90mm]{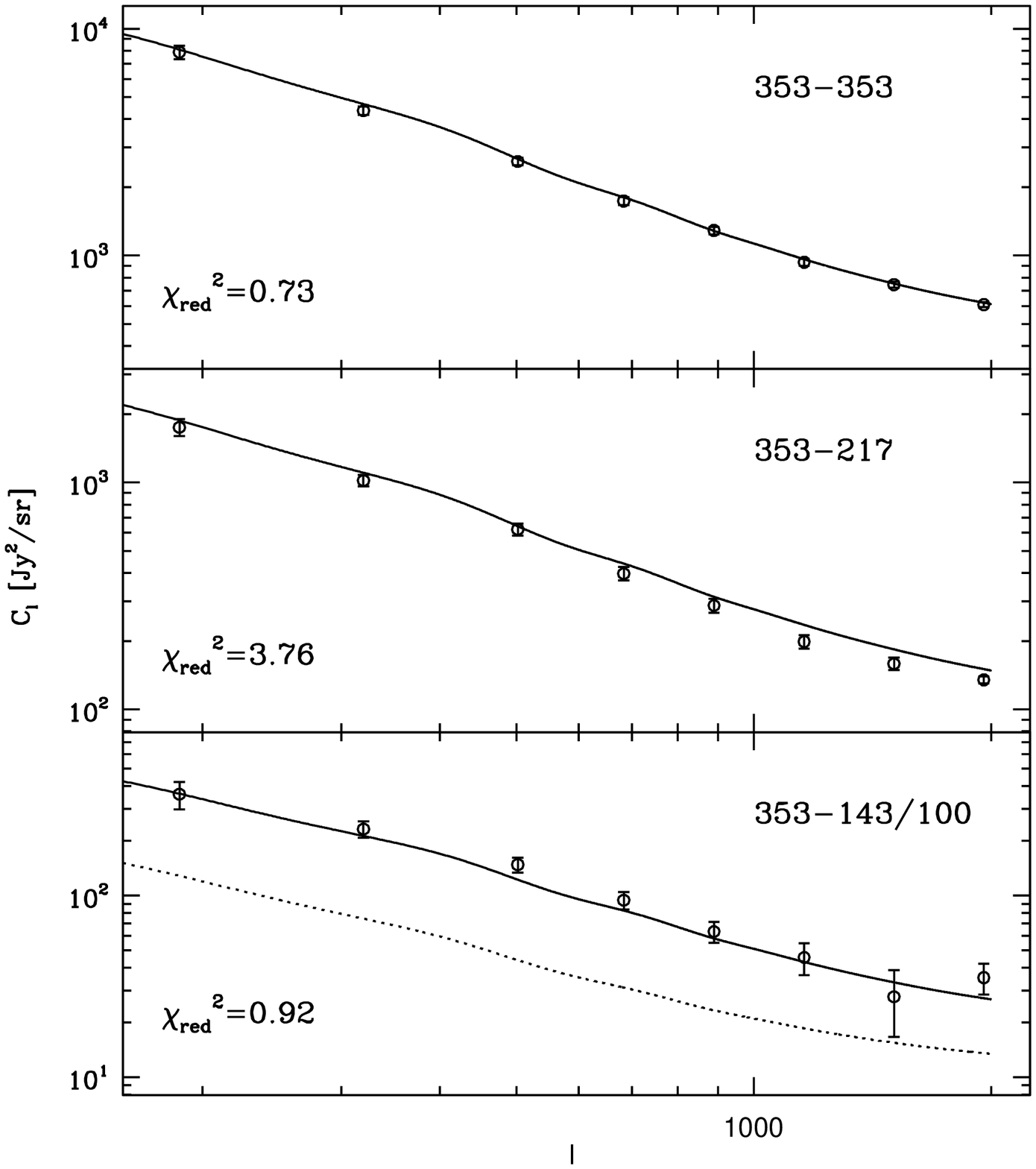}
\includegraphics[width=90mm]{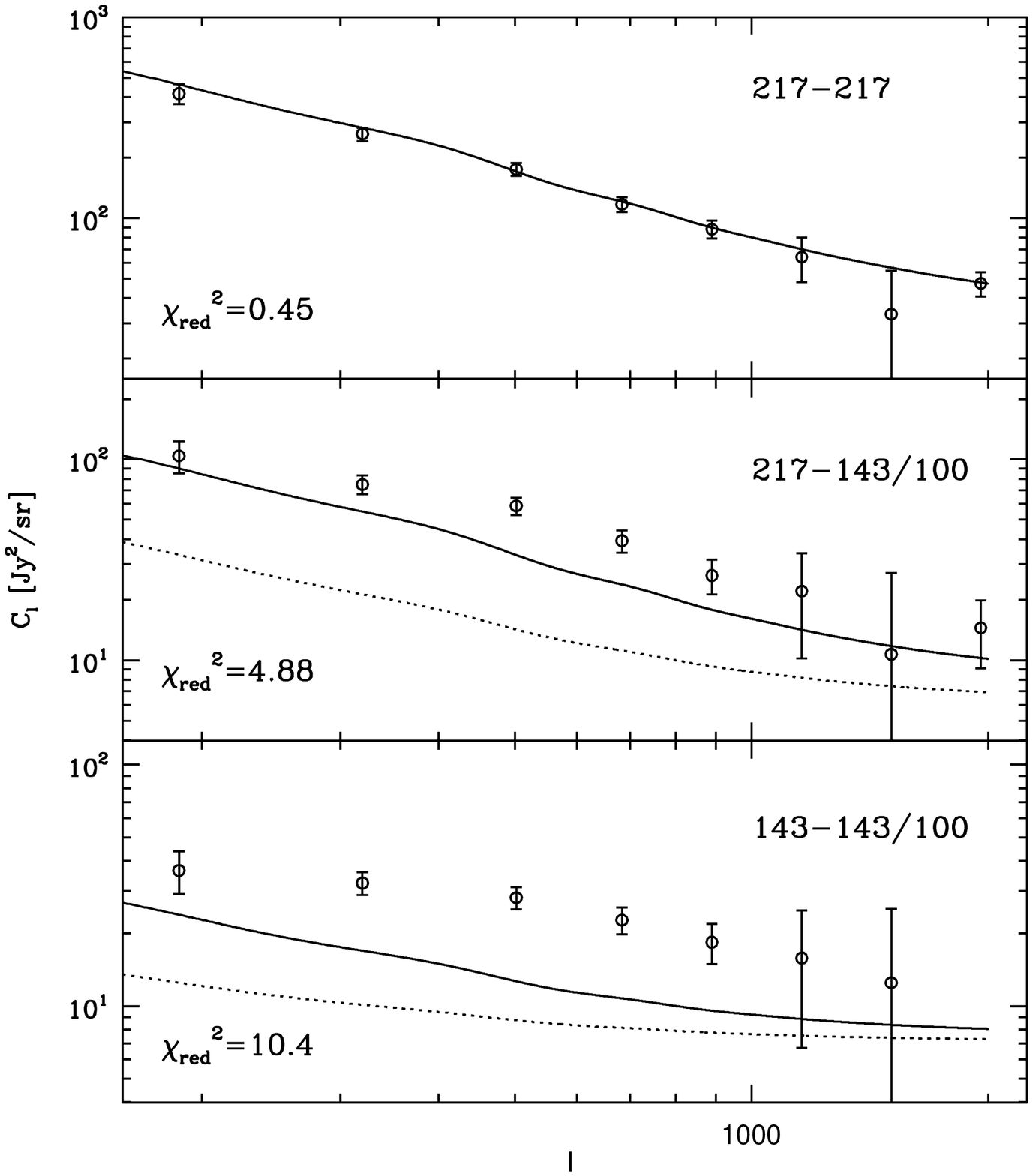}
\caption{Auto-- and cross--power spectra of CIB fluctuations measured
  by \Planck at the HFI frequencies \citep[][open
  points]{planck_xxx_2013}, compared with predictions of our best--fit
  model (solid lines). Dotted lines are for spectra involving the
  100\,GHz channel. Data and predictions include also contributions
  from radio and IR shot noise. The reduced $\chi^2$ of the fits are
  also provided.}
\label{fb1}
\end{figure}

\begin{figure}
\includegraphics[width=60mm]{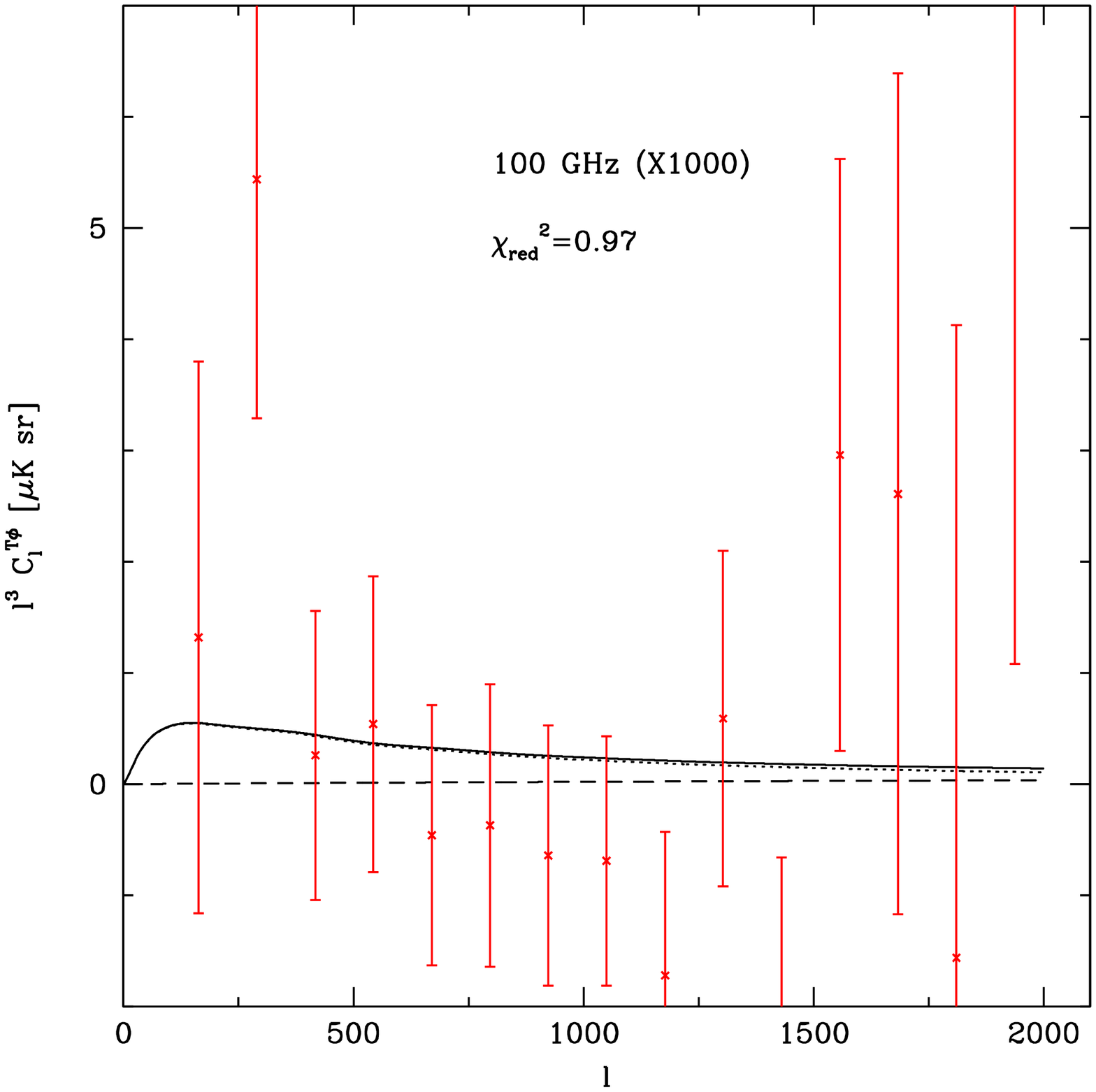}
\includegraphics[width=60mm]{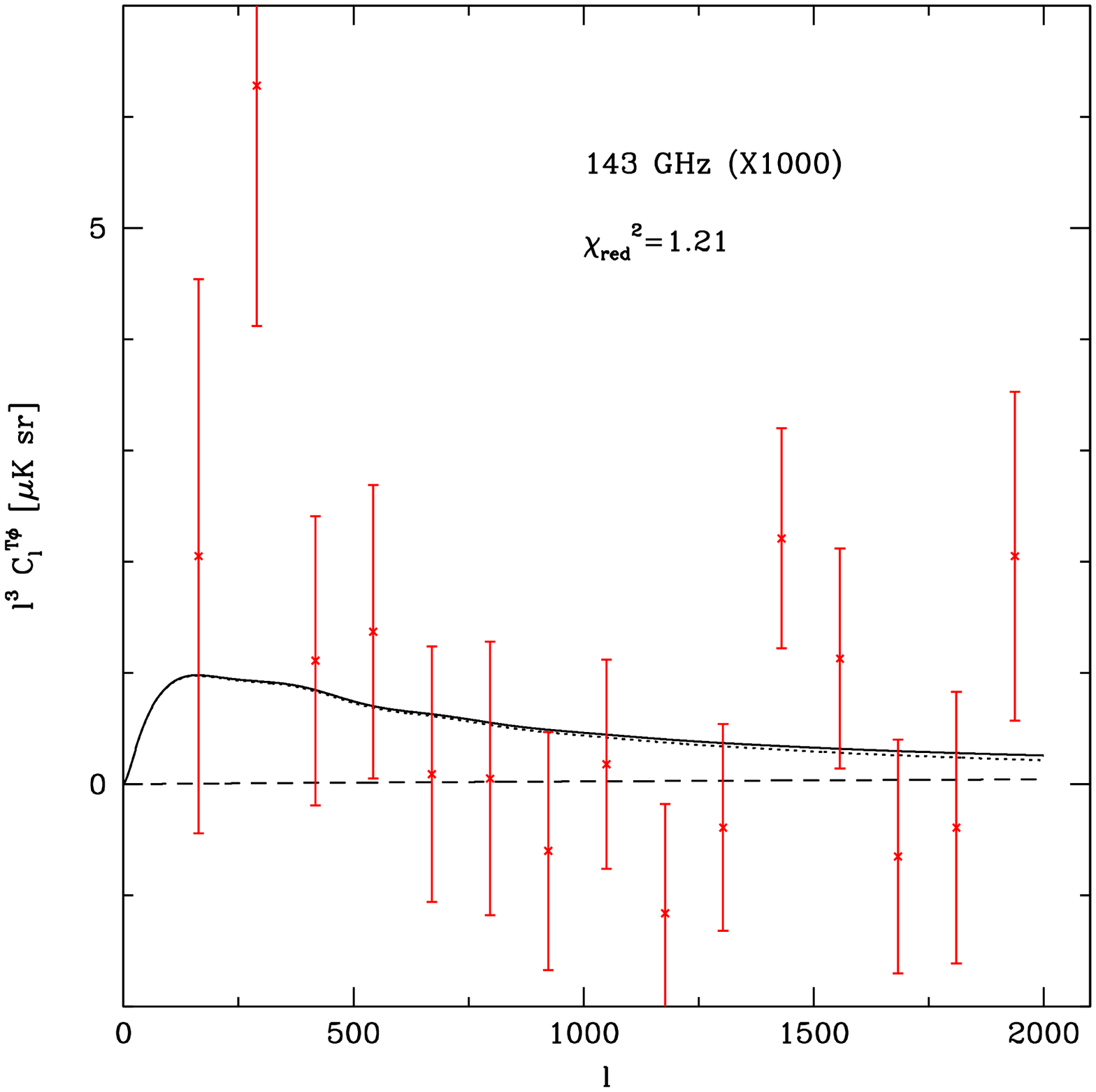}
\includegraphics[width=60mm]{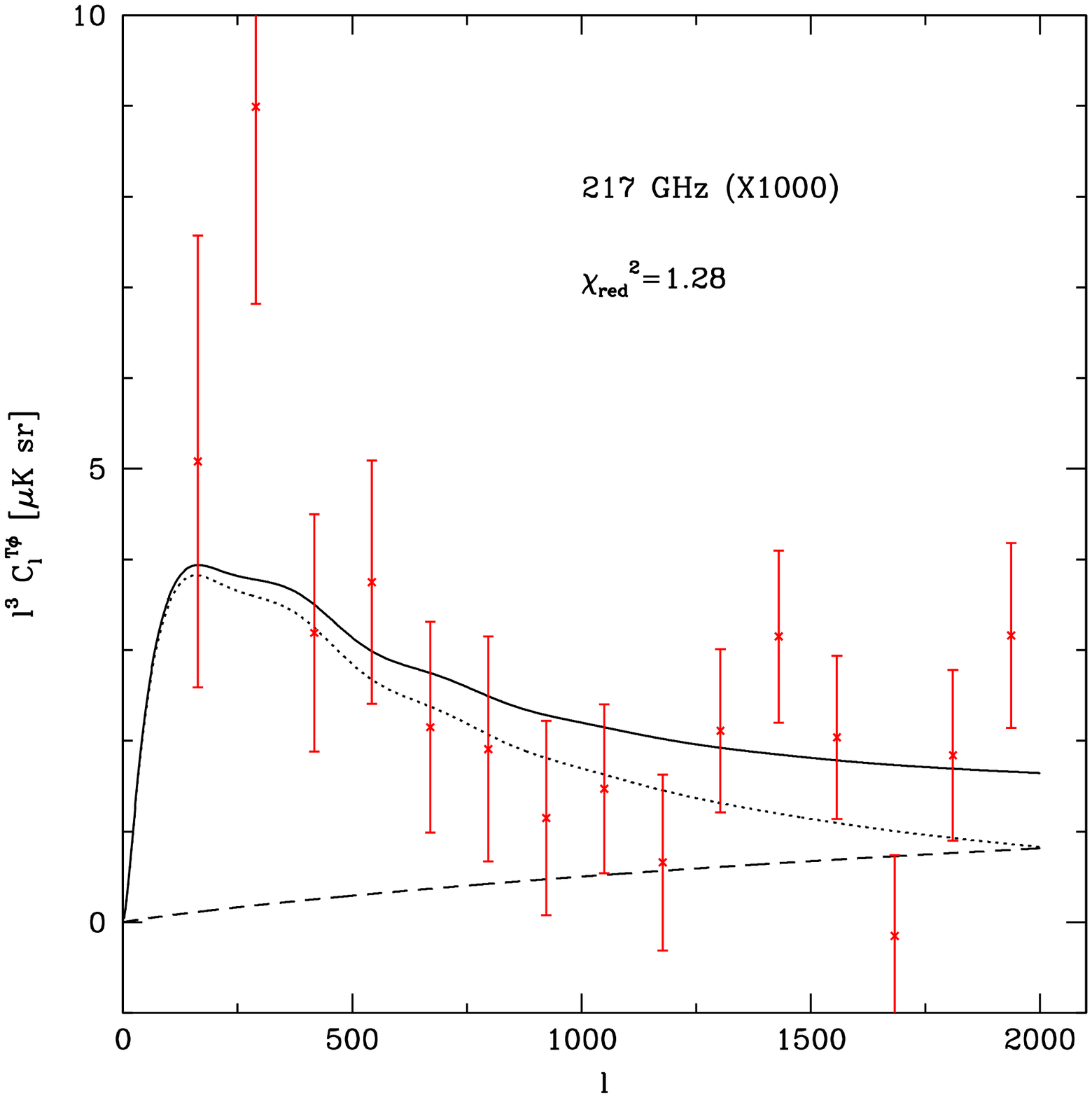}
\includegraphics[width=60mm]{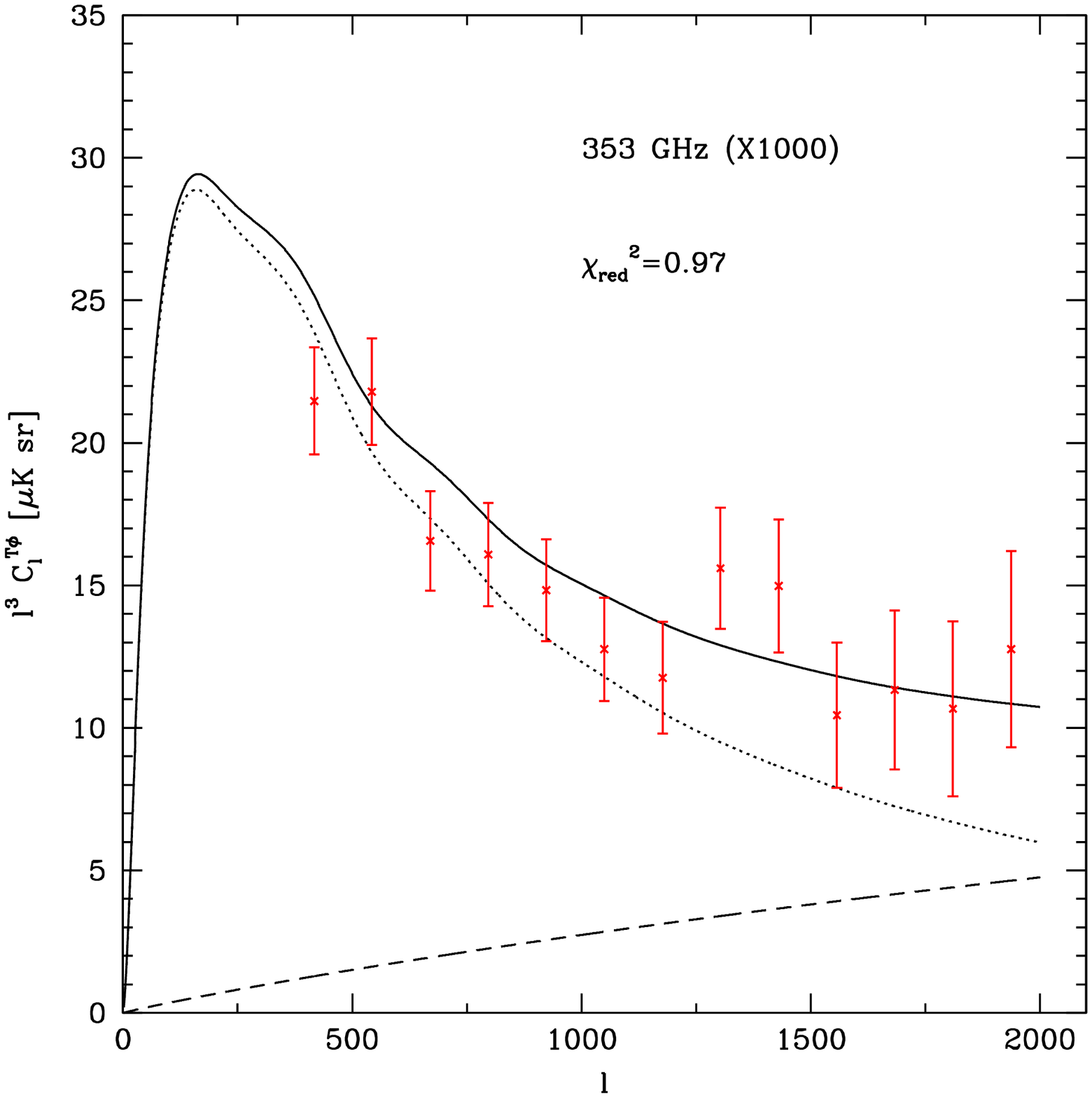}
\includegraphics[width=60mm]{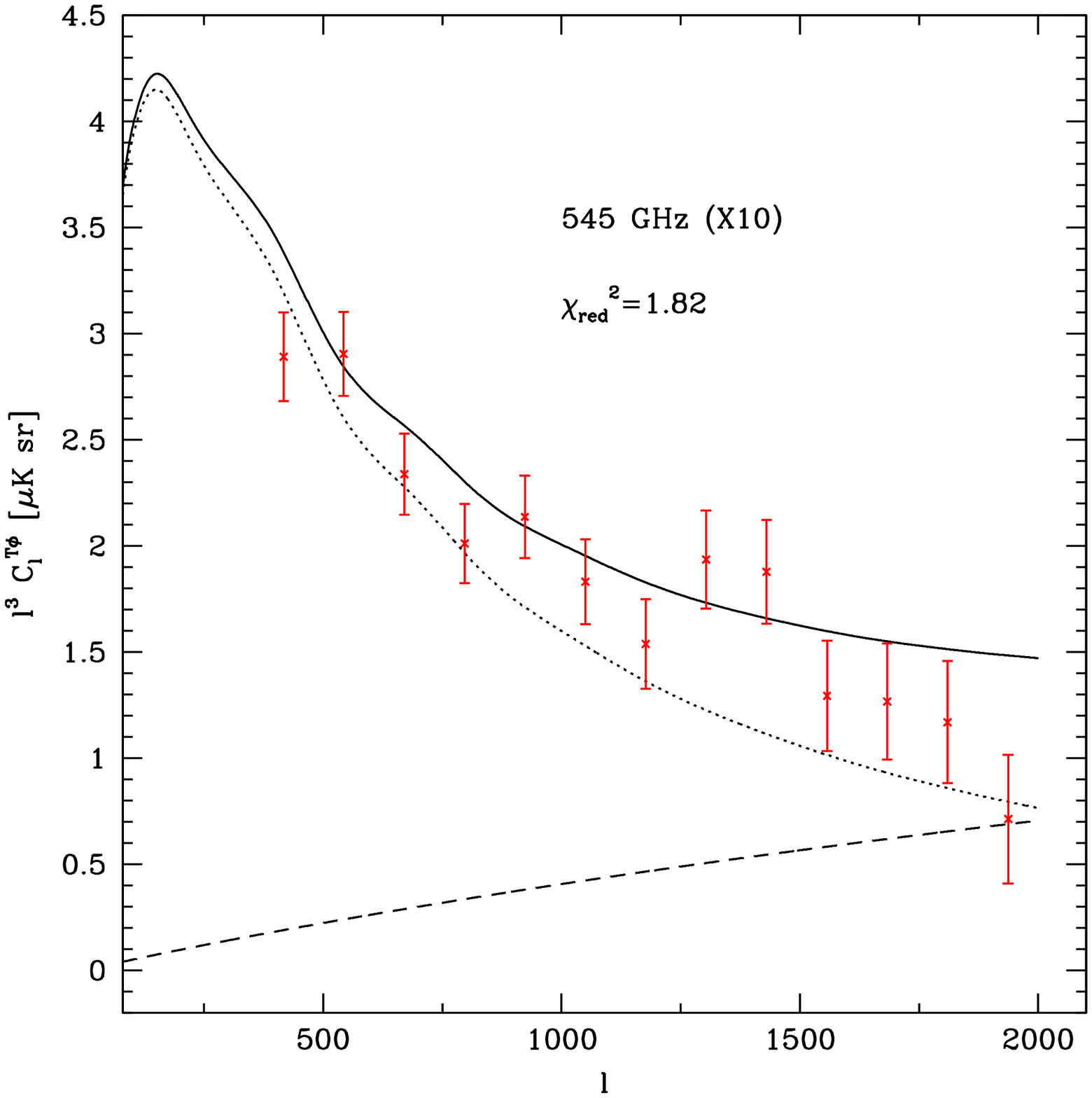}
\includegraphics[width=60mm]{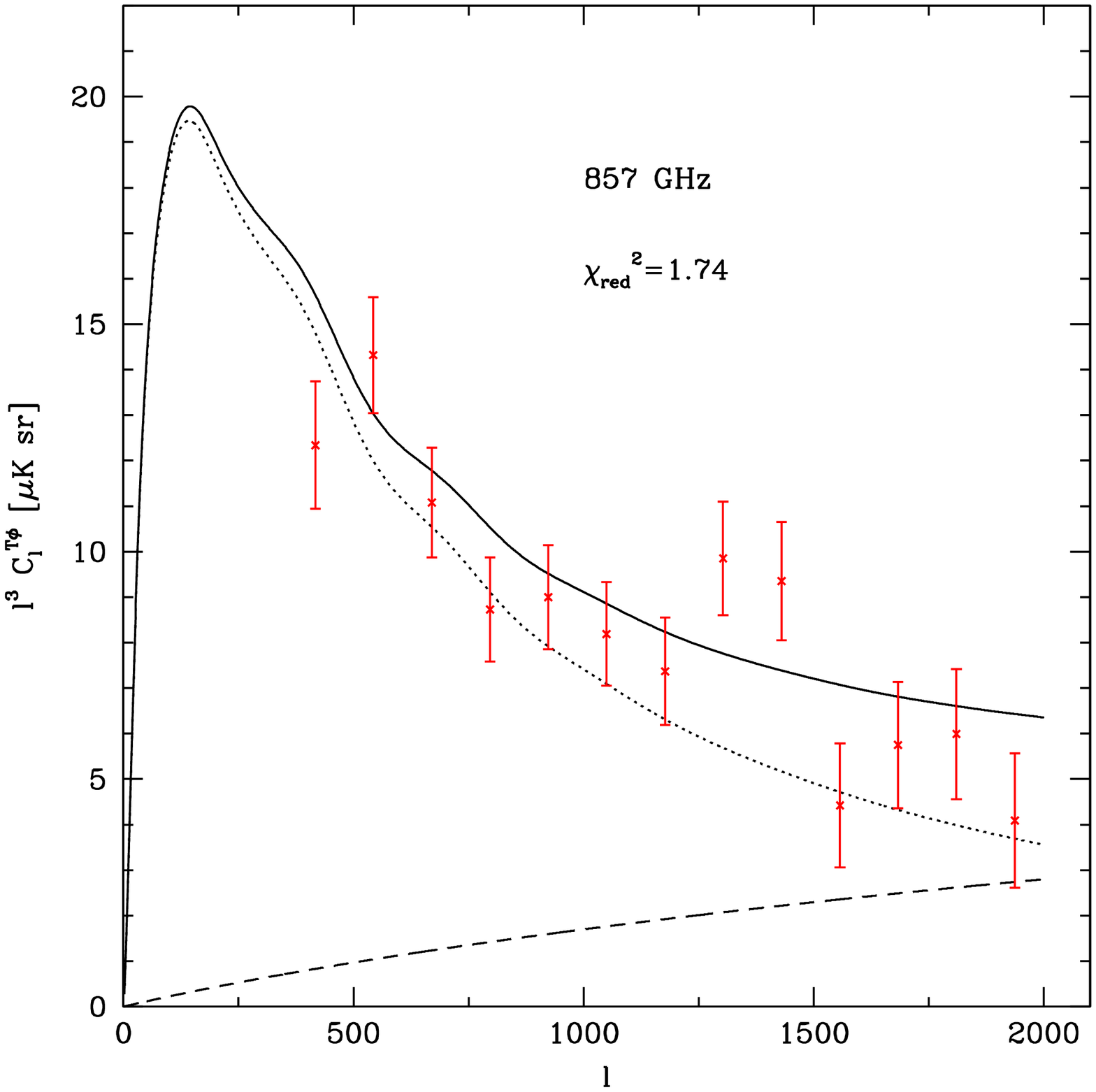}
\caption{CIB--Lensing power spectra from \citet{planck_xxx_2013},
  compared with predictions of our best--fit model: solid lines
  are for total spectra; dotted(dashed) lines are for the 2--(1--)halo
  term. The reduced $\chi^2$ of the fits are also provided.}
\label{fb2}
\end{figure}

In Figure\,\ref{fb1} we compare the model predictions with \Planck
observations. The fit of the model is in general good at the
high frequencies, both for auto-- and cross--spectra (it is to be
stressed that cross-spectra, $C_{\ell}^{(\CIB)}(\nu,\nu')$, have not
been used in the fitting procedure). The reduced chi--squared for
  the 857$\times$545 and 545$\times$545 spectra is quite
  large. However, due to the very small uncertainty on data, even for
  these cases the model can be considered a reasonable description of
  the data. Some discrepancies are found for all the
cross--spectra involving the 217\,GHz data and for the 143$\times$143
and 143$\times$217 spectra. The latter spectra are however
particularly problematic due to the subtraction of CMB anisotropies
and to spurious CIB and SZ signals that have to be corrected
\citep{planck_xxx_2013}.  Finally, Figure\,\ref{fb2} shows the
  CIB--lensing spectra from the model and from the data: the agreement
  is generally quite good for all the \Planck channels.
\begin{table*}
\begin{center}
\caption{Best--fit values for the parameters of the CIB model, and the
corresponding reduced chi--squared, $\chi_{\rm red}^2$.}
\begin{tabular}{cccccccc}
\hline
Frequency & [GHz] & 143 & 217 & 353 & 545 & 857 \\
\hline
$\log M_{\rm min}$ & [$M_{\odot}$] & 11.33$\pm$ 0.56 & 11.90 $\pm$
0.52 & 12.45 $\pm$ 0.33 & 12.10 $\pm$ 0.14 & 11.62 $\pm$ 0.27 \\ 
$\alpha_{\rm sat}$ & & 0.65$\pm$ 0.22 & 1.37 $\pm$ 0.26 & 1.21
$\pm$ 0.20 & 1.01 $\pm$ 0.04 & 0.86 $\pm$ 0.06 \\
$j_{\rm eff}$ & [Jy/Mpc/sr] & 15.3 $\pm$ 5.4 & 61. $\pm$ 13. &
191. $\pm$ 33. & 538. $\pm$ 38. & 1326. $\pm$ 160. \\
$\chi_{\rm red}^2$ & & 0.73 & 0.45 & 0.73 & 3.03 & 1.16 \\
\hline
\end{tabular}
\end{center}
\label{tb1} 
\end{table*}
%
%
\section{Shot--noise cross--power spectra and bispectra from extragalactic sources}
\label{appendix_sn}
The shot--noise power spectrum and bispectrum from extragalactic
sources at a fixed frequency $\nu$ can be computed by the well known
integrals:
\beq
C_{\rm sn}=\int_0^{S_c}\,{dN \over dS}S^2dS~~~~~~~~~~
b_{\rm sn}=\int_0^{S_c}\,{dN \over dS}S^3dS\,,
\label{ace1}
\eeq
where $S_c$ is the flux cut above which bright sources are
detected and removed, and $dN/dS$ are differential number counts of
sources. When different frequencies are considered, we use two
different approaches for computing the cross--power spectra and
bispectra, according to the class of extragalactic sources.

\begin{itemize}
\item For IR galaxies we use the same approach as in
  \citet{planck_xxx_2013}: the cross--spectra and bispectra for a
  single galaxy population can be approximated by
  \bea 
  C_{\rm sn}(\nu_1,\nu_2) & = &
 \int_{z=0}^7\int_0^{S_c(\nu_1)}\,
  H(S_{\nu_1}R_{\nu_1\nu_2}<S_c(\nu_2))
  {dN \over dS_{\nu_1}dz}S_{\nu_1}^2 R_{\nu_1\nu_2}(S_{\nu_1},z)\,
  dS_{\nu_1}dz
  \nonumber \\
  b_{\rm sn}(\nu_1,\nu_2,\nu_3) & = &
 \int_{z=0}^7\int_0^{S_c(\nu_1)}\,
  H(S_{\nu_1}R_{\nu_1\nu_2}<S_c(\nu_2))
  H(S_{\nu_1}R_{\nu_1\nu_3}<S_c(\nu_3))
  \times
  \nonumber \\
  & & \times {dN \over dS_{\nu_1}dz}S_{\nu_1}^3
  R_{\nu_1\nu_2}(S_{\nu_1},z)R_{\nu_1\nu_3}(S_{\nu_1},z)\,dS_{\nu_1}dz\,,
  \label{ace2}
  \eea
  where $R_{\nu_1\nu_2} (S_{\nu_1},z)$ is the mean colour between
  $\nu_1$ and $\nu_2$ in the considered flux density and redshift
  interval (i.e., the flux density at $\nu_2$ is written as
  $S_{\nu_2}=R_{\nu_1\nu_2}S_{\nu_1}$). We compute the mean colour
  from the \citet{bet11} model. $H(P_1)$ is equal to 1 when $P_1$
  is true and 0 otherwise.

\item The approach used for IR galaxies cannot be safely applied
  to radio sources because of the very large dispersion in the
  spectral shape of radio sources. When two different frequencies
  $\nu_1$ and $\nu_2$ are considered, we compute the cross--power
  spectra and bispectra of radio sources as a direct extension of
  Eq. (\ref{ace1}):
  \bea C_{\rm sn}(\nu_1,\nu_2) & = &
  \int_0^{S_c(\nu_1)}dS_{\nu_1}\,S_{\nu_1}
  \int_0^{S_c(\nu_2)}dS_{\nu_2}\,
  {d^2N \over dS_{\nu_1}dS_{\nu_2}}S_{\nu_2} \nonumber \\
  b_{\rm sn}(\nu_1,\nu_1,\nu_2) & = &
  \int_0^{S_c(\nu_1)}dS_{\nu_1}\,S^2_{\nu_1}
  \int_0^{S_c(\nu_2)}dS_{\nu_2}\, {d^2N \over
    dS_{\nu_1}dS_{\nu_2}}S_{\nu_2}\,,
\label{ace3}
\eea
where $d^2N/dS_{\nu_1}dS_{\nu_2}$ is the differential number of
sources with flux density in the interval $[S_{\nu_1},S_{\nu_1}+\Delta
S]$ at the frequency $\nu_1$ and with flux density in the interval
$[S_{\nu_2},S_{\nu_2}+\Delta S]$ at the frequency $\nu_2$. The shot
noise bispectra at the frequencies $\nu_1$, $\nu_2$ and $\nu_3$
require to compute differential number counts ($
d^3N/dS_{\nu_1}dS_{\nu_2}dS_{\nu_3}$) at the three different flux
density intervals, one for each frequency. This is too complex and
time consuming to be carried out in practice. For this reason we
decided to approximate the ``cross'' bispectra in the following
way. For a single population of radio sources, in the hypothesis
  of a full correlation between two frequencies $\nu_2$ and $
  \nu_3$ (i.e., $ S_{\nu_3}=S_{\nu_2}R_{\nu_2\nu_3}$), we have
\bea  
b_{\rm sn}(\nu_1,\nu_2,\nu_3)&=&
 \int_0^{S_c(\nu_1)}\,dS_{\nu_1}\,S_{\nu_1}
\int_0^{S_c(\nu_2)}\,dS_{\nu_2}\,S_{\nu_2}
\int_0^{S_c(\nu_3)}\,dS_{\nu_3}\,S_{\nu_3}
{d^3N \over dS_{\nu_1}dS_{\nu_2}dS_{\nu_3}} \nonumber \\
&=& \int_0^{S_c(\nu_1)}\,dS_{\nu_1}\,S_{\nu_1} 
\int_0^{S_c(\nu_2)}\,dS_{\nu_2}\,S^2_{\nu_2}{d^2N \over
  dS_{\nu_1}dS_{\nu_2}} R_{\nu_2\nu_3}
  H(S_{\nu_2}R_{\nu_2\nu_3}<S_c(\nu_3))=
b_{\rm sn}(\nu_1,\nu_2,\nu_2) R_{\nu_2\nu_3}\, 
\eea
where the mean colour $ R_{\nu_2\nu_3}$ has been assumed
independent of the flux density, and $
S_c(\nu_2)R_{\nu_2\nu_3}<S_c(\nu_3)$. A better approximation for the
``cross'' bispectra can be obtained by replacing $ R_{\nu_2\nu_3}$
with the ``decorrelation'' coefficient, $ C_{\rm
  sn}(\nu_2,\nu_3)/[C_{\rm sn}(\nu_2,\nu_2)C_{\rm
  sn}(\nu_3,\nu_3)]^{1/2}$, that provides a measure of the frequency
correlation of radio sources. Considering all the possible
combinations in frequency, we get
\beq 
b_{\rm sn}(\nu_1,\nu_2,\nu_3)\approx\bigg[\prod_{\overset{i=1,3}
{j=1,3;\,j\ne i}}\,\tilde{b}_{\rm sn}(\nu_i,\nu_i,\nu_j)\bigg]^{1/6}
~~~~~ {\rm with}~~~~~\tilde{b}_{\rm sn}(\nu_i,\nu_i,\nu_j)=
b_{\rm sn}(\nu_i,\nu_i,\nu_j){C_{\rm sn}(\nu_i,\nu_k) \over 
[C_{\rm sn}(\nu_i,\nu_i) C_{\rm sn}(\nu_k,\nu_k)]^{1/2}}
~~{\rm and}~k\ne i,j\,.
\label{ace4}
\eeq
This expression is an upper limit of the actual bispectrum, but we
expect it to be a good approximation for frequencies where the
correlation is high.  We applied Eq.\,(\ref{ace4}) to the CIB and we
found values typically 20--30\% higher than ones from
Eq. (\ref{ace2}), but very close when adjacent frequencies are
considered. We have also verified that the bias on the $f_{\rm nl}$
parameter does not change if we take $\tilde{b}_{\rm
  sn}(\nu_i,\nu_i,\nu_j)=b_{\rm sn}(\nu_i,\nu_i,\nu_j)$ in
Eq. (\ref{ace4}). This confirms the very small impact of the cross
bispectra of radio sources on the $f_{\rm nl}$ bias in the SEVEM
combined maps.
\end{itemize}
\newcommand{\Gaunt}{{G_{m_1 m_2 m_3}^{\ell_1 \ell_2 \ell_3}}}
\newcommand{\LensCIBCIB}{{\mathtt{CIBCIB-Lens}}}
\section{Covariance of the CIB-lensing bispectrum}
\label{appendix_cov_bisp}
We estimate the covariance matrix of the CIB--lensing bispectrum
taking into account higher order corrections for the cases where the
non-Gaussian signal becomes significant. The covariance of the
CIB--lensing bispectrum is given by
\begin{equation}
\mathtt{cov}\left(b^{(\LensCIB)}_{\ell_1 \ell_2 \ell_3}, b^{(\LensCIB)}_{\ell'_1 \ell'_2 \ell'_3}\right) = \langle b^{(\LensCIB)}_{\ell_1 \ell_2 \ell_3} b^{(\LensCIB)}_{\ell'_1 \ell'_2 \ell'_3}\rangle - \langle b^{(\LensCIB)}_{\ell_1 \ell_2 \ell_3} \rangle \langle b^{(\LensCIB)}_{\ell'_1 \ell'_2 \ell'_3}\rangle,
\label{the_covariance}
\end{equation}
where 
\begin{eqnarray}
b^{(\LensCIB)}_{\ell_{1}  \ell_{2}  \ell_{3}}\equiv \frac{1}{I_{\ell_{1} \ell_{2} \ell_{3}}}\sum_{m_1 m_2 m_3}\left(\begin{array}{ccc}
\ell_1 & \ell_2 & \ell_3 \\ m_1 & m_2 & m_3
\end{array}\right)\left(B^{(\LensCIB)}\right)_{\ell_1 \ell_2 \ell_3}^{m_1 m_2
  m_3} =  \frac{1}{I_{\ell_{1} \ell_{2} \ell_{3}}} \sum_{m_1 m_2 m_3}\left(\begin{array}{ccc}
\ell_1 & \ell_2 & \ell_3 \\ m_1 & m_2 & m_3
\end{array}\right) \tilde{a}^{(\CMB)}_{\ell_1 m_1} \tilde{a}^{(\CMB)}_{\ell_2 m_2} {a}^{(\CIB)}_{\ell_3 m_3},
\label{angle_averaged_bisp3}
\end{eqnarray}
${a}^{(\CIB)}_{\ell m}$ are the spherical harmonic coefficients of the
CIB map, $\tilde{a}^{(\CIB)}_{\ell m}$ are the spherical harmonic
coefficients of the CMB map, including the effect of the lensing up to
first order on the primordial ${a}^{(\CMB)}_{\ell m}$:
\begin{equation}
\tilde{a}^{(\CMB)}_{\ell m} =
a^{(\CMB)}_{\ell m} +\sum_{\ell' m' \ell'' m''}(-1)^m 
G_{\ell \ell' \ell''}^{m m' m''}
\Big[\frac{\ell'(\ell'+1)-\ell(\ell+1)+\ell''(\ell''+1)}{2}
a^{(\CMB)}_{\ell' m'}\phi_{\ell'' m''} \Big] \equiv a^{(\CMB)}_{\ell m} + \left(a \phi \right)_{\ell m}.
\label{almcmb2}
\end{equation}
The expected average of the CIB--lensing bispectrum is
\begin{eqnarray}
\nonumber
  \langle b^{(\LensCIB)}_{\ell_{1}  \ell_{2}  \ell_{3}} \rangle = b^{(\LensCIB)}_{\ell_{1}  \ell_{2}  \ell_{3}} = 
   \frac{\ell_1(\ell_1+1)-\ell_2(\ell_2+1)+\ell_3(\ell_3+1)}{2}C^{(\LensCIB,\nu)}_{\ell_3}\tilde{C}^{(\CMB)}_{\ell_1}  \\
    +\frac{\ell_2(\ell_2+1)-\ell_1(\ell_1+1)+\ell_3(\ell_3+1)}{2}C^{(\LensCIB,\nu)}_{\ell_3}\tilde{C}^{(\CMB)}_{\ell_2}.
\label{aa}
\end{eqnarray}
The correlation term in Eq. (\ref{the_covariance}) is given by:
\begin{eqnarray}
\langle b^{(\LensCIB)}_{\ell_1 \ell_2 \ell_3} b^{(\LensCIB)}_{\ell'_1 \ell'_2 \ell'_3}\rangle = \frac{1}{I_{\ell_{1} \ell_{2} \ell_{3}}I_{\ell'_{1} \ell'_{2} \ell'_{3}}} \sum_{m_1 m_2 m_3}\sum_{m'_1 m'_2 m'_3}\left(\begin{array}{ccc}
\ell_1 & \ell_2 & \ell_3 \\ m_1 & m_2 & m_3
\end{array}\right)\left(\begin{array}{ccc}
\ell'_1 & \ell'_2 & \ell'_3 \\ m'_1 & m'_2 & m'_3
\end{array}\right) \langle \tilde{a}^{(\CMB)}_{\ell_1 m_1} \tilde{a}^{(\CMB)}_{\ell_2 m_2} {a}^{(\CIB)}_{\ell_3 m_3}  \tilde{a}^{(\CMB)}_{\ell'_1 m'_1} \tilde{a}^{(\CMB)}_{\ell'_2 m'_2} {a}^{(\CIB)}_{\ell'_3 m'_3}\rangle.
\label{6order}
\end{eqnarray}
Considering 6 random variables, $x_1$, $x_2$, $x_3$, $x_4$, $x_5$,
$x_6$ with significant departures from Gaussianity due to their
bispectrum, and neglecting the trispectrum and higher order terms,
their sixth order moment is expanded as:
\begin{eqnarray}
\nonumber
& \langle x_1 x_2 x_3 x_4 x_5 x_6 \rangle  = \langle x_1 x_2 \rangle \langle x_3 x_4 \rangle \langle x_5 x_6 \rangle + \langle x_1 x_2 \rangle \langle x_3 x_5 \rangle \langle x_4 x_6 \rangle + \langle x_1 x_2 \rangle \langle x_3 x_6 \rangle \langle x_4 x_5 \rangle + \langle x_1 x_3 \rangle \langle x_2 x_5 \rangle \langle x_4 x_6 \rangle\\
\nonumber
&  + \langle x_1 x_3 \rangle \langle x_2 x_6 \rangle \langle x_4 x_5 \rangle + \langle x_1 x_3 \rangle \langle x_2 x_6 \rangle \langle x_4 x_5 \rangle + \langle x_1 x_4 \rangle \langle x_2 x_3 \rangle \langle x_5 x_6 \rangle + \langle x_1 x_4 \rangle \langle x_2 x_5 \rangle \langle x_3 x_6 \rangle + \langle x_1 x_4 \rangle \langle x_2 x_6 \rangle \langle x_3 x_5 \rangle \\
\nonumber
& + \langle x_1 x_5 \rangle \langle x_2 x_3 \rangle \langle x_4 x_6 \rangle + \langle x_1 x_5 \rangle \langle x_2 x_4 \rangle \langle x_3 x_6 \rangle +  \langle x_1 x_5 \rangle \langle x_2 x_6 \rangle \langle x_3 x_4 \rangle +  \langle x_1 x_6 \rangle \langle x_2 x_3 \rangle \langle x_4 x_5 \rangle +  \langle x_1 x_6 \rangle \langle x_2 x_4 \rangle \langle x_3 x_5 \rangle \\
\nonumber
& +  \langle x_1 x_6 \rangle \langle x_2 x_5 \rangle \langle x_3 x_4 \rangle \\
\nonumber
& + \langle x_1 x_2 x_3 \rangle \langle x_4 x_5 x_6 \rangle + \langle x_1 x_2 x_4 \rangle \langle x_3 x_5 x_6 \rangle +  \langle x_1 x_2 x_5 \rangle \langle x_3 x_5 x_6 \rangle +  \langle x_1 x_2 x_6 \rangle \langle x_3 x_4 x_5 \rangle +  \langle x_1 x_3 x_4 \rangle \langle x_2 x_5 x_6 \rangle\\
& +  \langle x_1 x_3 x_5 \rangle \langle x_2 x_4 x_6 \rangle + \langle x_1 x_3 x_6 \rangle \langle x_2 x_4 x_5 \rangle + \langle x_1 x_4 x_5 \rangle \langle x_2 x_3 x_6 \rangle + \langle x_1 x_4 x_6 \rangle \langle x_2 x_3 x_5 \rangle + \langle x_1 x_5 x_6 \rangle \langle x_2 x_3 x_4 \rangle 
\end{eqnarray}
Considering $\langle \tilde{a}^{(\CMB)}_{\ell_1 m_1} a^{(\CIB)}_{\ell_2
  m_2} {a}^{(\CIB)}_{\ell_3 m_3}\rangle = 0$, the correlation term
(Eq. \ref{6order}) is:
\begin{eqnarray}
\nonumber
 \langle b^{\LensCIB}_{\ell_1 \ell_2 \ell_3} b^{\LensCIB}_{\ell'_1
  \ell'_2 \ell'_3}\rangle & = &   
\frac{\tilde{C}^{(\CMB)}_{\ell_1}\tilde{C}^{(\CMB)}_{\ell_2}C^{(\CIB)}_{\ell_3}}{I_{\ell_{1} \ell_{2} \ell_{3}}I_{\ell'_{1} \ell'_{2} \ell'_{3}}}\left( 
  \delta_{\ell_1 \ell'_1}\delta_{\ell_2 \ell'_2}\delta_{\ell_3
    \ell'_3} + \delta_{\ell_1 \ell'_2}\delta_{\ell'_1
    \ell_2}\delta_{\ell_3 \ell'_3}
\right)+\\
 & + &  \frac{\langle b^{\LensCIB}_{\ell_1 \ell_2 \ell_3} \rangle 
\langle b^{\LensCIB}_{\ell'_1 \ell'_2 \ell'_3}\rangle}{I_{\ell_{1} \ell_{2} \ell_{3}}I_{\ell'_{1} \ell'_{2} \ell'_{3}}}\left(1+ 
{\delta_{\ell_3\ell'_3} \over 2\ell_3+1}+ 
{\delta_{\ell'_1\ell_2} \over 2\ell_2+1}+ 
{\delta_{\ell_2\ell'_2} \over 2\ell_2+1}+
{\delta_{\ell_1\ell'_2} \over 2\ell_1+1} + 
{\delta_{\ell_1\ell'_1} \over 2\ell_1+1}\right), 
\label{6order2}
\end{eqnarray}
and therefore, the covariance of the bispectrum is:
\begin{eqnarray}
\nonumber
\mathtt{cov}\left(b^{(\LensCIB)}_{\ell_1 \ell_2 \ell_3},
  b^{(\LensCIB)}_{\ell'_1 \ell'_2 \ell'_3}\right) & = & \frac{\tilde{C}^{(\CMB)}_{\ell_1}\tilde{C}^{(\CMB)}_{\ell_2}C^{(\CIB)}_{\ell_3}}{I_{\ell_{1} \ell_{2} \ell_{3}}I_{\ell'_{1} \ell'_{2} \ell'_{3}}}\left( \delta_{\ell_1 \ell'_1}\delta_{\ell_2 \ell'_2}\delta_{\ell_3 \ell'_3} + \delta_{\ell_1 \ell'_2}\delta_{\ell_1 \ell'_2}\delta_{\ell_3 \ell'_3}\right)+
\\
& + &  \frac{\langle b^{\LensCIB}_{\ell_1 \ell_2 \ell_3} \rangle 
\langle b^{\LensCIB}_{\ell'_1 \ell'_2 \ell'_3}\rangle}{I_{\ell_{1} \ell_{2} \ell_{3}}I_{\ell'_{1} \ell'_{2} \ell'_{3}}}\left( 
{\delta_{\ell_3\ell'_3} \over 2\ell_3+1}+ 
{\delta_{\ell'_1\ell_2} \over 2\ell_2+1}+ 
{\delta_{\ell_2\ell'_2} \over 2\ell_2+1}+
{\delta_{\ell_1\ell'_2} \over 2\ell_1+1} + 
{\delta_{\ell_1\ell'_1} \over 2\ell_1+1}\right), 
\label{6order3}
\end{eqnarray}
We consider the diagonal and non diagonal terms
\begin{eqnarray}
 \mathtt{cov}\left(b^{(\LensCIB)}_{\ell_1 \ell_2 \ell_3},
  b^{(\LensCIB)}_{\ell_1 \ell_2 \ell_3}\right) &=&
 \frac{2\tilde{C}^{(\CMB)}_{\ell_1}\tilde{C}^{(\CMB)}_{\ell_2}C^{(\CIB)}_{\ell_3}}{I_{\ell_{1} \ell_{2} \ell_{3}}^2}  + 
\frac{\langle b^{\LensCIB}_{\ell_1 \ell_2 \ell_3} \rangle \langle
b^{\LensCIB}_{\ell_1 \ell_2 \ell_3}\rangle}{I_{\ell_{1} \ell_{2} \ell_{3}}^2}\left({2 \over 2\ell_1+1}+
{2 \over 2\ell_2+1}+{1 \over 2\ell_3+1}\right)\,,\nonumber\\
 \mathtt{cov}\left(b^{(\LensCIB)}_{\ell_1 \ell_2 \ell_3},
  b^{(\LensCIB)}_{\ell'_1 \ell'_2 \ell'_3}\right) &=& 
\frac{\langle B^{\LensCIB}_{\ell_1 \ell_2 \ell_3} \rangle 
\langle B^{\LensCIB}_{\ell'_1 \ell'_2 \ell'_3}\rangle}{I_{\ell_{1} \ell_{2} \ell_{3}}I_{\ell'_{1} \ell'_{2} \ell'_{3}}} \left( 
{\delta_{\ell_3\ell'_3} \over 2\ell_3+1}+ 
{\delta_{\ell'_1\ell_2} \over 2\ell_2+1}+ 
{\delta_{\ell_2\ell'_2} \over 2\ell_2+1}+
{\delta_{\ell_1\ell'_2} \over 2\ell_1+1} + 
{\delta_{\ell_1\ell'_1} \over 2\ell_1+1}\right)\,.
\label{6order4B}
\end{eqnarray}
We have estimated that 
\begin{equation}
\frac{\langle b^{(\LensCIB)}_{\ell_1 \ell_2 \ell_3} \rangle^2I_{\ell_{1} \ell_{2} \ell_{3}}^2}{\tilde{C}^{(\CMB)}_{\ell_1}\tilde{C}^{(\CMB)}_{\ell_2}C^{(\CIB)}_{\ell_3}}\le 10^{-5}, 
\label{ineq1}
\end{equation}
depending on the frequency (see Table \ref{table}). Furthermore
\begin{equation}
\frac{\langle b^{(\LensCIB)}_{\ell_1 \ell_2 \ell_3} \rangle \langle b^{(\LensCIB)}_{\ell'_1 \ell'_2 \ell'_3} \rangle I_{\ell_{1} \ell_{2} \ell_{3}}I_{\ell'_{1} \ell'_{2} \ell'_{3}}}{\sqrt{\tilde{C}^{(\CMB)}_{\ell_1}\tilde{C}^{(\CMB)}_{\ell_2}C^{(\CIB)}_{\ell_3}\tilde{C}^{(\CMB)}_{\ell'_1}\tilde{C}^{(\CMB)}_{\ell'_2}C^{(\CIB)}_{\ell'_3}}}\le 10^{-5}, 
\label{ineq2}
\end{equation}
and the same replacing in Eq. (\ref{ineq2}) any of the pairs of
bispectra given in Eq. (\ref{6order4B}). This means that the off
diagonal elements of the covariance matrix are small, compared to the
diagonal. If we normalise it by the diagonal, it is then of the form
$\widehat{\mathtt{cov}} = \mathbb{1} +\varepsilon$ where $\mathbb{1}$
is the unit matrix and the hat denotes the normalisation,
$\widehat{\mathtt{cov}}_{ij} =
\mathtt{cov}_{ij}/\sqrt{\mathtt{cov}_{ii}\mathtt{cov}_{jj}}$. The
inverse can be approximated by a series expansion,
\begin{equation}
\widehat{\mathtt{cov}}^{-1} = (\mathbb{1} +\varepsilon)^{-1} = \mathbb{1} - \varepsilon + \mathcal{O}\left( \varepsilon^2 \right) ,
\end{equation}
i.e.\ the off diagonal elements of the inverse of the normalised
covariance matrix are also small, of the same order as the off
diagonal elements of the normalised covariance matrix. To understand
whether the off diagonal elements can become relevant we can thus look
at $\mathbb{1} - \varepsilon$, or since the signs of the off diagonal
elements are fairly random, directly at the covariance matrix, without
the need to perform an explicit inversion of an impossibly large
matrix.  Now, {\em if} the signs of the off diagonal elements are
random, then their sum grows only like the square-root of the number
of elements.  So even though there are many more off diagonal terms
than diagonal terms, their total contribution should still remain
sub-dominant.  To test whether this is the case, we have summed up the
off diagonal terms of the normalised matrix along rows, for $\ell_{\rm
  max}=2000$.  We find that the sum is always smaller than $0.1$. The
mean value of the sums is of the order of $10^{-5}$, with a standard
deviation of $1.2\times 10^{-3}$, which supports the assumption that
the signs are relatively random, and shows that the majority of the
row sums are small (indeed, 95\% are smaller, in absolute terms, than
$1.5\times 10^{-3}$ as the distribution of row-sum values is quite
non-Gaussian).  Therefore, the covariance can be approximated by the
Gaussian part:
\begin{eqnarray}
  \nonumber
  & \mathtt{cov}\left(b^{(\LensCIB)}_{\ell_1 \ell_2 \ell_3}, b^{(\LensCIB)}_{\ell'_1 \ell'_2 \ell'_3}\right) \simeq \frac{\tilde{C}^{(\CMB)}_{\ell_1}\tilde{C}^{(\CMB)}_{\ell_2}C^{(\CIB)}_{\ell_3}}{I_{\ell_{1} \ell_{2} \ell_{3}}I_{\ell'_{1} \ell'_{2} \ell'_{3}}}\left( \delta_{\ell_1 \ell'_1}\delta_{\ell_2 \ell'_2}\delta_{\ell_3 \ell'_3} + \delta_{\ell_1 \ell'_2}\delta_{\ell_1 \ell'_2}\delta_{\ell_3 \ell'_3}\right).
\label{the_covariance3}
\end{eqnarray}
\begin{table*}
  \begin{center}
    \caption{Maximum value for the ratio of the CIB--lensing
      bispectrum and its dispersion, $\max \Bigg( \frac{\langle B^{(\LensCIB)}_{\ell_1 \ell_2 \ell_3} \rangle^2}{\left( 1 + \delta_{\ell_1 \ell_2}\right)\tilde{C}^{(\CMB)}_{\ell_1}\tilde{C}^{(\CMB)}_{\ell_2}C^{(\CIB)}_{\ell_3}} \Bigg)$,
      for the \Planck raw maps at different frequencies including
      instrumental beams and noise. \label{table}}
    \begin{tabular}{c|cccccc}
      \hline
      \hline
      Frequency (GHz) & $\nu_{\CMB}$ = 100 & $\nu_{\CMB}$ = 143 & $\nu_{\CMB}$ = 217 & $\nu_{\CMB}$ = 353 & $\nu_{\CMB}$ = 545 & $\nu_{\CMB}$ = 857 \\
      \hline
       $\nu_{\CIB}$ = 143 & 4.49 $\times 10^{-6}$ & 4.50 $\times 10^{-6}$ & 4.49 $\times 10^{-6}$ & 4.46 $\times 10^{-6}$ & 2.49 $\times 10^{-6}$ & 2.61 $\times 10^{-8}$\\
       $\nu_{\CIB}$ = 217 & 8.84 $\times 10^{-6}$ & 9.74. $\times 10^{-6}$ & 9.47 $\times 10^{-6}$ & 8.77 $\times 10^{-6}$ & 4.46 $\times 10^{-6}$ & 4.38 $\times 10^{-8}$\\
       $\nu_{\CIB}$ = 353 & 1.45 $\times 10^{-5}$ & 2.25 $\times 10^{-5}$ & 2.19 $\times 10^{-5}$ & 1.26 $\times 10^{-5}$ & 5.92 $\times 10^{-6}$ &  5.20 $\times 10^{-8}$\\
       $\nu_{\CIB}$ = 545 & 1.29 $\times 10^{-5}$ & 1.83 $\times 10^{-5}$ & 1.78 $\times 10^{-5}$ & 1.18 $\times 10^{-5}$ & 5.94 $\times 10^{-6}$ & 5.64 $\times 10^{-8}$\\
       $\nu_{\CIB}$ = 857 & 1.36 $\times 10^{-5}$ & 2.14 $\times 10^{-5}$ & 2.08 $\times 10^{-5}$ & 1.17 $\times 10^{-5}$ & 4.76 $\times 10^{-6}$ & 5.47 $\times 10^{-8}$\\
      \hline
      \hline
    \end{tabular}
  \end{center}
\end{table*}
\end{appendix}
\end{document}